\documentclass[twocolumn]{aastex631}

\usepackage{amsmath}	
\usepackage{amssymb}	
\DeclareRobustCommand{\ion}[2]{%
\relax\ifmmode
\ifx\testbx\f@series
{\mathbf{#1\,\mathsc{#2}}}\else
{\mathrm{#1\,\mathsc{#2}}}\fi
\else\textup{#1\,{\mdseries\textsc{#2}}}%
\fi}

\usepackage{rotating}

\usepackage{graphicx}	
\usepackage{subfigure}

\received{\today}
\revised{\today}
\accepted{--}

\submitjournal{ApJS}

\shorttitle{DRAGNs in VLASS}
\shortauthors{Y.~A. Gordon et al.}
\graphicspath{{./}{}}

\begin{document}

\title{A \textit{Quick Look} at the $3\,$GHz Radio Sky.
II.
Hunting for DRAGNs in the VLA Sky Survey}

\correspondingauthor{Yjan~A. Gordon}
\email{yjan.gordon@wisc.edu}

\author[0000-0003-1432-253X]{Yjan~A. Gordon}
\affil{Department of Physics, University of Wisconsin-Madison, 
1150 University Ave, Madison, WI 53706, USA}


\author[0000-0001-5636-7213]{Lawrence Rudnick}
\affiliation{Minnesota Institute for Astrophysics, School of Physics and Astronomy, University of Minnesota, 116 Church Street SE,\\Minneapolis, MN 55455, USA}

\author[0000-0003-4873-1681]{Heinz Andernach}
\affiliation{Th\"uringer Landessternwarte, Sternwarte 5, D-07778 Tautenburg, Germany}
\affiliation{Permanent Address: Depto. de Astronomia, Univ. de Guanajuato,
Callej\'on de Jalisco s/n, Guanajuato, C.P. 36023, GTO, Mexico}

\author[0000-0003-0487-6651]{Leah~K. Morabito}
\affiliation{Centre for Extragalactic Astronomy, Department of Physics, Durham University, South Road, Durham, DH1 3LE, UK}
\affiliation{Institute for Computational Cosmology, Department of Physics, Durham University, South Road, Durham DH1 3LE, UK}

\author[0000-0001-6421-054X]{Christopher~P. O'Dea}
\affiliation{Department of Physics and Astronomy, University of Manitoba, 
Winnipeg, MB R3T 2N2, Canada}

\author[0000-0002-7726-2792]{Kaylan-Marie Achong}
\affiliation{Department of Physics, University of the West Indies, St. Augustine, Trinidad and Tobago}
\altaffiliation{Kaylan-Marie Achong and Caryelis Bayona-Figueroa are summer students at the National Radio Astronomy Observatory}
\affiliation{Department of Astronomy, University of Wisconsin-Madison, 
475 N. Charter St., Madison, WI 53703, USA}

\author[0000-0002-4735-8224]{Stefi~A. Baum}
\affiliation{Department of Physics and Astronomy, University of Manitoba, 
Winnipeg, MB R3T 2N2, Canada}

\author{Caryelis Bayona-Figueroa}
\affiliation{Departmento de Ciencias F\'isicas, Universidad de Puerto Rico, P.O. Box 23323, San Juan, PR 00931-3323, USA}
\altaffiliation{Kaylan-Marie Achong and Caryelis Bayona-Figueroa are summer students at the National Radio Astronomy Observatory}
\affiliation{Department of Astronomy, University of Wisconsin-Madison, 
475 N. Charter St., Madison, WI 53703, USA}

\author[0000-0003-0713-3300]{Eric~J. Hooper}
\affiliation{Department of Astronomy, University of Wisconsin-Madison, 
475 N. Charter St., Madison, WI 53703, USA}

\author[0000-0001-5649-938X]{Beatriz Mingo}
\affiliation{School of Physical Sciences, The Open University, Walton Hall, Milton Keynes, MK7 6AA, UK}

\author[0000-0001-9920-0210]{Melissa~E. Morris}
\affiliation{Department of Astronomy, University of Wisconsin-Madison, 
475 N. Charter St., Madison, WI 53703, USA}

\author[0000-0003-4227-4838]{Adrian~N. Vantyghem}
\affiliation{Department of Physics and Astronomy, University of Manitoba, 
Winnipeg, MB R3T 2N2, Canada}




\begin{abstract}
Active Galactic Nuclei (AGN) can often be identified in radio images as two lobes, sometimes connected to a core by a radio jet.
This multi-component morphology unfortunately creates difficulties for source-finders, leading to components that are a) separate parts of a wider whole, and b) offset from the multiwavelength cross identification of the host galaxy.
In this work we define an algorithm, \textsc{DRAGNhunter}, for identifying Double Radio Sources associated with Active Galactic Nuclei (DRAGNs) from component catalog data in the first epoch \textit{Quick Look} images of the high resolution ($\approx 3''$ beam size) Very Large Array Sky Survey (VLASS).
We use \textsc{DRAGNhunter} to construct a catalog of $>17,000$ DRAGNs in VLASS for which contamination from spurious sources is estimated at $\approx 11\,\%$.
A `high-fidelity' sample consisting of $90\,\%$ of our catalog is identified for which contamination is $<3\,\%$.
Host galaxies are found for $\approx 13,000$ DRAGNs as well as for an additional $234,000$ single-component radio sources.
Using these data we explore the properties of our DRAGNs, finding them to be typically consistent with Fanaroff-Riley class II sources and allowing us to report the discovery of $31$ new giant radio galaxies identified using VLASS.
\end{abstract}

\keywords{
\href{http://astrothesaurus.org/uat/1338}{Radio Astronomy (1338)},
\href{http://astrothesaurus.org/uat/1343}{Radio Galaxies (1343)},
\href{http://astrothesaurus.org/uat/508}{Extragalactic radio sources (508)},
\href{http://astrothesaurus.org/uat/654}{Giant radio galaxies (654)},
\href{http://astrothesaurus.org/uat/16}{Active galactic nuclei (16)}}


\vspace{10mm}
\section{Introduction}
\label{sec:intro}

Active Galactic Nuclei (AGN), where the accretion of matter onto a galaxy's central supermassive black hole is readily detectable, represent a key phase in a galaxy's evolution \citep[e.g.,][]{Ferrarese2000, Croton2006, Harrison2017}.
A few percent of AGN are able to launch two oppositely directed jets of relativistic plasma which produce powerful radio emission \citep[e.g.,][]{Padovani2017,Blandford2019, Hardcastle2020}.
These radio-loud AGN (RLAGN) have radio properties which likely depend on the black hole mass, spin and matter accretion rate \citep[e.g.,][]{Blandford2019}, as well as the interaction of the radio source with the gaseous environment that it propagates through \citep[e.g.,][]{Miley1980,Heinz1998, Hubbard2006, Sutherland2007, Morganti2021, ODea2021}. 

Early on, \citet{Fanaroff1974} observed a dichotomy in the morphology of extended radio emission from AGN, which they used to classify RLAGN.
Class I objects (FR Is) have their radio brightness peaks close to the nucleus, while class II objects (FR IIs) are `edge-brightened' with their brightness peaks being closer to the leading edge of the jet.
Subsequent developments in both observations \citep[e.g.,][]{Bridle1984} and theory \citep[e.g.,][]{Bicknell1985} have led to the following paradigm. 
The jets in lower radio power FR Is interact strongly with their environments and decelerate to non-relativistic velocities on scales of a few kpc \citep[e.g.,][]{Bicknell1985,Laing2014}. 
Because the jets in FR Is are non-relativistic on kpc scales, Doppler boosting effects are minimal and both jets are observed. 
The jets expand as they propagate outwards and become diffuse plumes, resulting in the brightness peaks being near the nucleus.
On the other hand, in the FR II sources the jets do not interact as strongly with the environment and remain relativistic all the way out  to the terminal shock at the end of the jet \citep[e.g.,][]{Laing1988,Garrington1988}. 
The Doppler boosting that occurs when the axis is close to the line of sight results in the jet pointing towards the observer being brighter than that in the opposite direction.
Conversely, when the jet axis is closer to the plane of the sky neither jet is significantly Doppler boosted with respect to the observer and often no or faint jets are seen. 
The terminal shock (or working surface) produces bright radio emission (called a hotspot) and thus the source brightness peaks near the outer edges of the source. 
At the hotspot, the jet plasma spreads sideways inflating a radio lobe \citep[e.g.,][]{Blandford1974,Norman1982, Begelman1989} producing the `classical double' radio structure. 
These double-lobed radio galaxies have become known as Double Radio sources associated with Active Galactic Nuclei, or DRAGNs \citep{Leahy1993, Leahy1995}.

The current era of blind radio continuum surveys over large swathes of the sky is producing observations of millions of radio galaxies \citep[][]{Condon1998, Norris2011, Norris2017, Gordon2021, Hale2021, Shimwell2022}.
The majority of RLAGN in blind surveys will appear as compact sources as a consequence of the limited angular resolution of the surveys and the predominance of intrinsically compact radio sources \citep{ODea1997, Reynolds1997, Alexander2000}.
Nonetheless, given the millions of RLAGN that will be observed, a large number of DRAGNs can be expected in survey imaging.

\begin{figure}
    \centering
    \subfigure{\includegraphics[width=0.7\columnwidth]{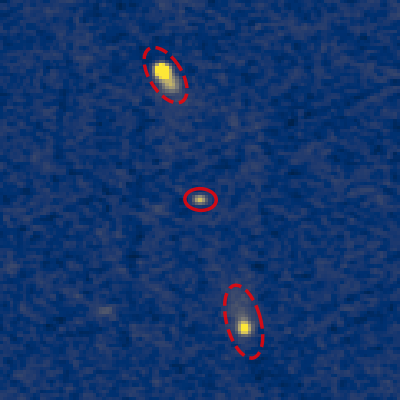}}
    \caption{Example of how double and triple radio sources can be split into multiple detections in radio `source' catalogs.
    The VLASS image of the source is shown by the blue/yellow colormap, with the cataloged components marked by red ellipses.
    The ellipse geometry in this figure is defined by the \textit{fitted} component geometry from the catalog of \citet{Gordon2021} with the major and minor axes multiplied by a factor of three to aid visibility in this figure.
    The ellipses with dashed lines show components that are resolved by VLASS, while the ellipse shown with a solid line is a point source and has a deconvolved size of zero.}
    \label{fig:multicomp}
\end{figure}

Identifying DRAGNs in survey data presents unique challenges.
Cataloging `sources' in any astronomical imaging is typically achieved with a `source-finding' algorithm that looks for regions of intensity above a predefined threshold (e.g. $5\times$ the rms noise).
In the case of DRAGNs however--and indeed more complex radio morphologies--this approach can lead to the two lobes being identified as separate `sources' even though they belong to the same physical object.
We show a textbook example of this complication in Figure \ref{fig:multicomp} using an image from the Karl G. Jansky Very Large Array (VLA) Sky Survey \citep[VLASS,][]{Lacy2020} and the associated catalog \citep{Gordon2021}.
In this particular example, the DRAGN is modelled as three distinct `sources' in the catalog (red ellipses in Figure \ref{fig:multicomp}).

Identification of DRAGNs in survey imaging is generally dependent on visual inspection \citep{Banfield2015, Vardoulaki2021, Gurkan2022}, or, increasingly, on machine learning \citep{Wu2019, Galvin2019, Scaife2021}.
The former approach is well suited to small data sets, but can be impractical for the large data sets of current- and next-generation wide-field surveys.
The latter approach is often dependent on large training sets of pre-identified and labelled morphological classifications.
Moreover, it is often necessary to specifically tailor machine learning algorithms to each data set as result of different survey characteristics, e.g., angular resolution, observing frequency, noise levels, etc.
While DRAGNs may fool source-finders, their morphology is still relatively simple, and as such one can imagine developing algorithms based on grouping `sources' together to identify DRAGNs.
Indeed, a number of such algorithms have been used to identify complex-morphology radio sources using survey catalog data \citep[e.g.][]{Magliocchetti1998, Best2005, Sadler2007, Ching2017}.
In this paper we define a new algorithm to detect DRAGNs specifically (as the `simplest' complex morphological type) in the VLASS catalog data with minimal contamination from more complex or unrelated sources.
This algorithm, which we are calling \textsc{DRAGNhunter}, is then used to build a catalog of DRAGNs in VLASS.
We combine this catalog with multiwavelength data to identify the likely host galaxies of these DRAGNs.

The layout of this paper is as follows.
In Section \ref{sec:data} we describe the catalog data we use and the \textsc{DRAGNhunter} algorithm with which we identify DRAGNs.
Section \ref{sec:relcomp} describes the reliability and completeness of our catalog of DRAGNs.
We identify host galaxies and redshifts for our DRAGNs where possible in Section \ref{sec:hosts}.
In Section \ref{sec:dragnproperties} we use the catalog we have produced to explore the general properties of DRAGNs in VLASS, with a focus on triple sources in Section \ref{sec:triples}.
A summary of this paper and a discussion of future work is presented in Section \ref{sec:summary}.
The data model of the catalog accompanying this article is described in Appendix \ref{apx:data-model}.
In order to differentiate between catalog entries and physical sources, throughout the rest of this paper we use the nomenclature \textit{`component'} to refer to a single  detection from a source-finder, and reserve \textit{`source'} to mean the physical object.
For example, the \textit{source} shown in Figure \ref{fig:multicomp} is a DRAGN composed of three \textit{components}.
Where referring to spectral index, $\alpha$, we use the convention where spectral index is related to flux density, $S$, by $S_{\nu} \propto \nu^{\alpha}$. A flat $\Lambda$CDM cosmology is adopted throughout, with: 
$h = 0.7$,
$H_{0} = 100\,h\,\text{km}\,\text{s}^{-1}\,\text{Mpc}^{-1}$,
$\Omega_{m}=0.3$
and $\Omega_{\Lambda}=0.7$.

\section{Identifying DRAGNs from Radio Component Data} \label{sec:data}

\subsection{VLASS Catalog Data}

VLASS is an ongoing survey to provide multi-epoch $\nu \sim 3\,$GHz mapping of the entire sky north of $-40^{\circ}$ in declination at high resolution \citep{Lacy2020}.
The first epoch of VLASS was completed in 2019, and rapidly-produced \textit{Quick Look} images covering $\approx34,000\,\text{deg}^{2}$ are publicly available.
These \textit{Quick Look} images have a typical rms noise level of $140\,\mu$Jy/beam \citep{Gordon2021}.
For this work we make use of the catalog of components in the \textit{Quick Look} images of VLASS epoch 1 presented in \citet{Gordon2021}.
Following the recommendations in Section 3 of \citet{Gordon2021}, we only consider components satisfying $\texttt{S\_Code}\neq \text{`E'}$, $\texttt{Quality\_flag}==(0|4)$ and $\texttt{Duplicate\_flag}<2$.
These criteria are designed to limit contamination by spurious detections arising from the limited quality of the VLASS \textit{Quick Look} images, as well potential duplicates resulting from overlaps between images \citep[for a full discussion of these criteria see][]{Gordon2021}.

The median beam size of VLASS in epoch 1 is $2.''9$, the smallest of any near-all-sky radio continuum survey to date.
While a number of narrower field surveys use smaller beams, e.g., the VLA-COSMOS surveys \citep{Schinnerer2004, Schinnerer2007, Smolcic2017} and the LOFAR-deep high-definition fields \citep{Sweijen}, such very-high-resolution projects only cover of the order a few square degrees of the sky.
The combination of high angular resolution and near-all-sky coverage of VLASS makes the survey ideally suited to identifying large numbers of sources in the radio sky and differentiating those that are genuinely compact from sources that have extended radio morphologies.

Unless dominated by hotspot emission, the lobes of DRAGNs are expected to have extended radio morphologies, rather than appear point-like in high resolution imaging.
The ability of VLASS to cleanly differentiate compact and extended radio morphologies can thus be exploited to find radio detections that are more likely to be a radio lobe than a radio core. Furthermore, extended-morphology radio sources generally have steeper radio spectra than the point-like sources, likely resulting from the contribution of flat-spectrum radio cores to the unresolved source population  \citep{Gordon2021, Norris2021}; this can be utilized with future releases of VLASS single epoch images \citep{Lacy2022}.

\subsection{Finding Pairs of Lobes}
\label{ssec:lobepairs}

\begin{figure}
    \centering
    \includegraphics[width=0.99\columnwidth]{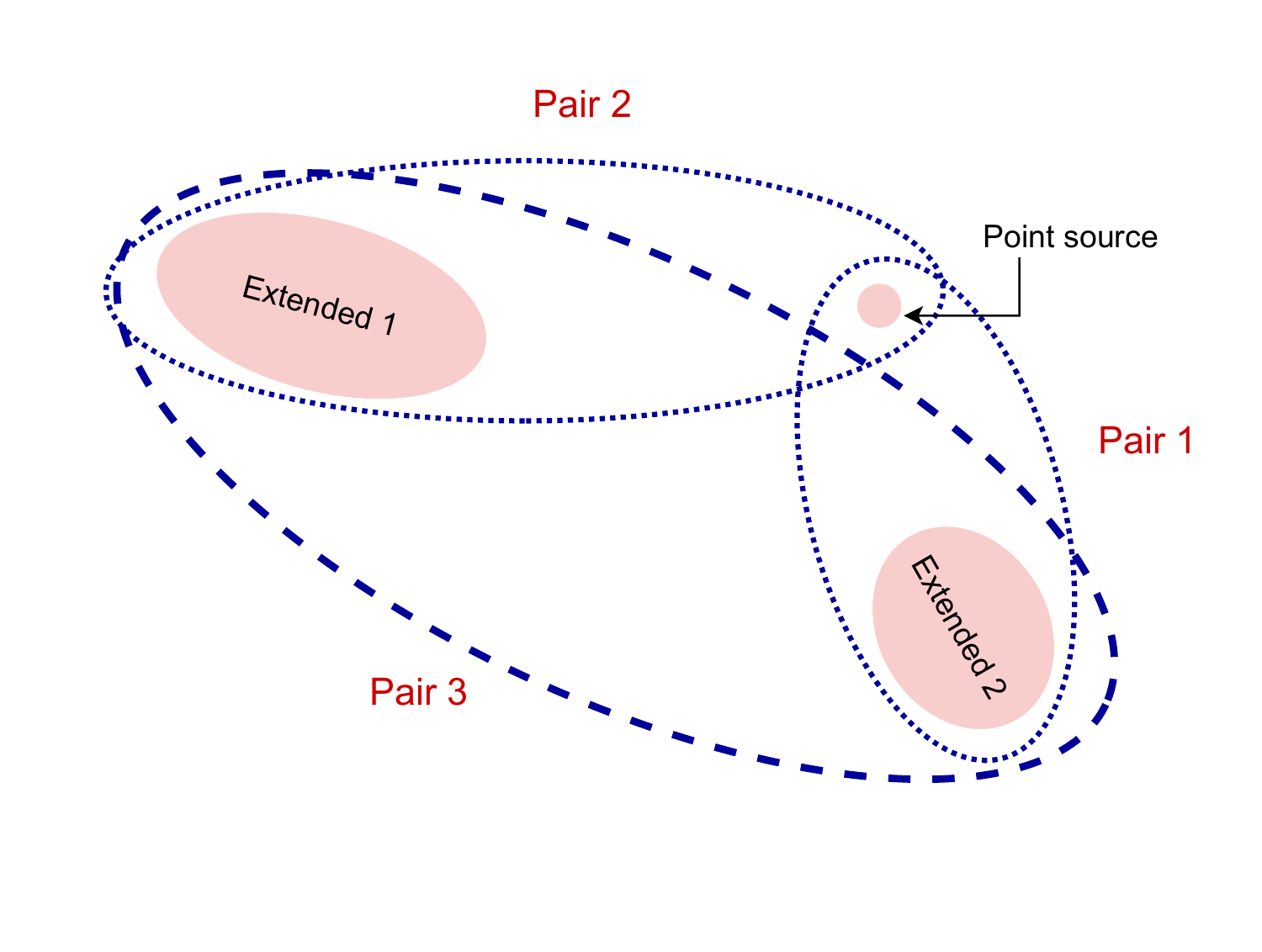}
    \caption{A representation of the approach \textsc{DRAGNhunter} takes to looking for radio doubles.
    Three model radio components are shown in this schematic, two of which are extended, while one is a point source at the resolution of the image.
    In this scenario the two extended components may represent two lobes of a radio galaxy and the point source may be an associated core or entirely unrelated.
    If one were to just search for the nearest neighbour blindly, then the two pairs represented by the dotted blue ellipses (Pair 1 and Pair 2) would be selected.
    However, by excluding point sources from the pair finding \textsc{DRAGNhunter} will select Pair 3 (blue dashed ellipse) instead.}
    \label{fig:pair_schematic}
\end{figure}

The algorithm we use to search for DRAGNs, \textsc{DRAGNhunter}, is primarily searching for nearest-neighbor pairs of likely radio \textit{lobes}, rather than just any pairing of detected radio \textit{components}.
This distinction is important as searching for just pairs of radio \textit{components} will likely result in some pairings that do not represent the full radio source (e.g. a pair consisting of a radio core and a radio lobe, see Figure \ref{fig:pair_schematic}) or that are completely unrelated to one another.
Whilst it is impossible to select only radio lobes just from radio component catalog data, the component geometry can be used to find those components that are more extended, and thus more likely to represent radio lobe detections.

In \citet{Gordon2021} we showed that reliability of the flux density measurements in the VLASS quick look component catalog lessens at $S_{\text{peak}}<3\,$mJy/beam.
Therefore we only include components brighter than $3\,$mJy/beam in this work.
We wish to identify extended components as candidate lobes.
Given the median beam size in VLASS epoch 1 is $2.''9$, we consider components with a major axis after deconvolution from the beam, $\Psi$, greater than $3''$ to be cleanly extended.
While the subtraction of the beam from the image would allow us to measure extents well below the beam size, this of course comes with an increase in the relative uncertainty in the measured size -- the mean relative error in angular size for components with $\Psi<3''$ is $\approx 7\,\%$, whereas for those with $\Psi>3''$ the mean relative error is $\approx 4\,\%$.
Larger components thus represent a conservative selection of candidate lobes where the flux density measurement is considered reliable and extent of the component is cleanly resolved.
Naturally, relaxing these criteria would allow for the detection of more DRAGNs but likely at lower confidence (see Section \ref{sec:relcomp}).
A search for pairs of such lobe candidates returns $80,325$ unique `nearest neighbour' pairs, i.e., the same pair is not repeated with the component order swapped.
In cases where a component is associated with multiple pairs by virtue of being the nearest neighbor of at least one other candidate lobe, we flag the pair with the smallest angular separation as the preferred pair, resulting in a sample of $72,832$ pairs of candidate lobes. 

\begin{figure}
    \centering
    \includegraphics[width=0.99\columnwidth]{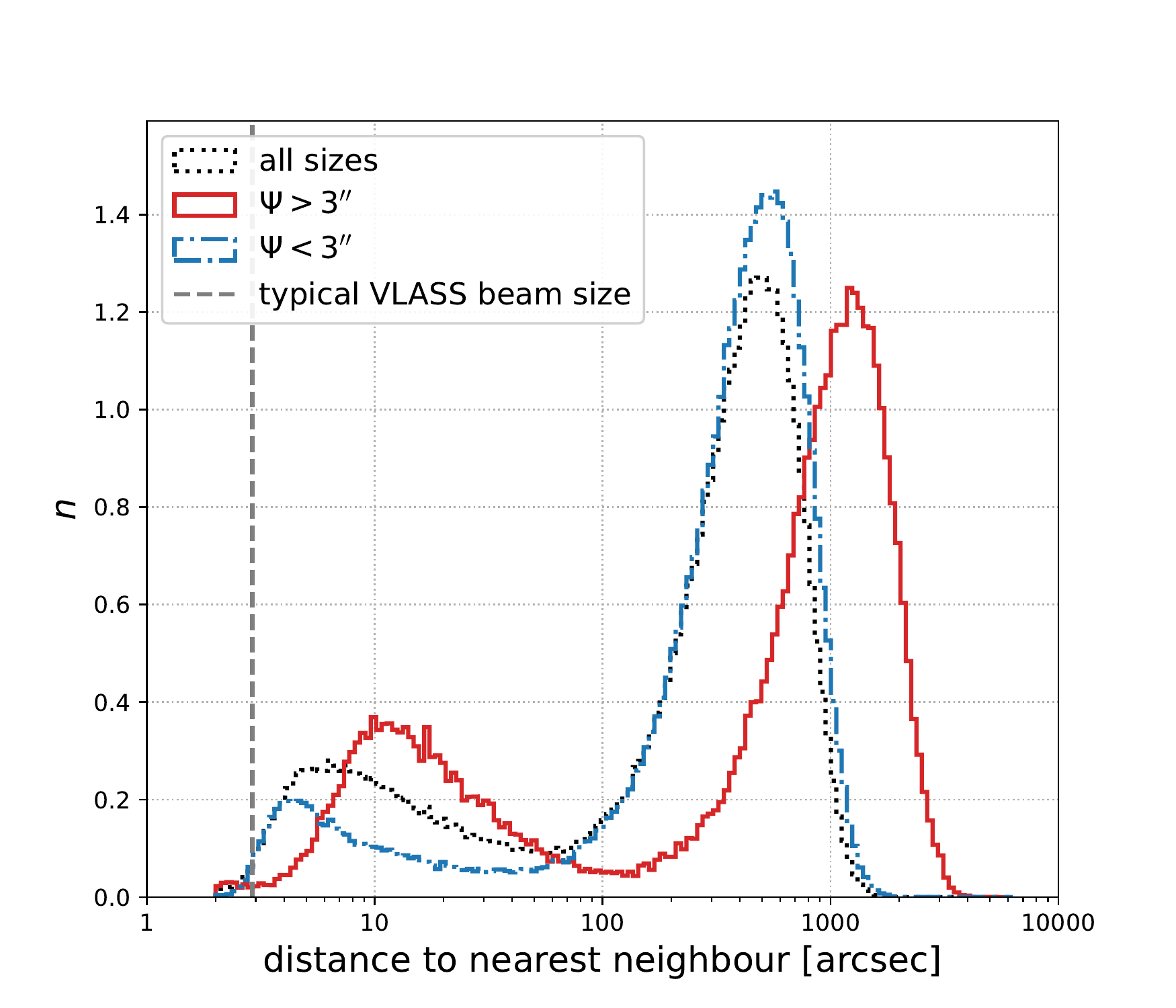}
    \caption{Distribution of angular distance between the nearest neighbour for three different populations of VLASS components with $S_{\text{peak}}>3\,$mJy/beam.
    The red solid line shows those with a deconvolved angular size, $\Psi$, $> 3''$ (candidate lobes).
    For comparison the blue dot-dashed line shows the distribution for components with $\Psi < 3''$ (not considered as candidate lobes).
    The black dotted line shows the nearest neighbor distances without applying a size cut, and the grey dashed vertical line represents the typical VLASS beam size.
    }
    \label{fig:pair_seps}
\end{figure}

In Figure \ref{fig:pair_seps} we show the distribution of angular separation to the nearest neighbour for three different selections of components with $S_{\text{peak}}>3\,$mJy/beam.
The black dotted line shows components of all sizes, the blue dash-dotted line shows components with $\Psi < 3''$ and the red solid line shows components with $\Psi > 3''$.
All three populations show a clear bimodal distribution with a peak at large angular separations that is dominated by random VLASS detections and a peak at smaller angular separations that mostly results from genuinely associated radio components \citep[see also the radio two-point correlation fucntion, e.g.,][]{Cress1996, Blake2002, Gordon2021}.
Notably, the small angular separation peak constitutes a larger fraction of the population for extended components than it does for compact components.
The relative positions of distribution peaks in Figure \ref{fig:pair_seps} is an effect of source density--lower on-sky densities will drive the nearest neighbour distributions to higher angular separations. 

\begin{figure}
    \centering
    \includegraphics[width=0.99\columnwidth]{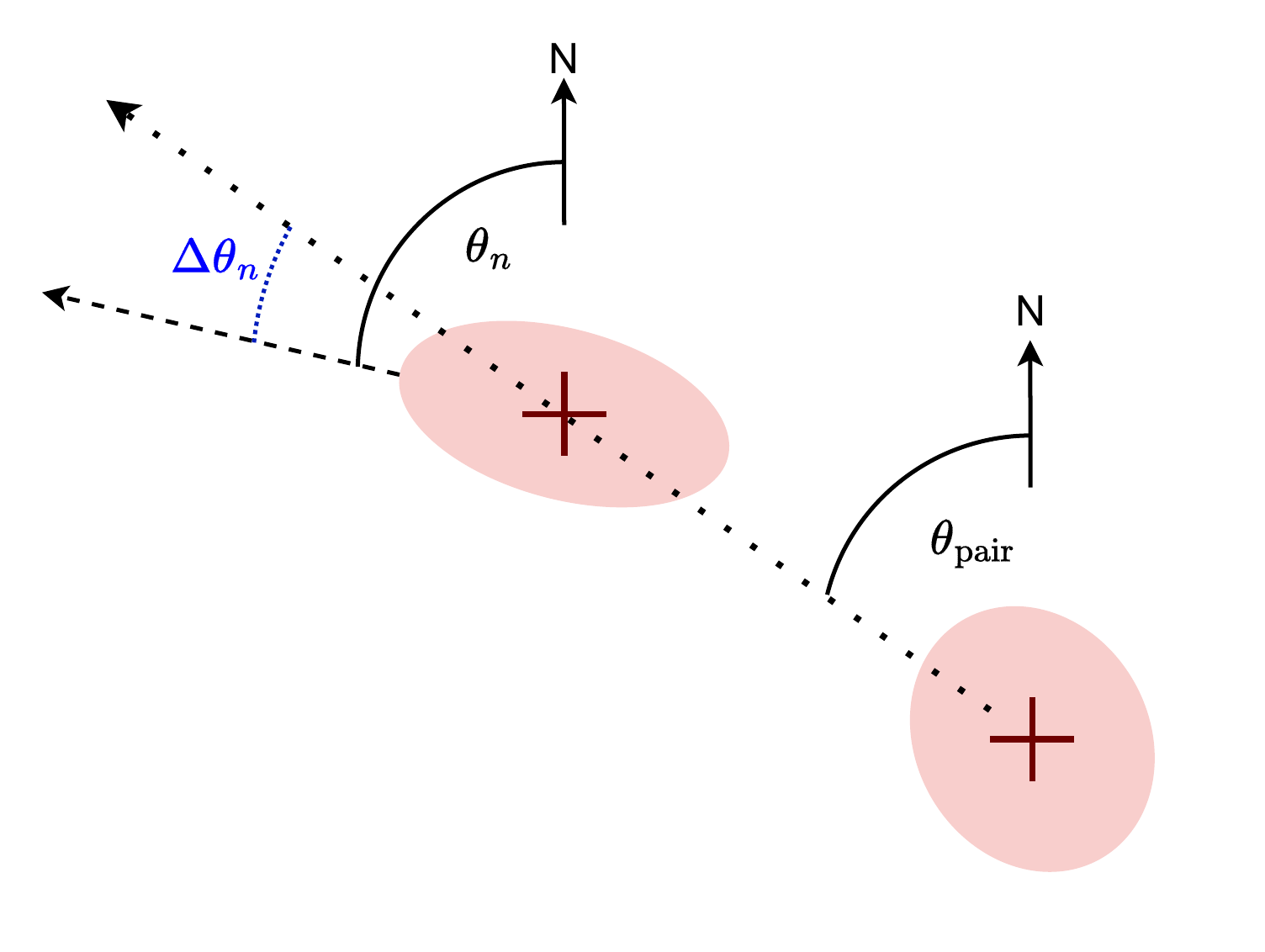}
    \caption{Schematic showing how component (mis)alignment is determined by \textsc{DRAGNhunter}.
    For any pair of components, the position angle of the axis of the pair, $\theta_{\text{pair}}$, provides a reference by which to measure the alignment of the two components (pink ellipses, red crosses mark the central position of the components).
    Each component has its own position angle, $\theta_{\text{n}}$, which can be compared to $\theta_{\text{pair}}$ giving an alignment of component $n$ relative to the pair axis, $\Delta\theta_{\text{n}}$ (shown in blue) with a value between $0$ and $90^{\circ}$. 
    For clarity $\theta_{\text{n}}$ and $\Delta\theta_{\text{n}}$ are only shown for the component on the left hand side in this figure, but measurements for both components are determined.}
    \label{fig:pair_anatomy}
\end{figure}

Even pairs of components with low angular separations will suffer some contamination from random associations.
In order to improve the reliability of our data selection even further we make use of the mean misalignment of the components in the pair relative to the axis of the pair (see Figure \ref{fig:pair_anatomy}).
Here, we define the mean misalignment as:
\begin{equation}
    \label{eq:meanmis}
    \Delta \theta_{\text{mean}} = \frac{\Delta \theta_{1} + \Delta \theta_{2}}{2},
\end{equation}
where $\Delta \theta_{n}$ is the relative misalignment (between $0$ and $90^{\circ}$) of the position angle of component $n$, $\theta_{n}$, relative to the position angle of the pair, $\theta_{\text{pair}}$, given by:
\begin{equation}
    \label{eq:misalign}
    \Delta \theta_{n} = \lvert \theta_{n} - \theta_{\text{pair}} \rvert.
\end{equation}
At larger angular separations one would expect the components of the majority of true DRAGNs to be relatively well aligned with the pair axis, since the emission will arise from the originating jets or the trailing lobe structures. 

\begin{figure}
    \centering
    \includegraphics[width=0.99\columnwidth]{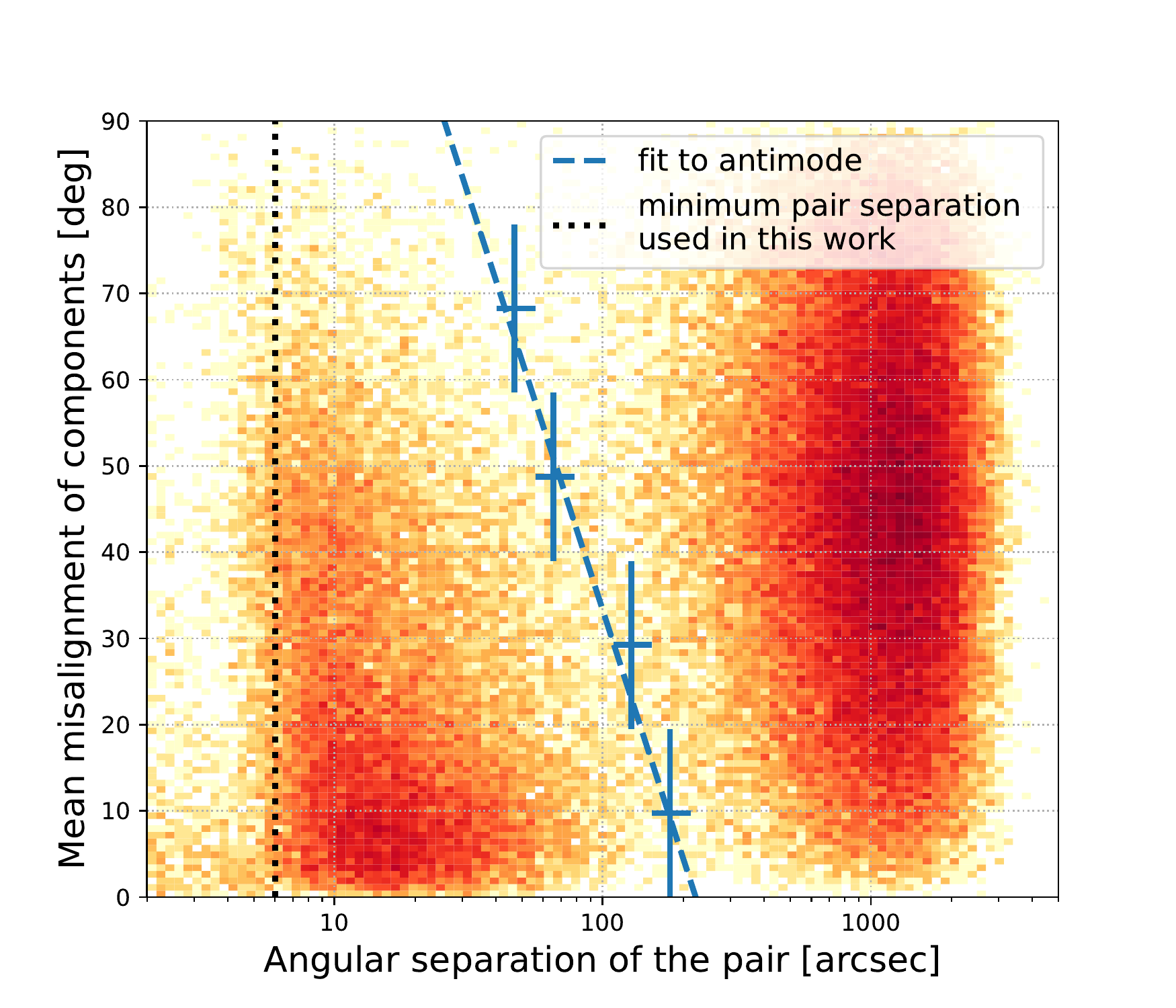}
    \caption{The mean misalignment of the lobes of candidate DRAGNs as a function of their angular separation.
    For component pairs separated by less than $\approx 200''$ the components typically have position angles similar to the axis of the pair, whereas larger pairs show a random distribution of component alignments relative to the pair axis.
    The blue points highlight the pair separation antimodes for bins of mean alignment, the error bars on these points represent the bins used.
    The blue dashed line is a least squares fit to the antimodes that we use to define a sample of likely real double radio sources (see Eq. \ref{eq:sepalignfit}).
    The black dashed line represents the minimum pair separation we use to select our DRAGNs in this work.
    }
    \label{fig:sep_v_align}
\end{figure}

In Figure \ref{fig:sep_v_align} we plot the mean component misalignment of our pair sample as a function of pair separation, demonstrating that the antimode (local minima) of the pair separation distribution moves to smaller values as the misalignment of the pairs increases.
Taking the antimode of the pair separation for pairs in different mean misalignment bins (blue crosses in Figure \ref{fig:sep_v_align}), we derive a linear fit given by:
\begin{equation}
    \label{eq:sepalignfit}
    \frac{\Delta\theta_{\text{mean}}}{\text{deg}} < -96.01\log_{10}\frac{d}{\text{arcsec}} + 225.32,
\end{equation}
to aid in selecting real double sources (blue dashed line in Figure \ref{fig:sep_v_align}), where $d$ is the angular separation of the pair components and $\Delta\theta$ is the mean misalignment of the pair.
Figure \ref{fig:sep_v_align} shows that most of the pairs in the left-hand population have pair separations with
\begin{equation}
    \label{eq:minsep}
    d > 6''.
\end{equation}
A pair separation of $6''$ is $\approx2\times$ the VLASS beam size and represents a clean separation of two extended components.
The small population of pairs with $d<6''$ do not follow the general trend of increasing mean misalignment with decreasing pair separation seen in the rest of the pairs, and likely represent pairings of components with poorly constrained measurements.
We therefore select pairs satisfying Equation \ref{eq:sepalignfit} and \ref{eq:minsep} as our DRAGNs.
These criteria select $17,724$ DRAGNs represented by data points to the right of the black dotted line and underneath the blue dashed line in Figure \ref{fig:sep_v_align}.

\subsection{Core finding}
\label{ssec:corefinding}

The key aspect of the double finding we perform is that it searches specifically for pairs of extended radio components.
One necessary consequence of pre-selecting such candidate lobes is that compact components that may represent radio cores are initially excluded from association with the radio sources found by the pair finding. 
In order to attempt to recover these missed cores, we search for candidate radio cores in the population of radio components that were not considered to be candidate radio lobes, i.e., $S_{\text{peak}}>3\,$mJy/beam and $\Psi < 3''$.
We search for such candidate cores within $30''$ or half the pair separation (whichever is the lesser) of the central position of our DRAGNs.

\begin{figure}
    \centering
    \includegraphics[width=\columnwidth]{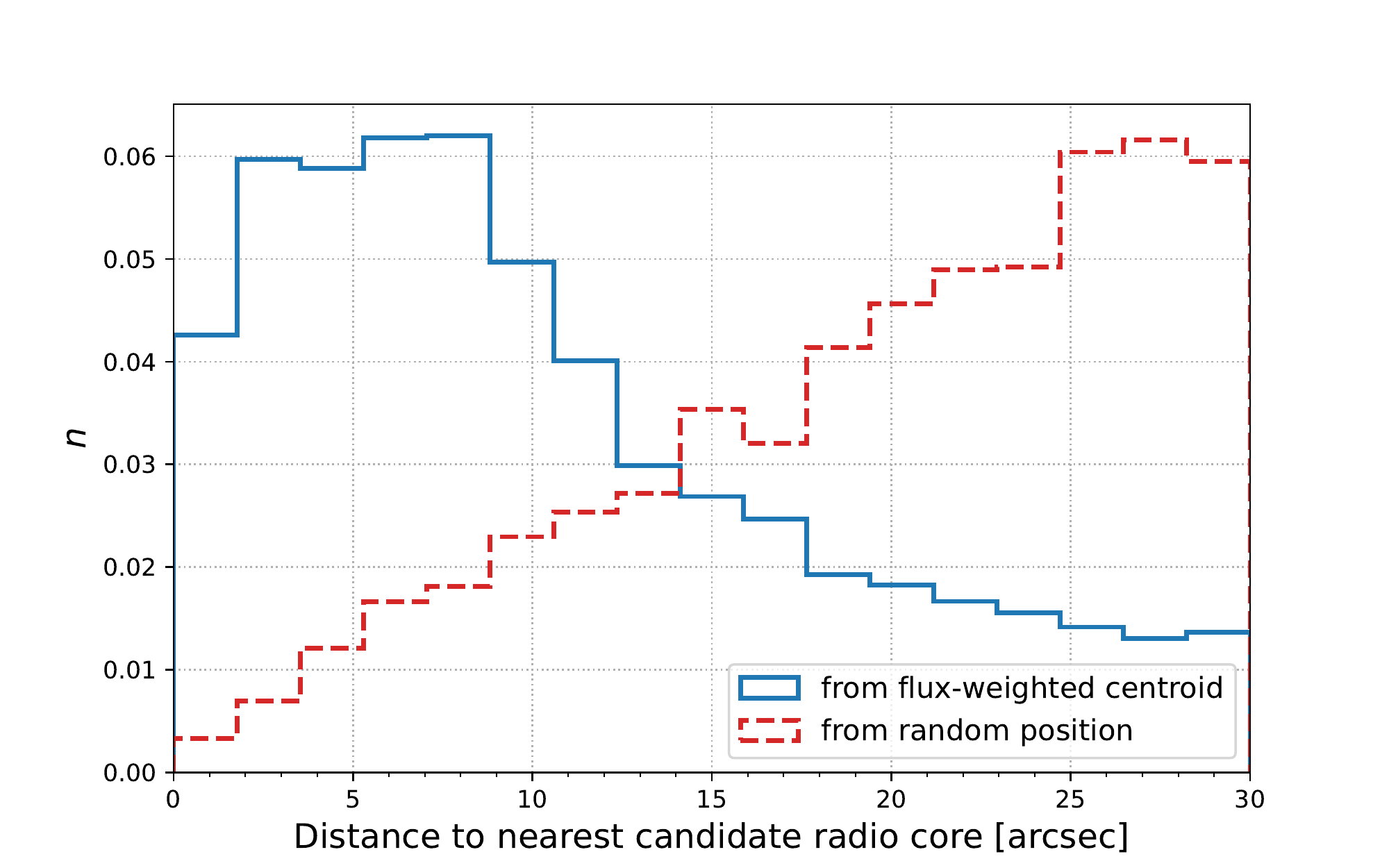}
    \caption{The normalised (by area under the lines) distributions of the angular distance to the nearest core candidate from the flux-weighted central position of the DRAGN (blue solid line), and from random sky coordinates (red dashed line).
    }
    \label{fig:corecount}
\end{figure}

In $\approx90\%$ of cases no core is found.
Of the remaining $10\%$, only one core candidate is found in the majority of cases, with $<1\%$ of our candidate DRAGNs being associated with more than one core candidate.
Visual inspection of DRAGNs with multiple core candidates shows these cases to generally be the result of sidelobes produced by bright sources. 
In Figure \ref{fig:corecount} we show the angular separation to the nearest candidate core from the flux-weighted centroid of the DRAGN (blue) and from a random sky coordinate (red).
The random sky coordinates are obtained by subtracting 1 degree from the declination of the real positions.
Where core candidates are found, they are generally within $\approx 10''$ of the flux weighted central position of the DRAGN.
The strong peak in the core candidate distribution at small separations from the positions of the candidate DRAGNs is indicative of this sample being dominated by real DRAGNs with a detected core.
In the rare cases where a DRAGN is associated with multiple core candidates, the closest candidate to the central position of the DRAGN is adopted as the core ID.
In total, $1,836$ of our DRAGNs have a core identification.

Having identified all of the VLASS components associated with our DRAGNs, we also have a list of VLASS components that are not a part of any of our DRAGNs.
Although some of these will be associated with more complex radio structures, the majority will be simple-morphology sources.
For those components unaffiliated with any of our DRAGNs we select those with \text{$S_{\text{peak}} > 3\,$mJy/beam} as a reference set of single-component radio sources with which to compare our DRAGNs throughout the rest of this work.
This sample contains $577,651$ sources.

\subsection{Key Measurements}

With a catalog of DRAGNs in hand, we can derive key observable properties of the DRAGNs based on their constituent components.
Perhaps the most fundamental property of any radio source is its total flux density, $S$.
For our DRAGNs we estimate this by taking the sum of the total flux densities of all the constituent components, i.e., $S_{\text{DRAGN}} = S_{\text{Lobe 1}} + S_{\text{Lobe 2}} + S_{\text{Core}}$.

\begin{figure}
    \centering
    \includegraphics[width=\columnwidth]{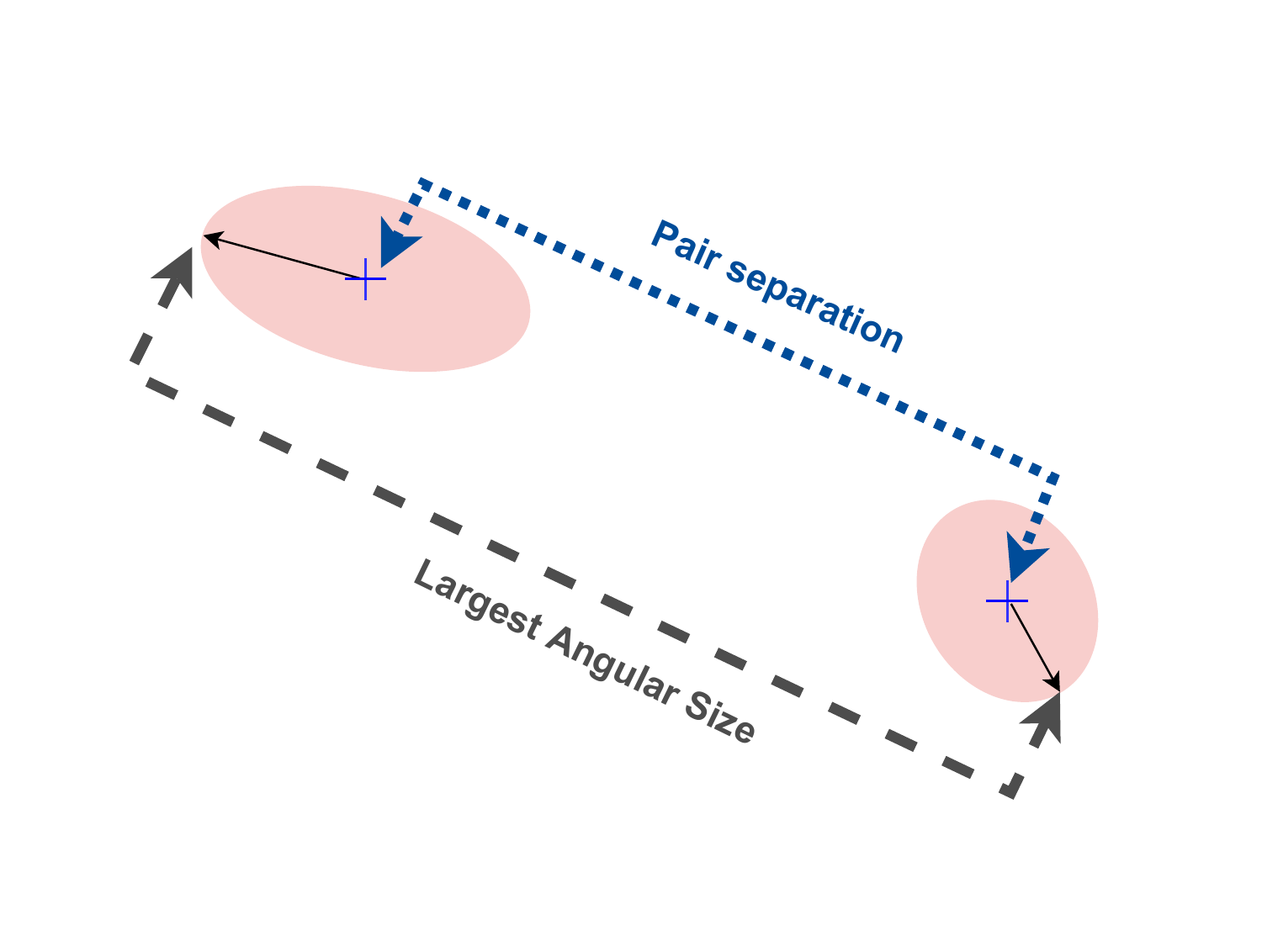}
    \caption{Schematic showing how we determine the Largest Angular Size (LAS, grey dashed line) of our DRAGNs.
    The pink ellipses represent the geometry of the two lobe components after deconvolution from the VLASS beam.
    Our LAS measurement differs from the pair separation (blue dotted line) of the two lobe components by extending the pair size by the semi major size of the components (black arrows) in direction of the component position angle away from the pair centre.}
    \label{fig:las}
\end{figure}

Another essential measurement to make when characterising DRAGNs is the largest angular size (LAS) of the radio source.
To determine the LAS of a DRAGN, one could take the separation of the two lobe components as a proxy for the LAS.
However, this would more accurately represent the distance between the radio component centroids rather than the full extent of the radio structure.
A more robust approach would be to measure the angular extent subtended by the radio source above some signal-to-noise threshold.
Such an approach would require making additional measurements from the image of the source rather than being easily calculable from the catalog data used by \textsc{DRAGNhunter}.
Instead, we choose a compromise approach to estimate the LAS of our DRAGNs.
We define the extreme coordinates of the radio double to be the lobe component coordinates offset by their semi major axis size in the direction away from the pair centre given by the lobe component position angles (see Figure \ref{fig:las}), and take the distance between these coordinates to be the LAS.
For our single-component sources, we take the LAS to be the measurement of deconvolved major axis size as listed in the component catalog.
The distributions of integrated flux density and LAS for our DRAGNs are shown in Figure \ref{fig:sizeflux}, demonstrating that larger DRAGNs are generally brighter.

\section{Sample Reliability and Completeness}
\label{sec:relcomp}

\subsection{Reliability of DRAGN detections}
\label{ssec:reliability}

\subsubsection{Overall Sample Reliability}

In order to assess the reliability of our algorithm in selecting DRAGNs, a validation sample of 500 random DRAGNs are visually inspected. 
The errors reported on our fractional estimates (here, and throughout) are binomial uncertainties as described in \citet{Cameron2011}.
Using our validation sample, we find that overall $89.0_{-1.6}^{+1.2}\,\%$ of the selected `DRAGNs' are genuine radio doubles, and example real DRAGNs identified are shown in Figure \ref{fig:egdoubles}.
Approximately $11\,\%$ of the time the sources identified by \textsc{DRAGNhunter} are not the radio doubles the algorithm is designed to select.
This is to be expected given the range of complex radio morphologies that exist and the limitations of using only catalog data produced by a source finding program rather than the image data directly.

\begin{figure}
    \centering
    \includegraphics[width=\columnwidth]{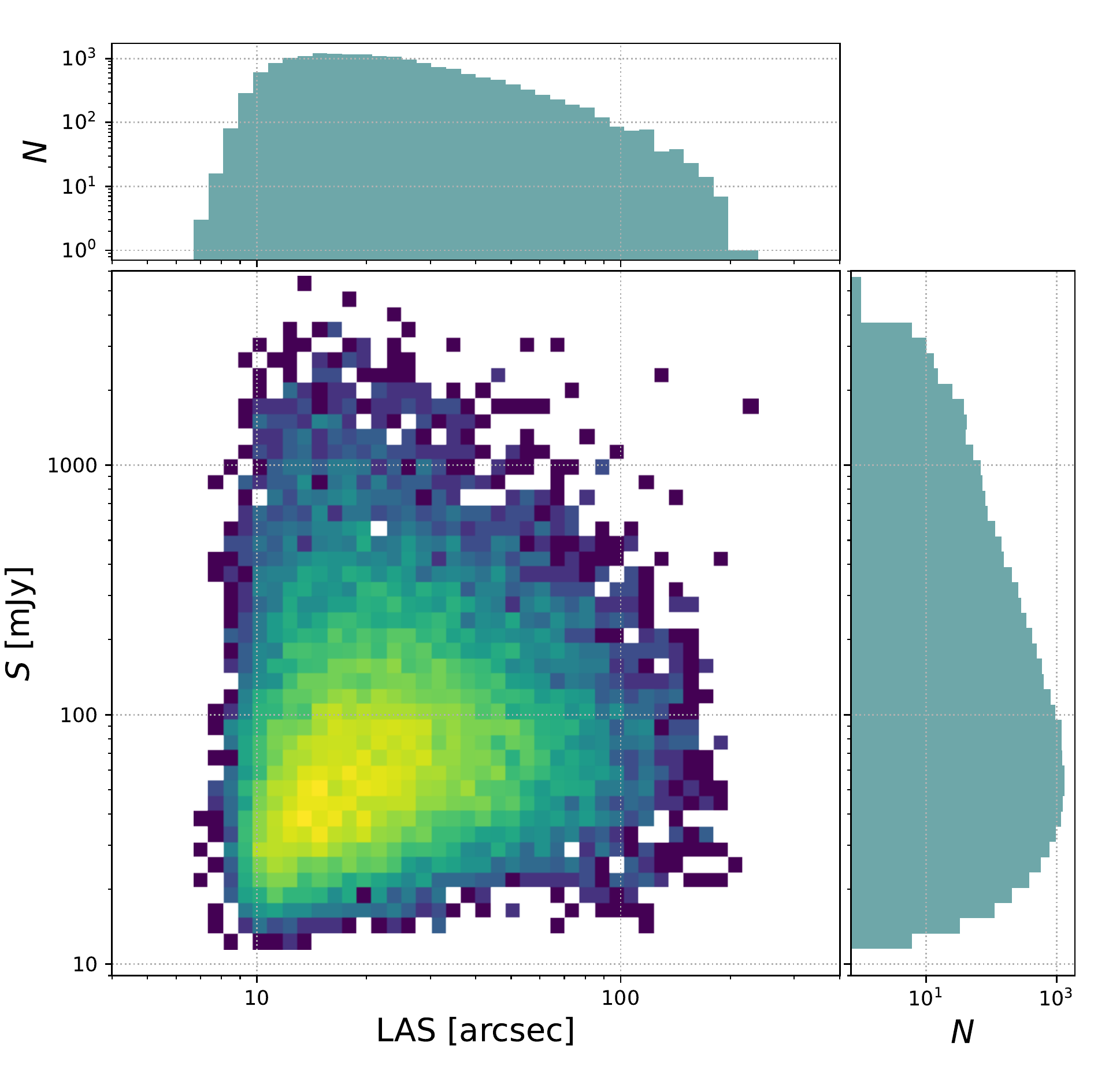}
    \caption{The distributions of largest angular size (LAS) and integrated flux density ($S$) for our DRAGNs.}
    \label{fig:sizeflux}
\end{figure}

\begin{figure*}
    \centering
    \subfigure{\includegraphics[width=0.32\columnwidth]{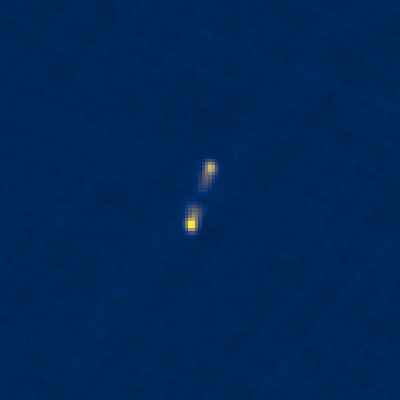}}
    \vspace{-3mm}
    \subfigure{\includegraphics[width=0.32\columnwidth]{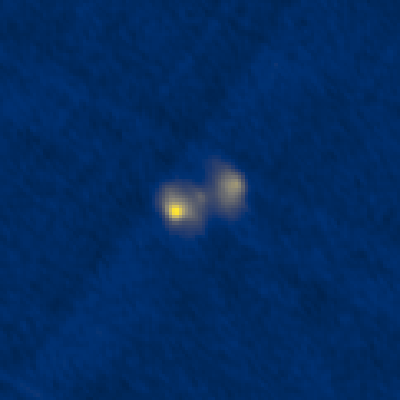}}
    \subfigure{\includegraphics[width=0.32\columnwidth]{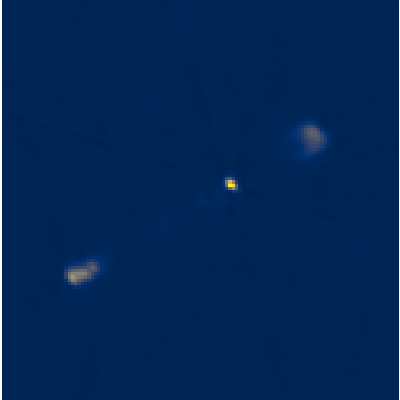}}
    \subfigure{\includegraphics[width=0.32\columnwidth]{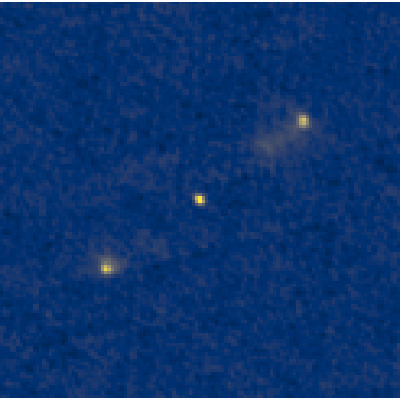}}
    \subfigure{\includegraphics[width=0.32\columnwidth]{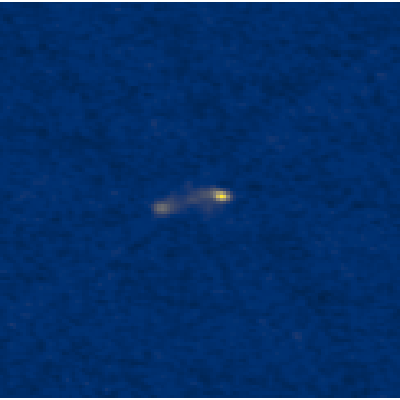}}
    \subfigure{\includegraphics[width=0.32\columnwidth]{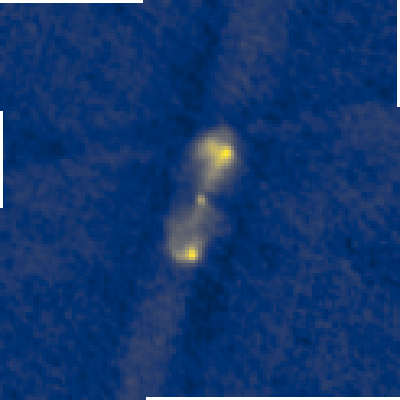}}
    
    \subfigure{\includegraphics[width=0.32\columnwidth]{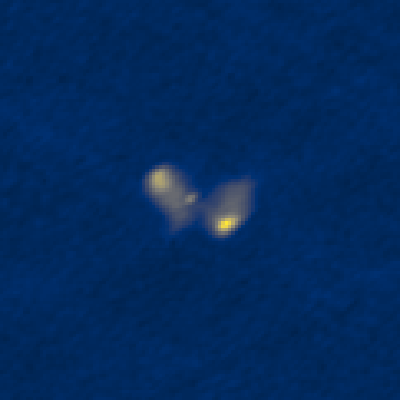}}
    \vspace{-3mm}
    \subfigure{\includegraphics[width=0.32\columnwidth]{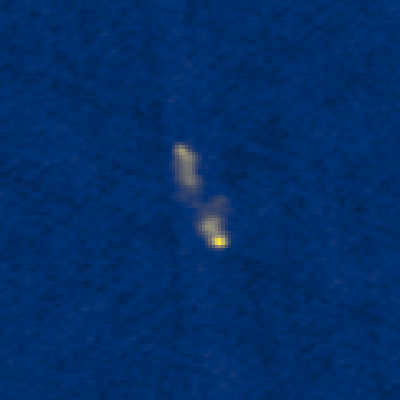}}
    \subfigure{\includegraphics[width=0.32\columnwidth]{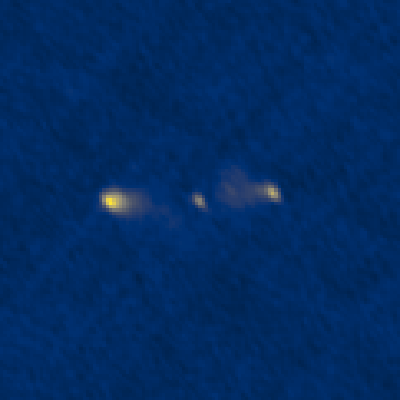}}
    \subfigure{\includegraphics[width=0.32\columnwidth]{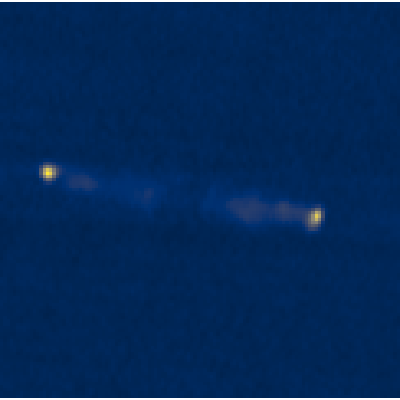}}
    \subfigure{\includegraphics[width=0.32\columnwidth]{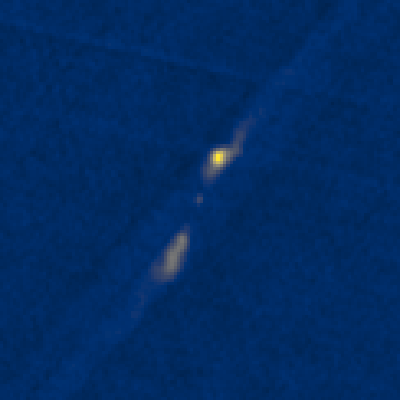}}
    \subfigure{\includegraphics[width=0.32\columnwidth]{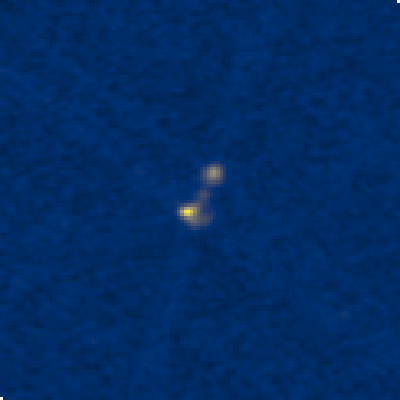}}
    
    \subfigure{\includegraphics[width=0.32\columnwidth]{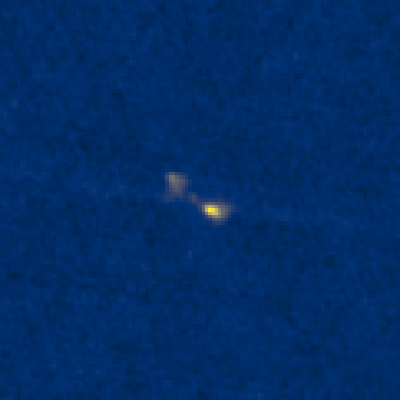}}
    \vspace{-3mm}
    \subfigure{\includegraphics[width=0.32\columnwidth]{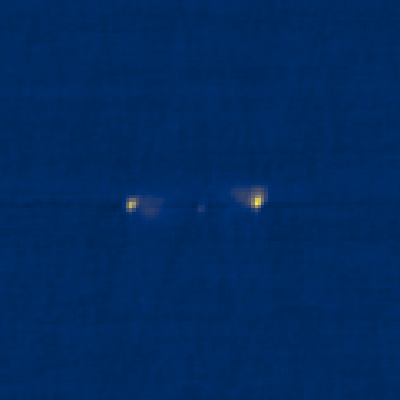}}
    \subfigure{\includegraphics[width=0.32\columnwidth]{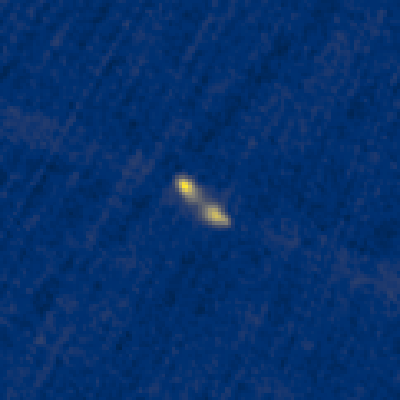}}
    \subfigure{\includegraphics[width=0.32\columnwidth]{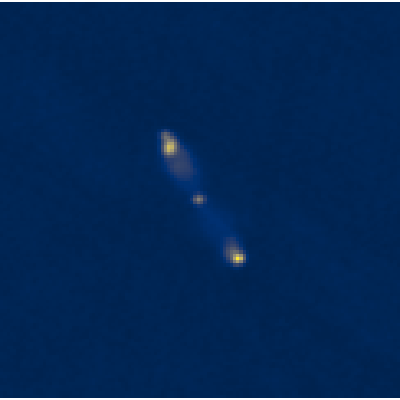}}
    \subfigure{\includegraphics[width=0.32\columnwidth]{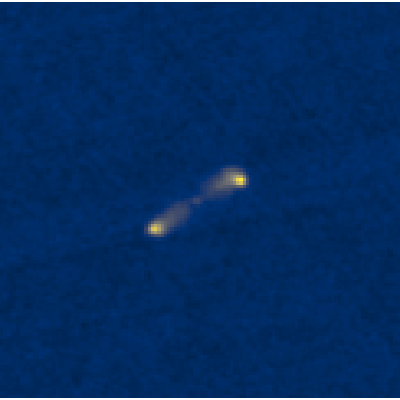}}
    \subfigure{\includegraphics[width=0.32\columnwidth]{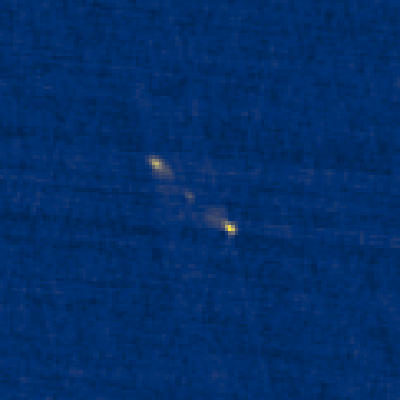}}
    
    \subfigure{\includegraphics[width=0.32\columnwidth]{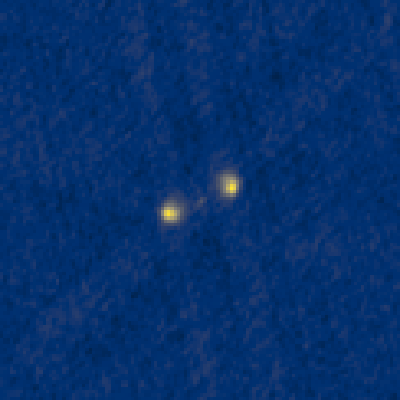}}
    \subfigure{\includegraphics[width=0.32\columnwidth]{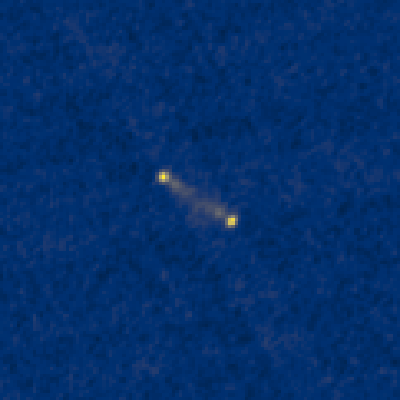}}
    \subfigure{\includegraphics[width=0.32\columnwidth]{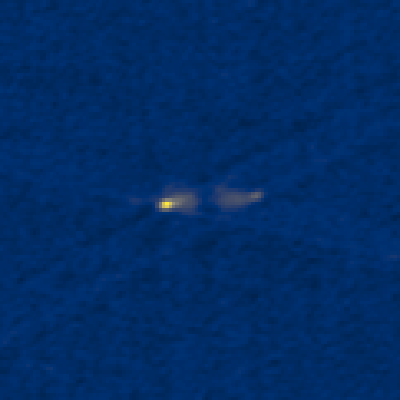}}
    \subfigure{\includegraphics[width=0.32\columnwidth]{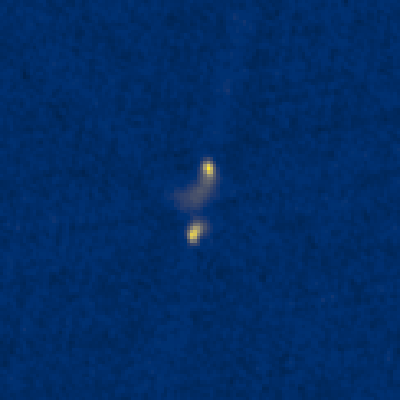}}
    \subfigure{\includegraphics[width=0.32\columnwidth]{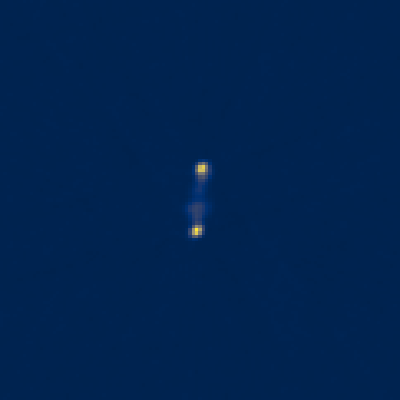}}
    \subfigure{\includegraphics[width=0.32\columnwidth]{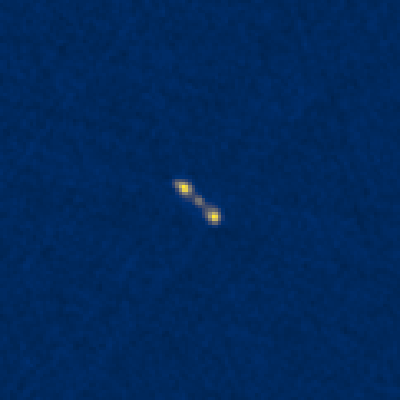}}
    \caption{
    Postage stamp cutouts ($2'\times2'$) of 24 examples of genuine DRAGNs.
    In many cases a core is evident in the imaging even if not identified by \textsc{DRAGNhunter}, a result of our minimum $3\,$mJy/beam brightness threshold.
    Genuine radio doubles like these make up $89\,\%$ of our DRAGN selection
    }
    \label{fig:egdoubles}
\end{figure*}

\begin{figure*}
    \centering 
    \subfigure[]{\includegraphics[width=0.45\columnwidth]{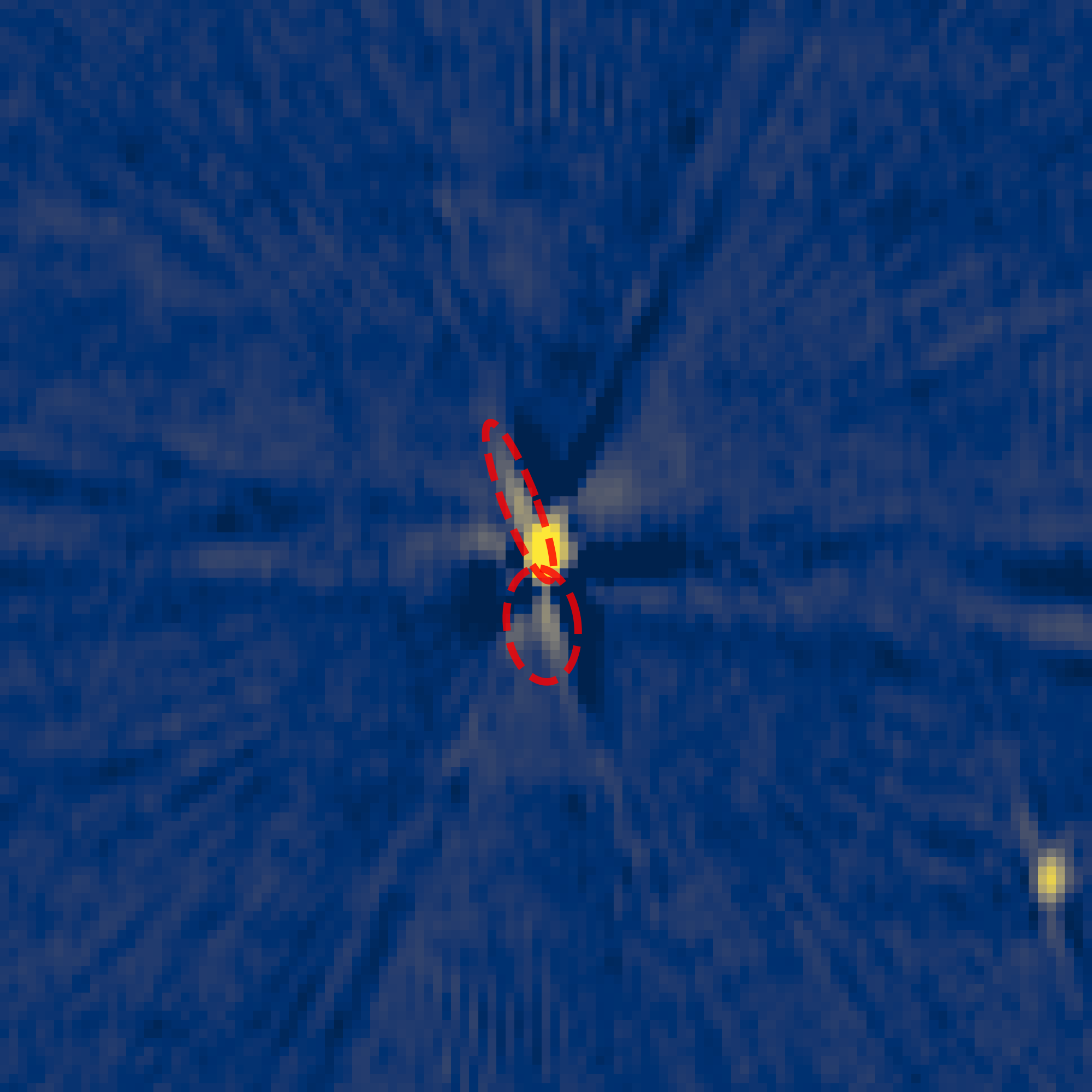}}
    \subfigure[]{\includegraphics[width=0.45\columnwidth]{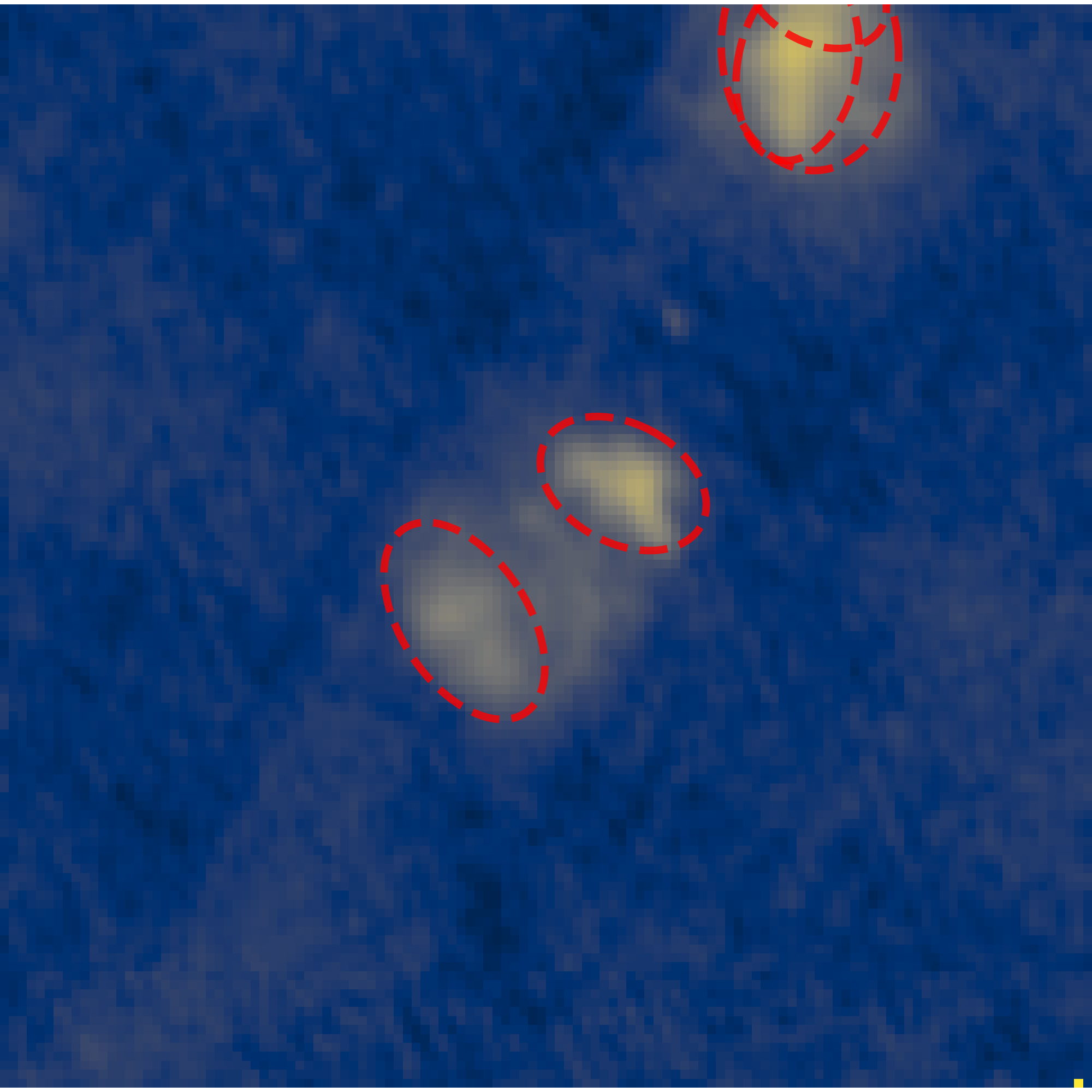}}
    \subfigure[]{\includegraphics[width=0.45\columnwidth]{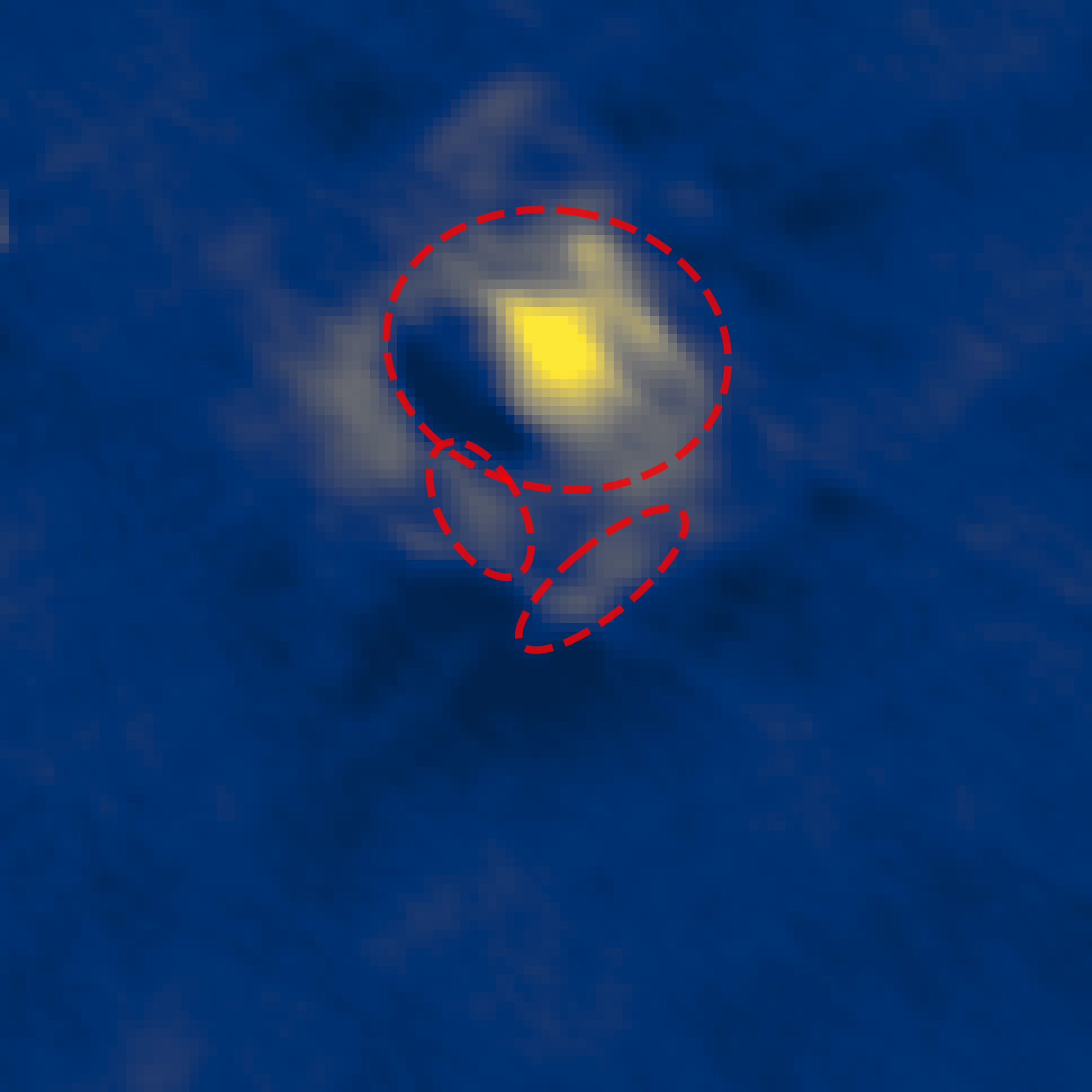}}
    \subfigure[]{\includegraphics[width=0.45\columnwidth]{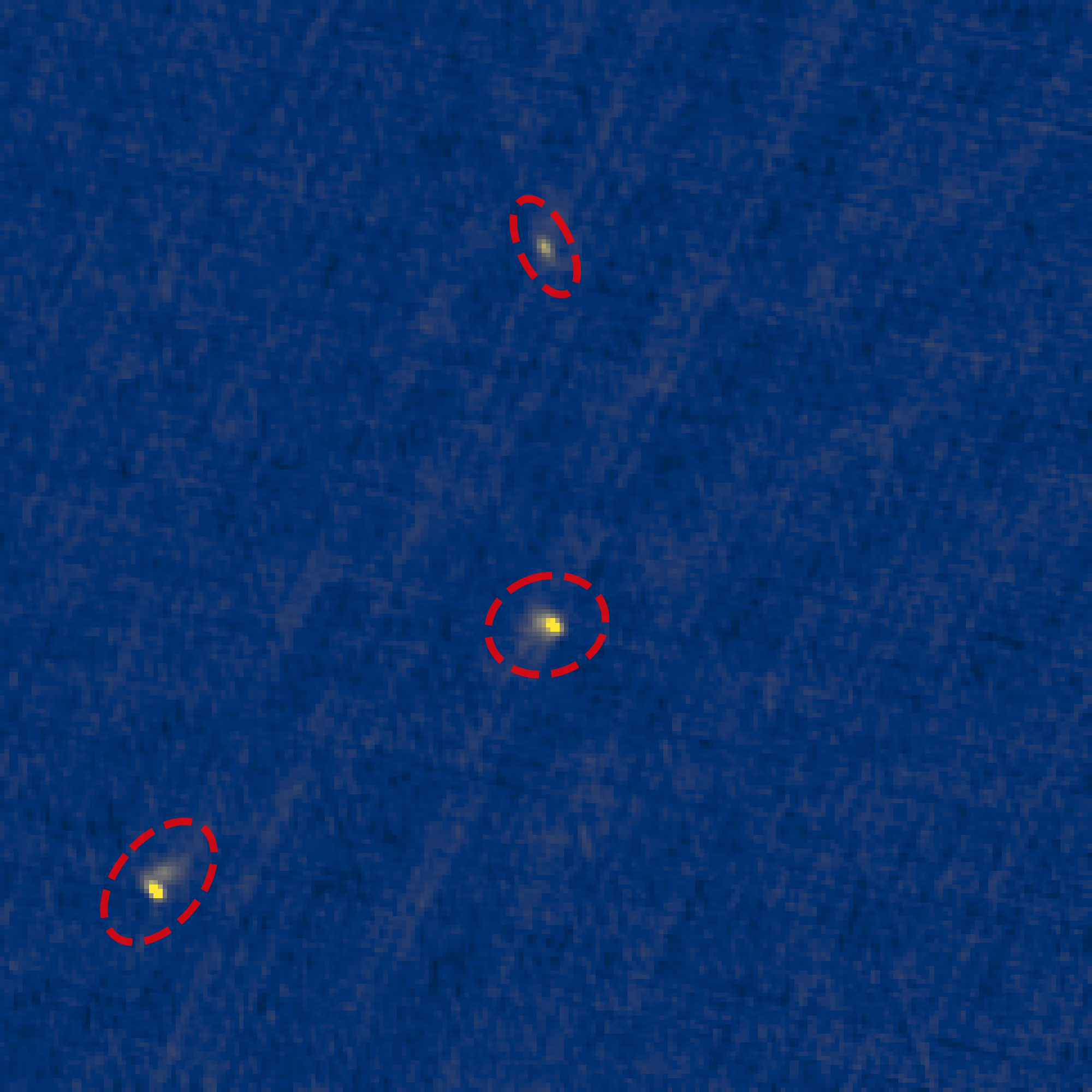}}
    \caption{
    Four example postage stamp cutouts where \textsc{DRAGNhunter} has identified extended emission that is not the result of two distinct radio lobes.
    Panel a is a bright object with visible sidelobes where two of the sidelobes have been spuriously included in the component catalog and paired together by \textsc{DRAGNhunter}.
    In panel b part of an extended radio galaxy but not the whole source has been selected as a DRAGN.
    Panel c shows a supernova remnant where part of the continuum emission has been detected as multiple components.
    In Panel d, the component in the centre and south east of the image constitute a genuine DRAGN.
    However, in this case \textsc{DRAGNhunter} has paired the component in the centre of the image with the unrelated component at the north of the image as this is the closer pairing of candidate lobes.
    Panels a-c are $2'\times2'$ cutouts while panel d is $4'\times 4'$.
    The red dashed lines show the position of components considered as candidate lobes by \textsc{DRAGNhunter}.
    }
    \label{fig:egfail}
\end{figure*}

There are three distinct types of spurious detections that \textsc{DRAGNhunter} produces.
First, image artifacts (such as sidelobes around bright sources) can contaminate the component catalog used as an input for \textsc{DRAGNhunter}.
Where this happens, these spurious detections can be selected as either one or both of the candidate lobes in a pairing of components (e.g. Figure \ref{fig:egfail}a).
Second, large diffuse structures that are not separate lobes of a DRAGN may be grouped together by \textsc{DRAGNhunter}.
In panel b of Figure \ref{fig:egfail} we show an example where the two candidate lobes identified by \textsc{DRAGNhunter} are actually substructure within a single lobe of the DRAGN.
Similarly, in panel c of Figure \ref{fig:egfail} we show a supernova remnant where \textsc{DRAGNhunter} mistakenly identifies part of the emission as two lobes of DRAGN.
Third, in some cases a genuine lobe may be paired with an interloping candidate lobe (e.g. Figure \ref{fig:egfail}d) because it is closer than the real counterpart.
The risk of this type of false association increases with the angular size of the DRAGN. 
The on-sky density of candidate lobes (components with $S_{\text{peak}}>3\,$mJy/beam and $\Psi > 3''$) is $3.5\,\text{deg}^{-2}$.
For DRAGNs with $\text{LAS} < 30''$ ($>70\,\%$ of our sample) contamination from interloping candidate lobes is estimated to be $<0.1\,\%$, while for  DRAGNs with $\text{LAS} < 60''$ ($>90\,\%$ of our DRAGNs) this type of contamination rises to $0.3\,\%$.

\subsubsection{Parameter Space Differences Between Real and Spurious Detections}

\begin{figure*}
    \centering
    \subfigure[]{\includegraphics[width=0.99\columnwidth]{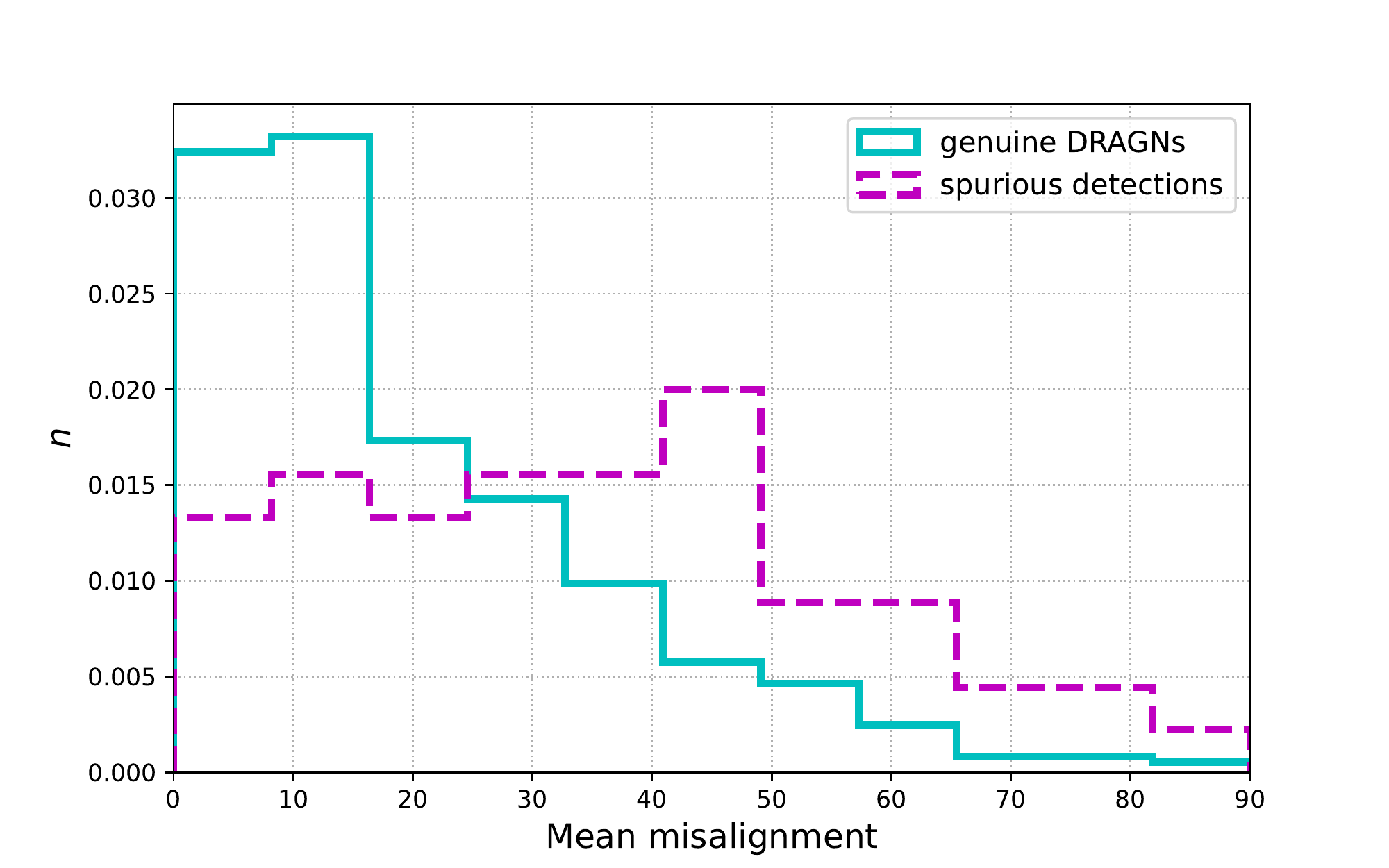}}
    \subfigure[]{\includegraphics[width=0.99\columnwidth]{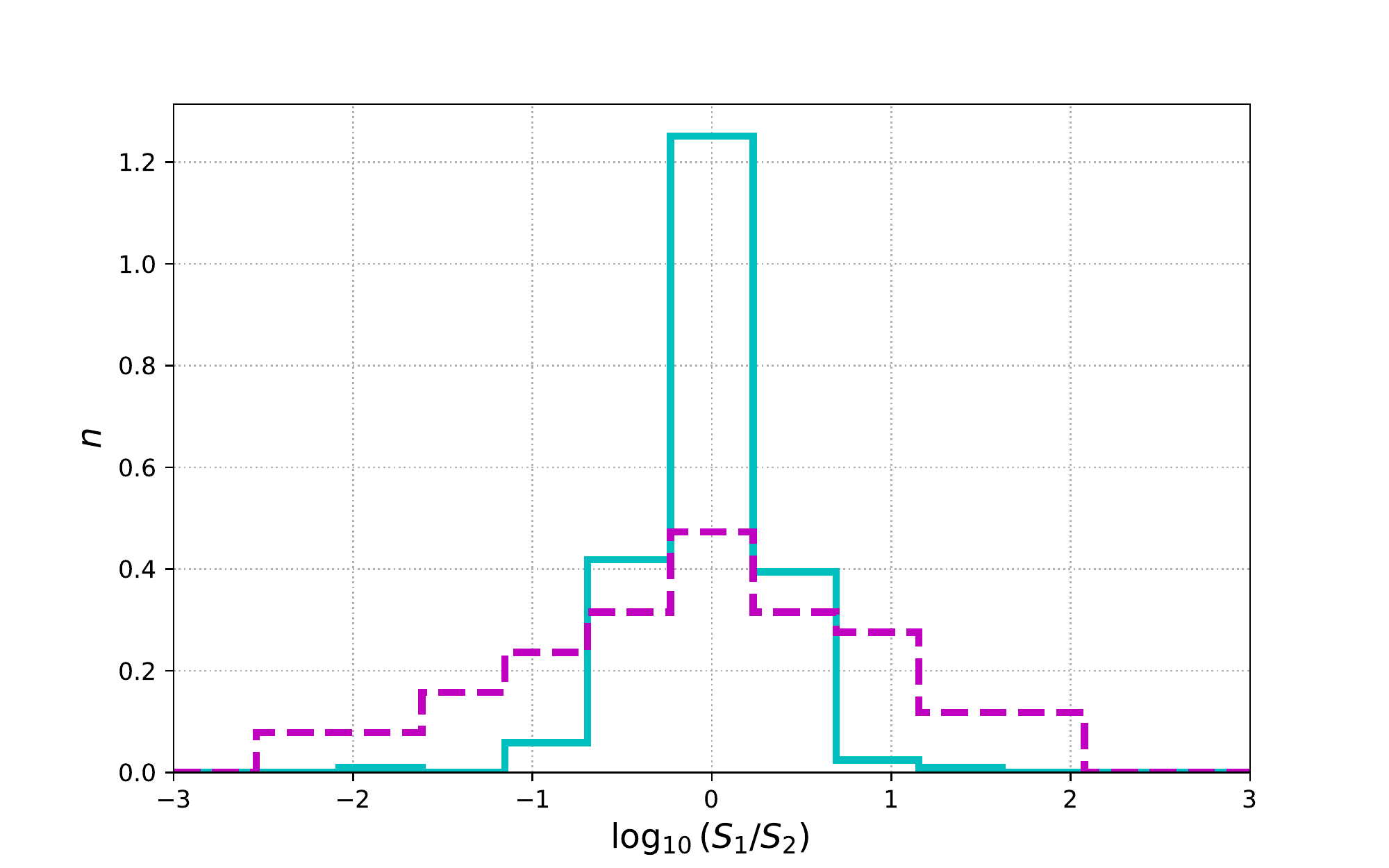}}
    \subfigure[]{\includegraphics[width=0.99\columnwidth]{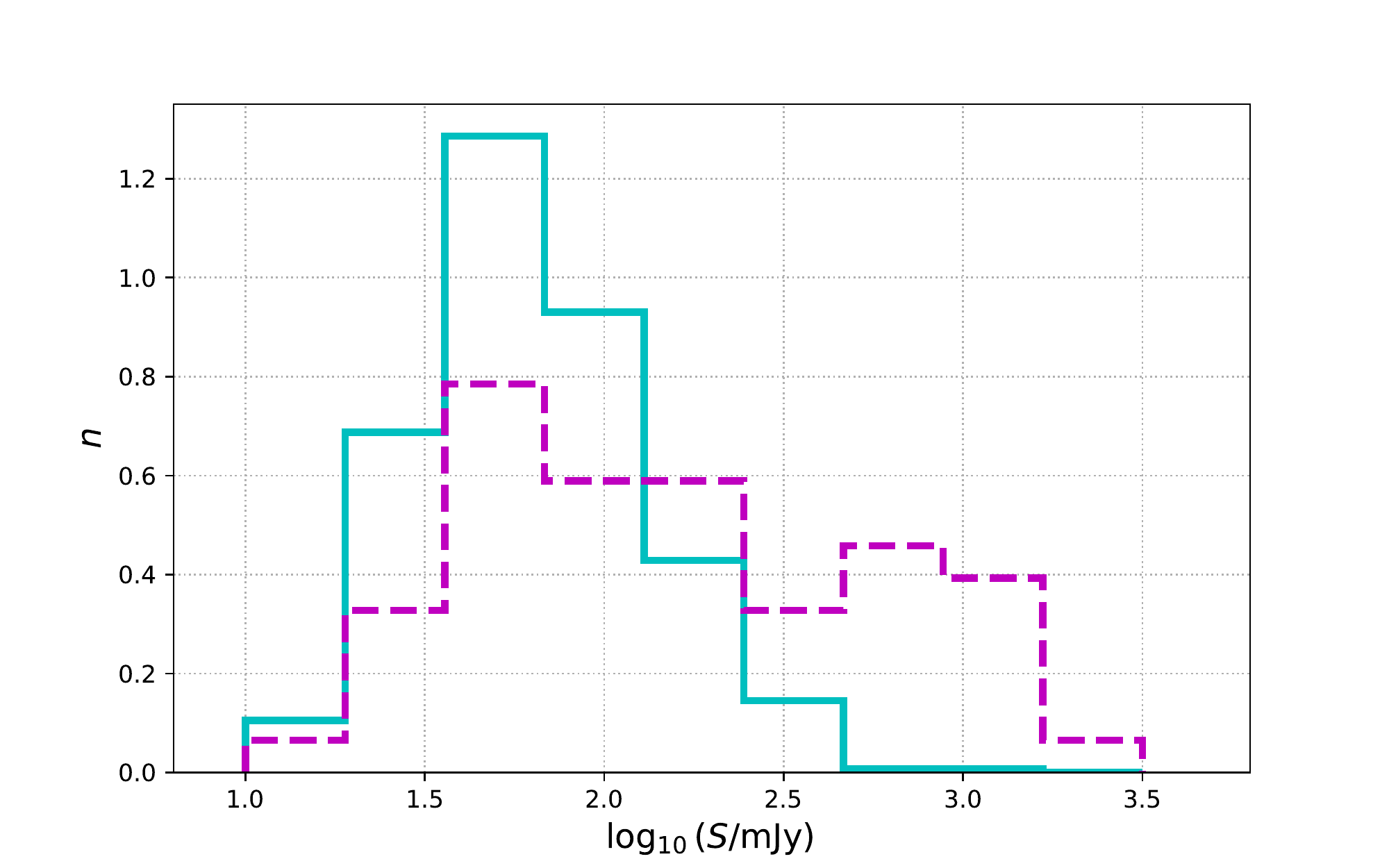}}
    \subfigure[]{\includegraphics[width=0.99\columnwidth]{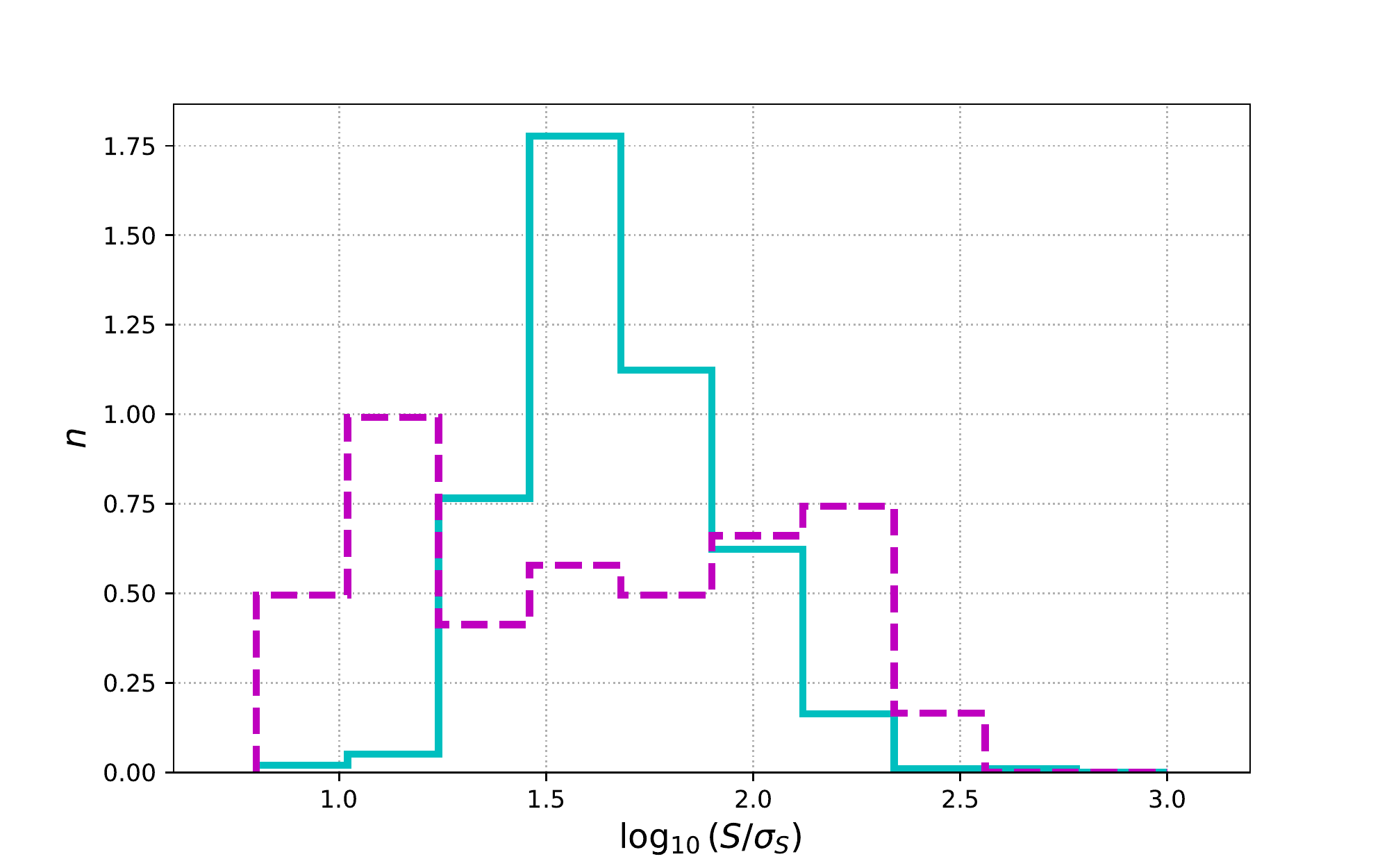}}
    \subfigure[]{\includegraphics[width=0.99\columnwidth]{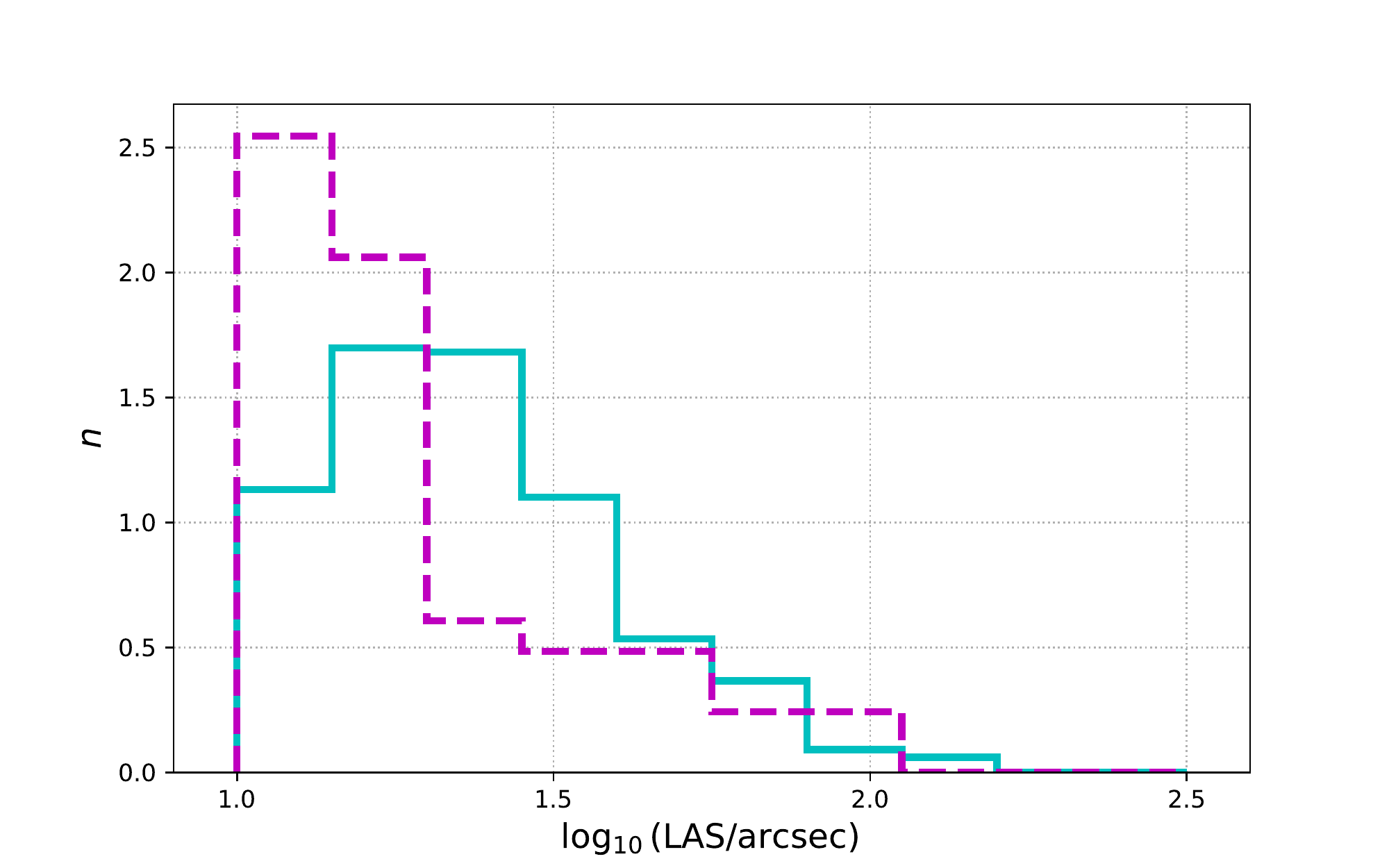}}
    \subfigure[]{\includegraphics[width=0.99\columnwidth]{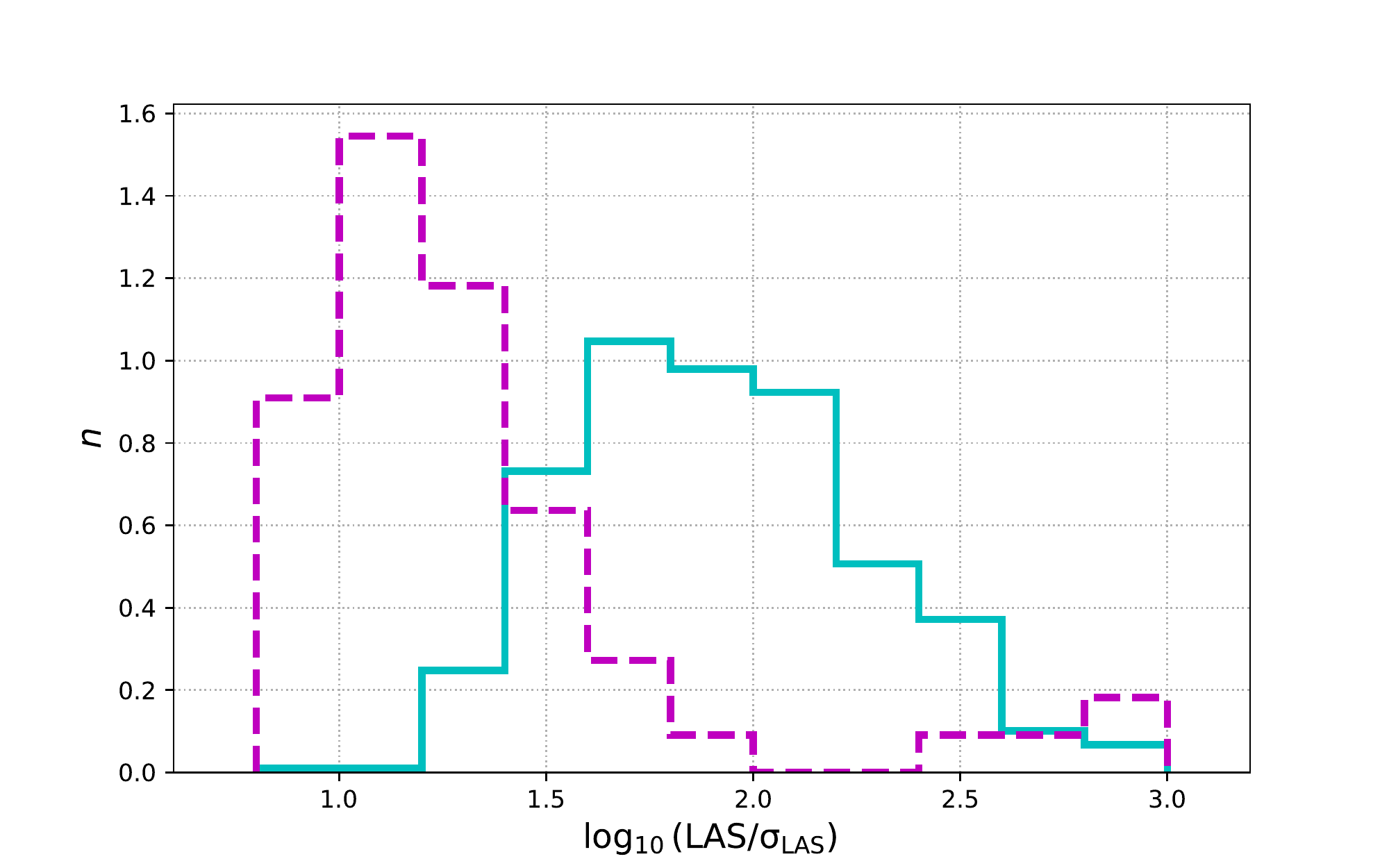}}
    \caption{Distributions (normalised by area under the histogram) of mean component misalignment (a), component flux ratio (b), Total source flux density (c), S/N in source flux (d), source LAS (e) and S/N in LAS (f) for sources identified as genuine (cyan solid line) and spurious (magenta dashed line) in our validation sample of 500 DRAGNs.}
    \label{fig:goodbadparams}
\end{figure*}

It is impossible to identify and remove all spurious detections without visual inspection of the entire catalog of $>17,000$ objects.
However, knowing where \textsc{DRAGNhunter} fails allows for more robust selection criteria to be used in cases where sample fidelity is more important than sample completeness.
Figure \ref{fig:goodbadparams} shows comparative distributions (normalised by the area of the histogram) of a number of properties for genuine DRAGNs ($N=445$) and spurious sources ($N=55$) in our validation sample.
For cases where \textsc{DRAGNhunter} is finding either part of a double (but not the whole) or the more complex morphologies of, e.g., supernova remnants, there is little reason to expect the components to be well aligned with the pair axis.
Indeed, panel a of Figure \ref{fig:goodbadparams} shows that spurious detections are more likely to have higher mean misalignments of their components: the median value of mean misalignment for genuine DRAGNs is $15^{\circ}$, while for spurious detections it is $35^{\circ}$.
Selecting DRAGNs where the mean misalignment of the candidate lobes is less than $30^{\circ}$ improves the sample reliability to $93.6_{-1.6}^{+1.1}\,\%$. 
However, only $\approx70\,\%$ of our DRAGNs satisfy this criterion, so there is a cost in sample completeness.

Where unassociated components have been paired together by \textsc{DRAGNhunter}, there is no reason to expect the flux densities of the individual components to be correlated.
Panel b of Figure \ref{fig:goodbadparams} shows the distribution of $S_{1}/S_{2}$, where $S_n$ is the flux of the component associated with lobe $n$.
As expected, the spread of flux density ratios is larger for the spurious sources, where the standard deviation of $\log_{10}(S_{1}/S_{2})$ is $0.98$, than for genuine DRAGNs, which have a standard deviation of $0.34$ in $\log_{10}(S_{1}/S_{2})$.
Nearly all of the genuine DRAGNs have components with flux densities within a factor of 10 of each other.
Selecting only DRAGNs with $0.1 < S_{1}/S_{2} < 10$, improves the reliability of the DRAGN selection to $92.6_{-1.2}^{+0.8}\,\%$.
Approximately $95\,\%$ of the DRAGNs in our catalog have flux ratios lying in this range, making $S_{1}/S_{2}$ a very useful metric for identifying spurious sources.

The skew of the flux density distribution of spurious sources toward higher values relative to the sample of genuine DRAGNs shown in panel c of Figure \ref{fig:goodbadparams} is consistent with contamination from bright source and sidelobe pairings.
Such contaminating sources should also have small total angular extents.
Panel e of Figure \ref{fig:goodbadparams} suggests most contaminants do in fact have relatively small values of LAS.
However, while simply cutting sources with small angular sizes may improve the sample reliability, it will of course cut all the genuine DRAGNs with small angular sizes as well.
Panel f of Figure \ref{fig:goodbadparams} shows a cleaner distinction between genuine doubles and spurious detections in terms of the signal-to-noise ratio (S/N) of the LAS measurement.
The median $\text{LAS}/\sigma_{\text{LAS}}$ for genuine DRAGNs is $\approx 80$, whereas for spurious detections it is $\approx 16$.
Approximately $94\,\%$ of our catalog of DRAGNs have $\text{LAS}/\sigma_{\text{LAS}} > 20$, and when only considering sources satisfying this criterion the reliability of our validation sample is $95.8_{-1.2}^{+0.8}\,\%$.

Both $S_{1}/S_{2}$ and $\text{LAS}/\sigma_{\text{LAS}}$ can be used to produce higher purity samples of DRAGNs at a relatively small cost in sample completeness.
Applying cuts in both of these parameters can improve the sample fidelity further still, whilst still only having a relatively low impact on the sample completeness.
Of the DRAGNs in our catalog, $90\,\%$ satisfy both $0.1 < S_{1}/S_{2} < 10$ and \text{$\text{LAS}/\sigma_{\text{LAS}} > 20$}. 
The reliability of such sources is estimated to be $97.5_{-1.0}^{+0.5}\,\%$ based on our validation sample.
However, this estimate is based on a small number ($55$) of spurious detections in our validation sample.
In order to confirm that the flux density ratio of the lobe components and signal-to-noise of the LAS estimate are indeed good metrics to select reliable DRAGNs, we randomly select a further $100$ DRAGNs from our catalog that satisfy $0.1 \leq S_{1}/S_{2} \leq 10$ and $\text{LAS}/\sigma_{\text{LAS}} \geq 20$.
Visually inspecting these $100$ reveals two spurious detections, consistent with our estimate based on our validation sample.
In our catalog of DRAGNs we flag the $\approx 10\,\%$ of entries with either $S_{1}/S_{2} < 0.1$ or $S_{1}/S_{2} > 10$ or $\text{LAS}/\sigma_{\text{LAS}} < 20$ as potential contaminants (see the catalog data model in Appendix \ref{apx:data-model} for details).
This ability to select a large number of DRAGNs with high reliability will be of use to those wishing, for example, to create training sets for machine learning algorithms designed to identify DRAGNs in radio images.

\subsection{Sample Completeness}
\label{ssec:completeness}

\subsubsection{DRAGNs in VLASS Missed by DRAGNhunter}

\begin{figure}
    \centering
    \subfigure[]{\includegraphics[width=0.49\columnwidth]{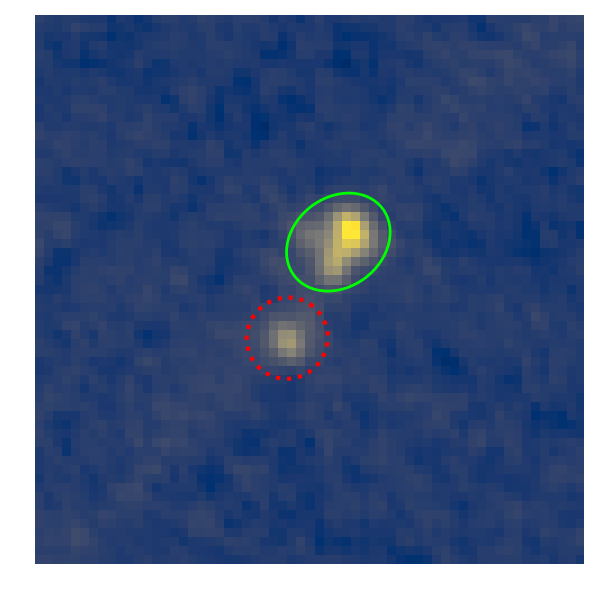}}
    \subfigure[]{\includegraphics[width=0.49\columnwidth]{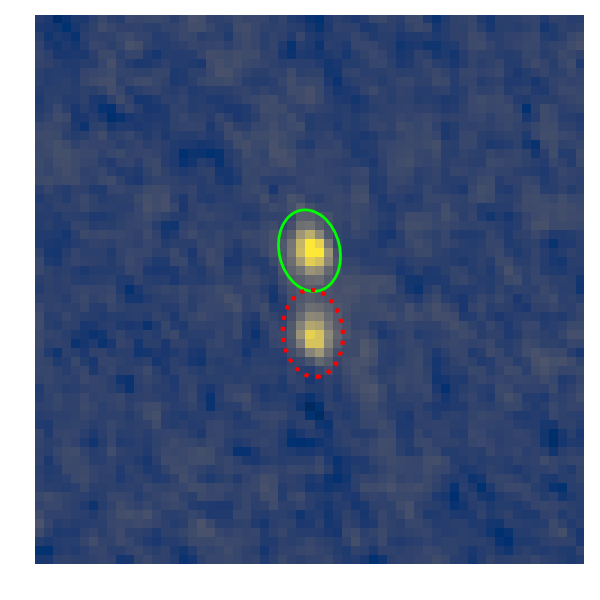}}
    \subfigure[]{\includegraphics[width=0.49\columnwidth]{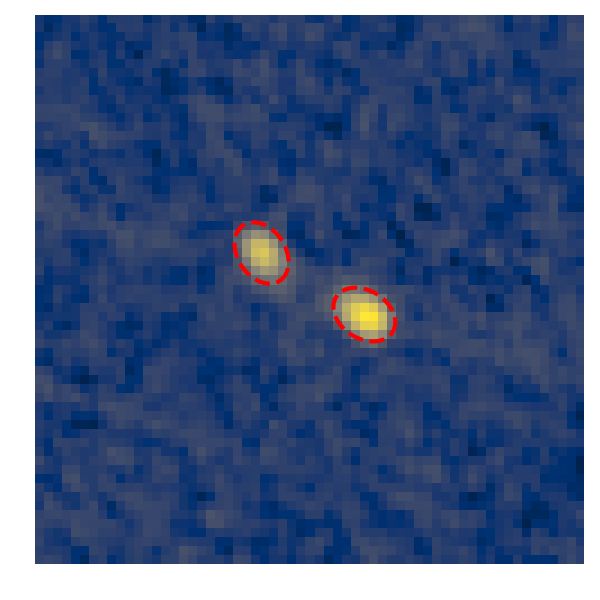}}
    \subfigure[]{\includegraphics[width=0.49\columnwidth]{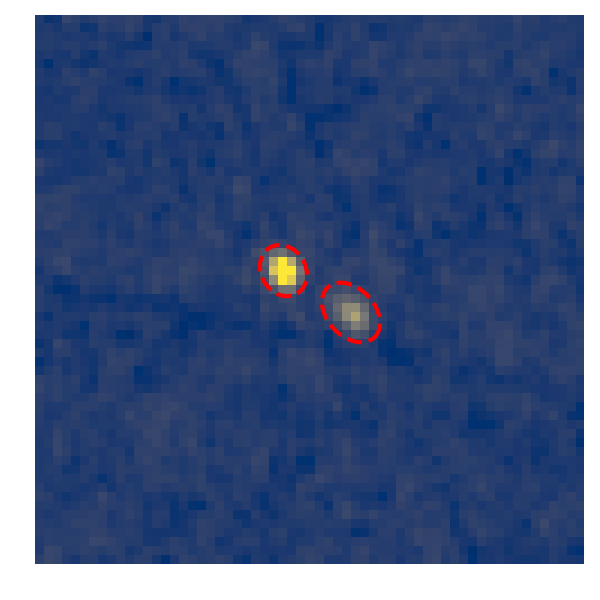}}
    \caption{Example DRAGNs in VLASS not identified by \textsc{DRAGNhunter}.
    The overlaid components (colored ellipses) show why these examples are not not selected by \textsc{DRAGNhunter}.
    Green ellipses with a solid line show components that satisfy the initial selection criteria as \textit{candidate lobes} (see Section \ref{ssec:lobepairs}).
    Red ellipses are components that are not identified as \textit{candidate lobes} either as a result of being too faint (dotted line) or too compact (dashed line).
    These postage stamps are $1' \times 1'$ and the ellipse sizes are set to $2\times$ the \textit{fitted} component size for clarity.}
    \label{fig:egmissing}
\end{figure}

In order to check how well \textsc{DRAGNhunter} recovers DRAGNs from the VLASS data, $50$ VLASS \textit{Quick Look} images ($1^{\circ} \times 1^{\circ}$) with the cataloged components overlaid are visually inspected.
This inspection reveals $\approx 2$ DRAGNs per square degree that are visible in the image but not picked up by \textsc{DRAGNhunter} (examples given in Figure \ref{fig:egmissing}).
Checking the catalog entries of such sources quickly demonstrates that in these cases one or both of the lobes have components that do not satisfy our original criteria for consideration as a candidate lobe (see Section \ref{ssec:lobepairs}).
Panels a and b of Figure \ref{fig:egmissing} show example `missing' DRAGNs where one component is identified as a candidate lobe (green ellipse) but the other has a peak flux density of $<3\,$mJy/beam and is therefore too faint to be considered.
Panels c and d of Figure \ref{fig:egmissing} show example DRAGNs where both components have angular sizes of $< 3''$ after deconvolution from the VLASS beam.
These components are therefore too compact to be identified as candidate lobes in this work.
All of the DRAGNs where both components satisfy the candidate lobe criteria in the $50\,\text{deg}^{2}$ of \textit{Quick Look} images inspected are identified by \textsc{DRAGNhunter}.

\textsc{DRAGNhunter} thus does a good job at identifying brighter DRAGNs with clear extended lobes.
Where \text{DRAGNhunter} fails is mostly on the fainter sources and those with smaller lobes.
The fainter sources should be picked up more readily by relaxing the minimum brightness limit we employ in this work, and this may be appropriate for the \textit{Single Epoch} VLASS images as they become available.
The VLASS \textit{Single Epoch} images will be of a higher quality than the \textit{Quick Look} images as a result of the use of self-calibration and deeper cleaning during image production \citep{Lacy2022}.
Consequently there should be fewer image quality issues at low signal-to-noise than in the \textit{Quick Look} images.


\subsubsection{Comparisons with Previous Catalogs}

Estimating the completeness of our sample of DRAGNs requires a `ground truth' catalog of all the existing DRAGNs that could be detected in the VLASS images at the sensitivities used by \textsc{DRAGNhunter}.
As yet, such data does not exist.  As an alternative, we compare the on-sky density of double sources identified here with other samples of doubles from the Faint Images of the Radio Sky at Twenty cm survey \citep[FIRST,][]{Becker1995} and the LOFAR Two Metre Sky Survey \citep[LoTSS,][]{Shimwell2017, Shimwell2019}.

\begin{figure*}
    \centering
    \includegraphics[width=1.7\columnwidth]{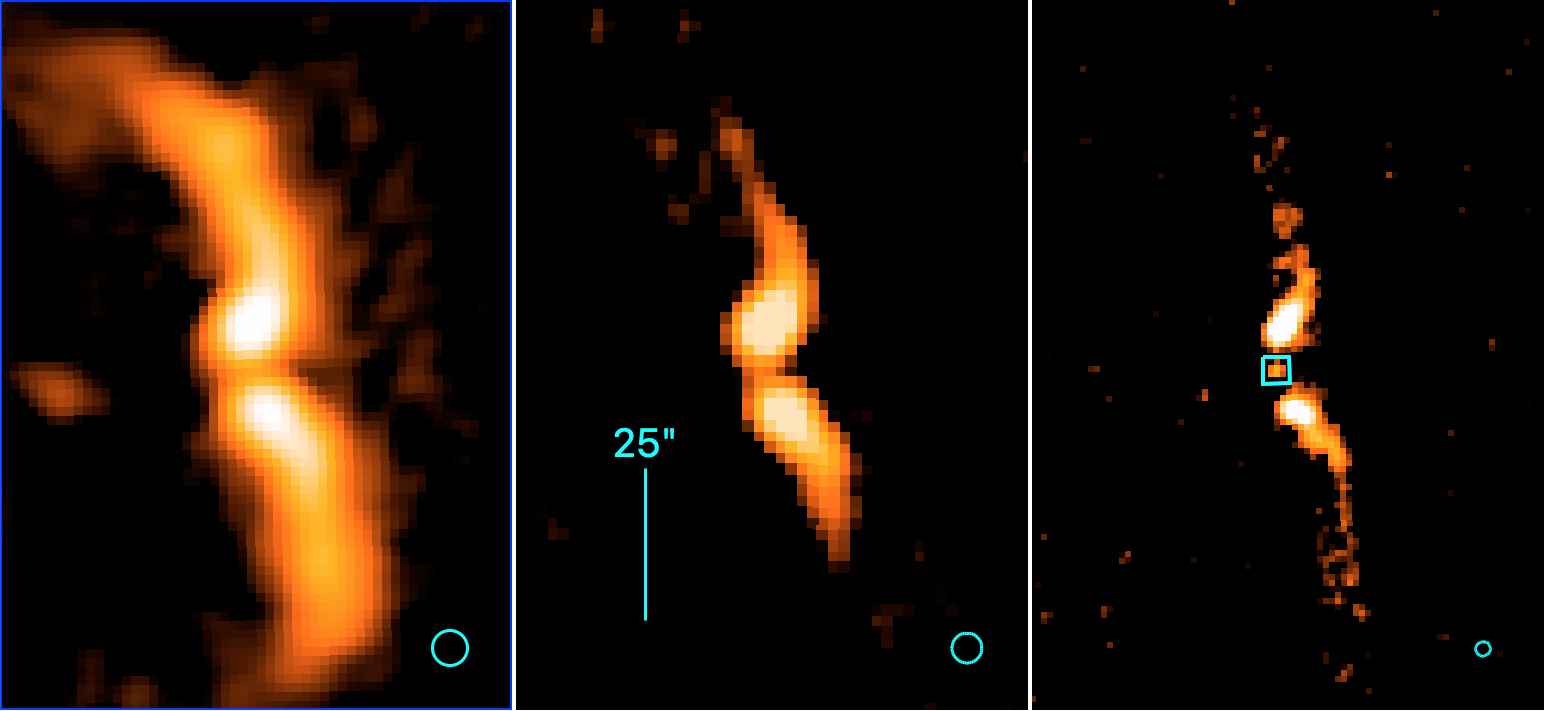}
    \caption{Radio images of the DRAGN J115915.34$+$491729.1 from LoTSS (left), FIRST (middle) and VLASS (right) on the same angular scale.
    The VLASS image shows the radio core (highlighted by the cyan box) that is not seen in the other two surveys, but misses the larger scale low surface brightness emission seen clearly in LoTSS.
    The ellipse in the lower right of each panel shows the beam size and all the images are log scaled.
    The images are each stretched until the background noise just becomes visible. The ranges are $-0.6\,$mJy/beam to $135\,$mJy/beam (LoTSS),  $-0.1\,$mJy/beam to $12.3\,$mJy/beam (FIRST) and  $0.46\,$mJy/beam 0 $3.1\,$mJy/beam (VLASS).
    }
    \label{fig:vlass-first-lotss}
\end{figure*}

It is important to note that both FIRST and LoTSS are expected to capture more diffuse emission than VLASS because of the lack of short uv spacings in its VLA $3\,$GHz, B- and BnA-configuration data.  In Figure \ref{fig:vlass-first-lotss} we show an example DRAGN as seen by VLASS, FIRST and LoTSS.
The LoTSS image clearly captures more extended emission than VLASS and even FIRST.
Additionally, older lobes and plumes in DRAGNs are likely to have steeper spectra and will be preferentially missed at high frequencies. The integrated flux densities from the maps in Figure \ref{fig:vlass-first-lotss} correspond to  $\alpha = -0.8$.  If we instead use the fluxes from the component catalogues, we find steeper values, of $\alpha \approx -1.1$. 
In the following comparisons, we use a more conservative value of $\alpha = -0.7$ to scale between the surveys, but given the presence of steeper emission, especially in the component catalogs,  and expected spectral curvature, the derived completeness values for VLASS likely represent a lower limit to the completeness for what should actually be visible  in the $3\,$GHz radio sky.

For a comparison with FIRST, we extrapolate from \citet{Proctor2011}, who identified  $\approx 90,000$ double and triple sources. 
They used the April 2003 version of the FIRST component catalog \citep{White1997, Becker2003} covering $\approx 9,000\,\text{deg}^{2}$ down to \text{$S_{1.4\,\text{GHz}}\approx 1\,$mJy,} for a source density of $\approx 10\,\text{deg}^{-2}$. 
To compare this to the VLASS results, we first correct for the spectral index, assuming a typical spectral index for radio lobes of $\alpha = -0.7$. 
We then correct for the different sensitivities in the catalogs; $98\,\%$ of our DRAGNs are brighter than $20\,$mJy (see Figure \ref{fig:sizeflux}), corresponding to $S_{1.4\,\text{GHz}} \gtrsim 34\,$mJy, or $17\,$mJy per component. 
This represents only $\approx 10\%$ of FIRST sources, leading to an expected source density of $\approx 1.16\,\text{deg}^{-2}$ just over twice the \textsc{DRAGNhunter} source density of $\approx 0.51\,\text{deg}^{-2}$.
If we look at only the brightest DRAGNs, with $S_{3\,\text{GHz}}>100\,$mJy, the corresponding densities are $0.17\,\text{deg}^{-2}$ in VLASS using \textsc{DRAGNhunter}, compared to  $0.20\,\text{deg}^{-2}$ from \cite{Proctor2011}.

We also compared the \textsc{DRAGNhunter} source densities with those from LoTSS.  
\citet{Mingo2019} cataloged $3,511$ \text{FR Is} and \text{FR IIs} with $S_{150\,\text{MHz}}\gtrsim 1\,$mJy across $424\,\text{deg}^{2}$.
Again, using our $S_{3\,\text{GHz}} = 20\,$mJy comparison, this is equivalent to the $487$ \text{FR Is} and \text{FR IIs} with $S_{150\,\text{MHz}} > 163\,$mJy. This corresponds to  $1.15\,\text{deg}^{-2}$, comparable to the above estimates from FIRST, and twice the density observed in VLASS.  The corresponding numbers for $S_{3\,\text{GHz}}>100\,$mJy are  $0.28\,\text{deg}^{-2}$ from \cite{Mingo2019}, compared to the $0.17\,\text{deg}^{-2}$ in VLASS.  

Comparisons with data from both FIRST and LoTSS suggest that our catalog of DRAGNs is $\approx 45\,\%$ complete at $S_{3\,\text{GHz}}>20\,$mJy. At $S_{3\,\text{GHz}}>100\,$mJy we recover $85\,\%$ of what we might expect based on FIRST, but only $60\,\%$ of what the LoTSS numbers suggest, likely due to the combined effects of uv coverage, steeper spectrum emission than used in the calculations and the non-detection of many FRI sources.


\section{Multiwavelength Counterparts to Radio Sources}
\label{sec:hosts}

\subsection{Host Candidates}
\label{ssec:wisecandidates}

To understand the physics underpinning the evolution of DRAGNs, it is necessary to identify the galaxy hosting the radio source.
Not only does the multiwavelength cross-identification provide information about the galaxy hosting the AGN, but is essential in order to obtain a redshift estimate required to determine, e.g., the luminosity distance of the radio source.
We search for potential counterparts to our DRAGNs that have been detected by the Wide-field Infrared Survey Explorer telescope \citep[WISE,][]{Wright2010}, as WISE provides mid-infrared coverage of the entire sky.
The typical point-spread function of WISE is $6.''1$ in its bluest filter (W1, $3.4\,\mu\text{m}$), and an astrometric precision of better than $0.''5$ is achieved even for faint sources.
To this end, we use the AllWISE catalog \citep{Cutri2012, Cutri2013}, which is around $95\%$ complete at \text{$\text{W1} < 17.1\,$mag (Vega)} to identify host candidates for our DRAGNs.

\begin{figure*}
    \centering
    \subfigure[DRAGNs]{\includegraphics[width=0.66\columnwidth]{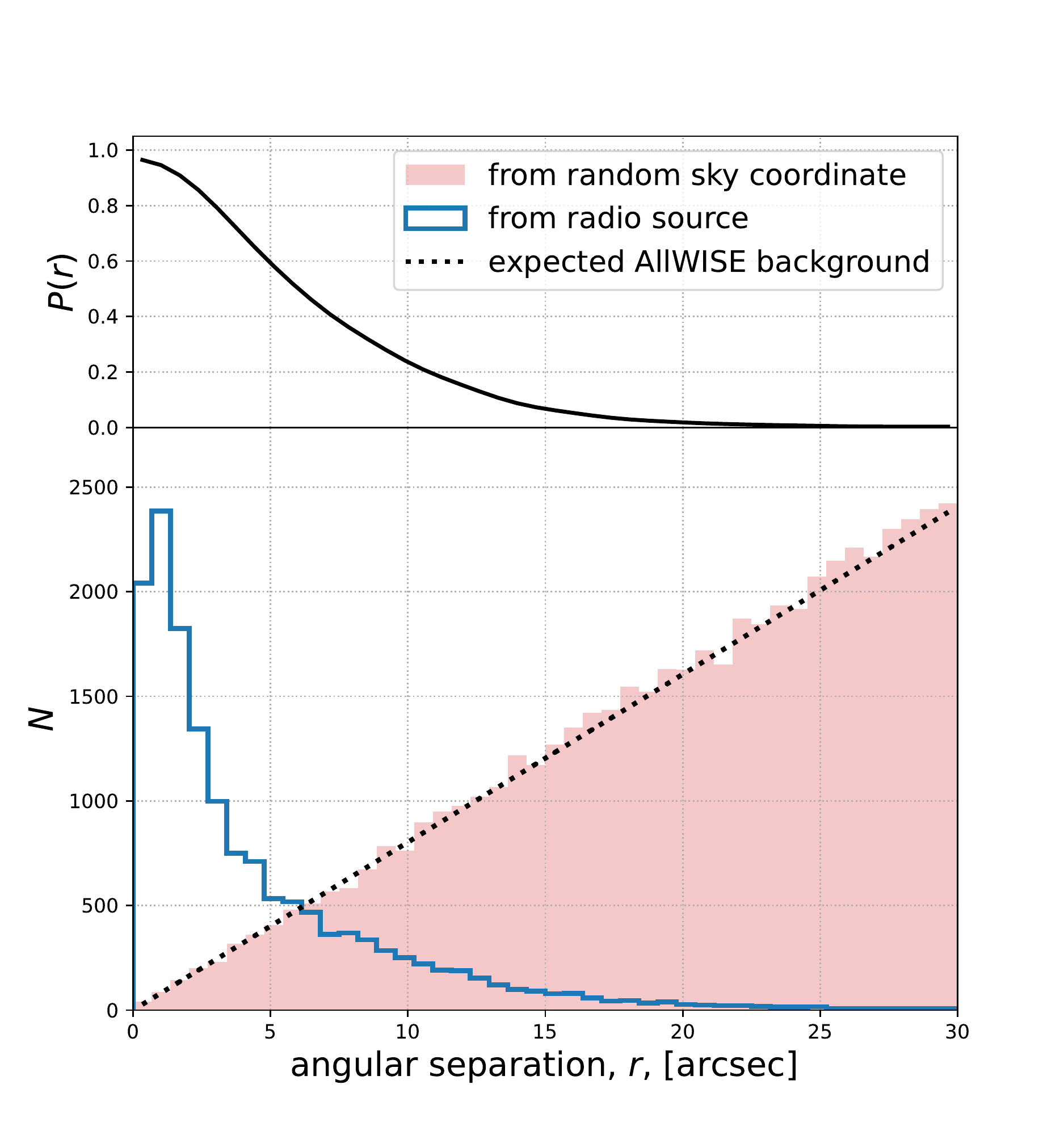}}
    \subfigure[DRAGNs with cores]{\includegraphics[width=0.66\columnwidth]{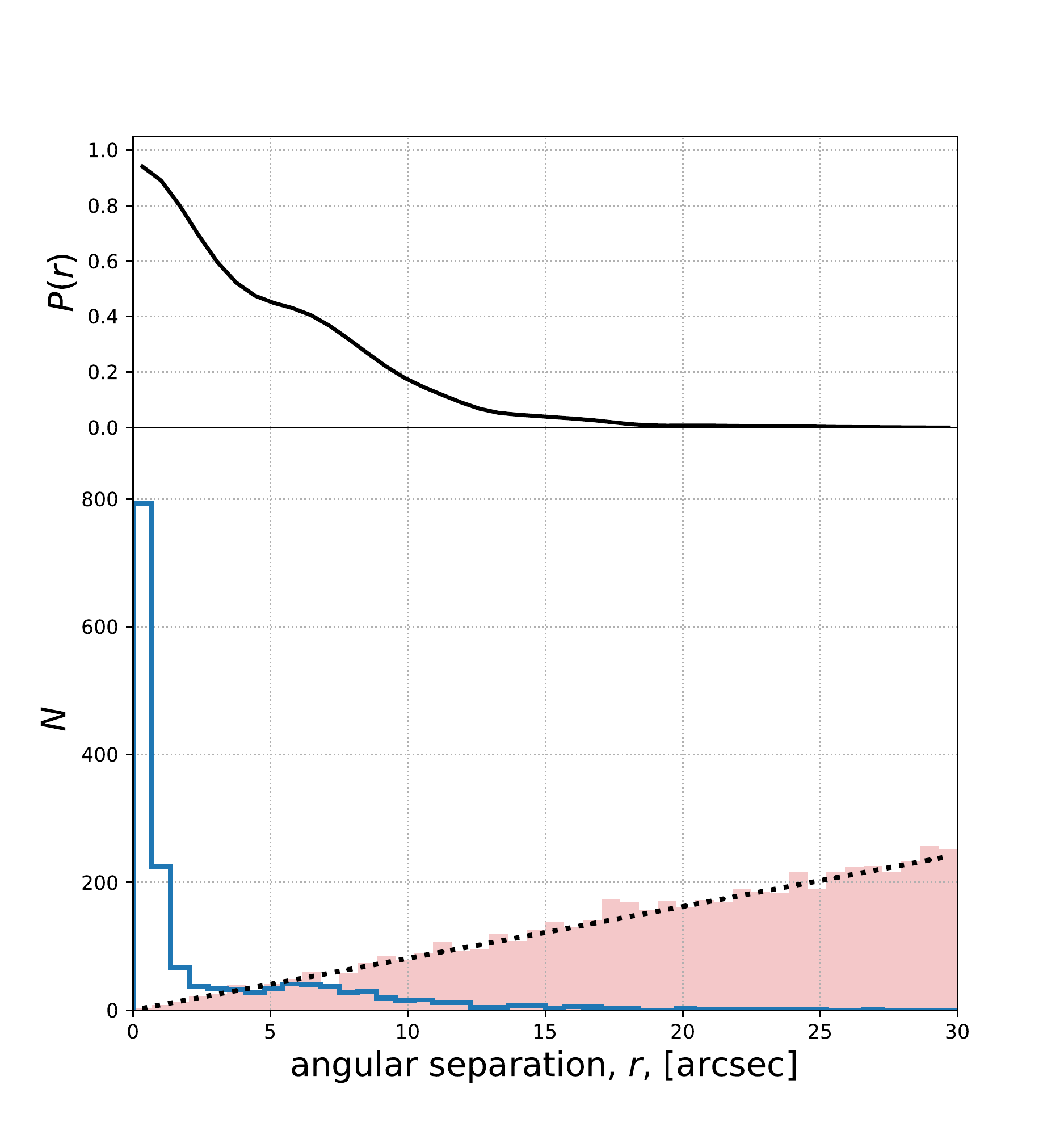}}
    \subfigure[Single-component sources]{\includegraphics[width=0.66\columnwidth]{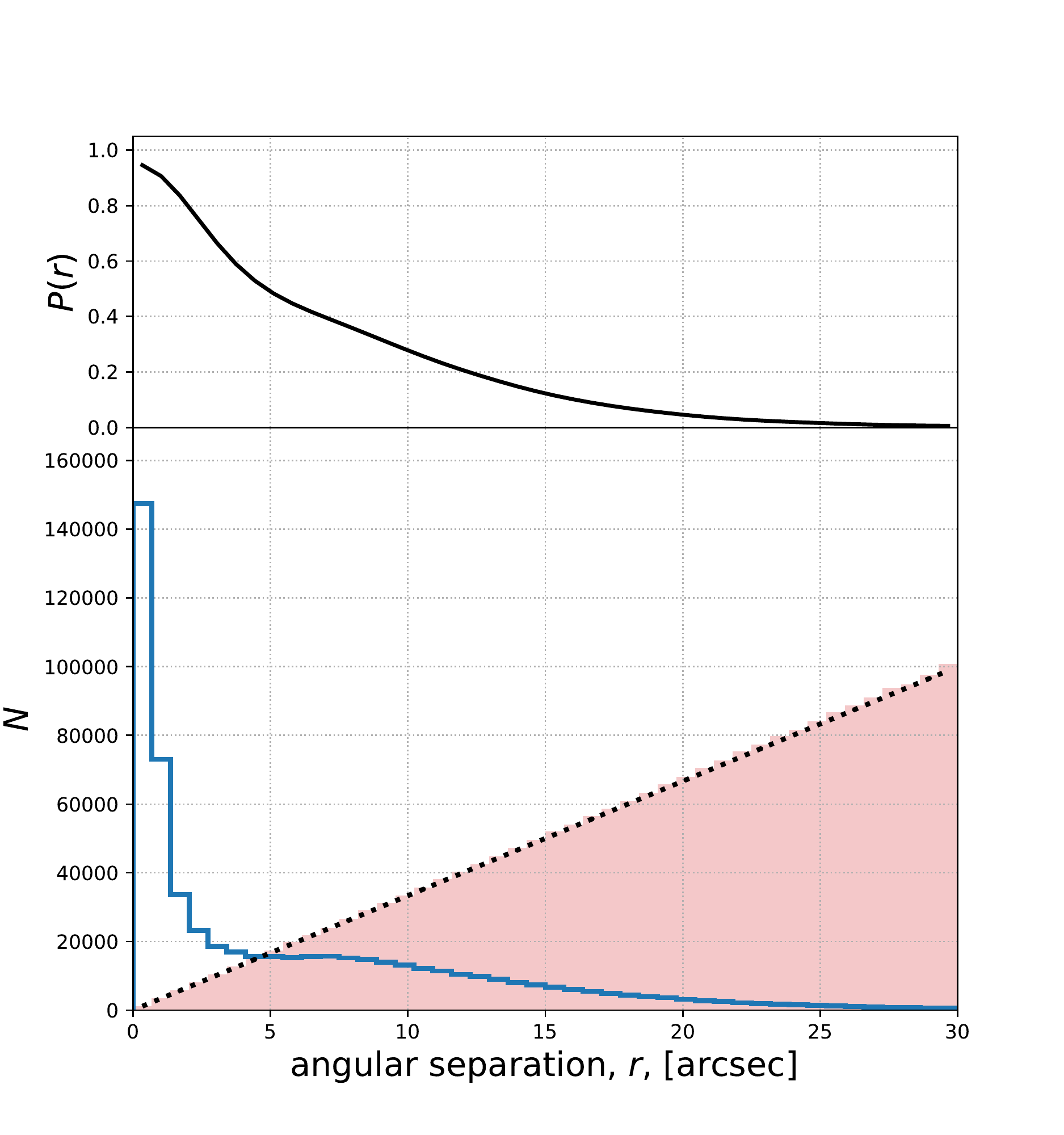}}
    \caption{The lower part of each panel compares the angular separation between radio sources and the nearest AllWISE source (blue solid line) and the number of AllWISE matches as a function of separation from the same number of random sky coordinates (pink solid histogram).
    The expected number of random matches given the AllWISE source density is shown as a black dotted line.
    The upper part of each panel shows the estimate of $P(r)$ determined by equation \ref{eq:pofr} for this data.
    Panel a shows the comparison for our DRAGNs, panel b shows the same comparison only using DRAGNs with cores, and panel c shows the comparison done for single-component sources.
    The `hump' in the $P(r)$ curve at $r\approx 6''$ (most clearly visible in panel b) results from the typical separation at which the \textit{nearest} random AllWISE match will be found.}
    \label{fig:hostsep}
\end{figure*}

We query the AllWISE catalog for sources within $30''$ of the coordinates of the DRAGN.
Where possible, the position of the radio core is taken as the coordinates of the DRAGN, but where a core has not been detected we use the flux-weighted central coordinates of the two lobes as there is an expectation for the brighter lobe to be closer to the host galaxy than the fainter lobe \citep[][see also Section \ref{ssec:fluxsize_symmetry}]{delaRosa2019}.
In Figure \ref{fig:hostsep} (panel a) we show the angular separation to the nearest AllWISE source from our DRAGNs (blue solid line).
For reference we show the number of sources detected when querying from random sky coordinates as a pink solid histogram, as well as the expected background count assuming the AllWISE source density (black dotted line).
Following the approach outlined in \citet{Galvin2020}, the expected background count, $B$, between given match offset radii, $r$ and $r+dr$, using the AllWISE source density, $\rho\approx 17,000\,\text{deg}^{-2}$, is estimated by:
\begin{equation}
    \label{eq:backgroundcount}
    B = N\, \rho\, 2 \pi r\, dr,
\end{equation}
where $N$ is the number of coordinates being searched around.

For any particular angular separation, $r$, one can estimate an approximate probability, $P$, that a real match to the DRAGN will have such an angular offset by taking
\begin{equation}
    \label{eq:pofr}
    P(r) = \frac{N_{\text{match}}(r)}{N_{\text{match}}(r)+N_{\text{background}}(r)},
\end{equation}
where $N_{\text{match}}$ is the number of genuine AllWISE associations with our DRAGNs, and $N_{\text{background}}$ is the number of expected contaminating sources.
In practice we approximate $N_{\text{match}}$ at any given separation using the distribution shown in blue in Figure \ref{fig:hostsep} and $N_{\text{background}}$ using equation \ref{eq:backgroundcount}.
Crucially, by plotting $P(r)$, Figure \ref{fig:hostsep} allows us to obtain a first-order approximation of the positional accuracy, $\sigma_{\text{pos}}$ of our DRAGNs.
Taking the full-width half-maximum of the $P(r)$ curve shown in Figure \ref{fig:hostsep} a, we expect our DRAGNs to have a typical positional uncertainty of $\sigma_{\text{pos}} \approx 6''$.

Panels b and c of Figure \ref{fig:hostsep} show the same comparison of angular separations to AllWISE sources shown in panel a, but for DRAGNs with cores (panel b) and single component sources (panel c).
For single component sources and DRAGNs with cores the expected position of the host galaxy is known and the width of the $P(r)$ curve is driven by astrometric (im)precision.
For DRAGNs where a core is not identified the location of the host is less well constrained, and it is this lack of information rather than any astrometric imprecision that dominates the width of the $P(r)$ curve for DRAGNs.
The adopted positional uncertainty for our DRAGNs thus represents the typical uncertainty in the location of the host for our sample of DRAGNs as a whole and not a hard limit on where we expect to find a host.
In Section \ref{ssec:LR} we use this information to help identify the most likely host out of all potential candidates for each of our DRAGNs.

\subsection{Likelihood Ratio Identifications}
\label{ssec:LR}

\begin{figure}
    \centering
    \includegraphics[width=\columnwidth]{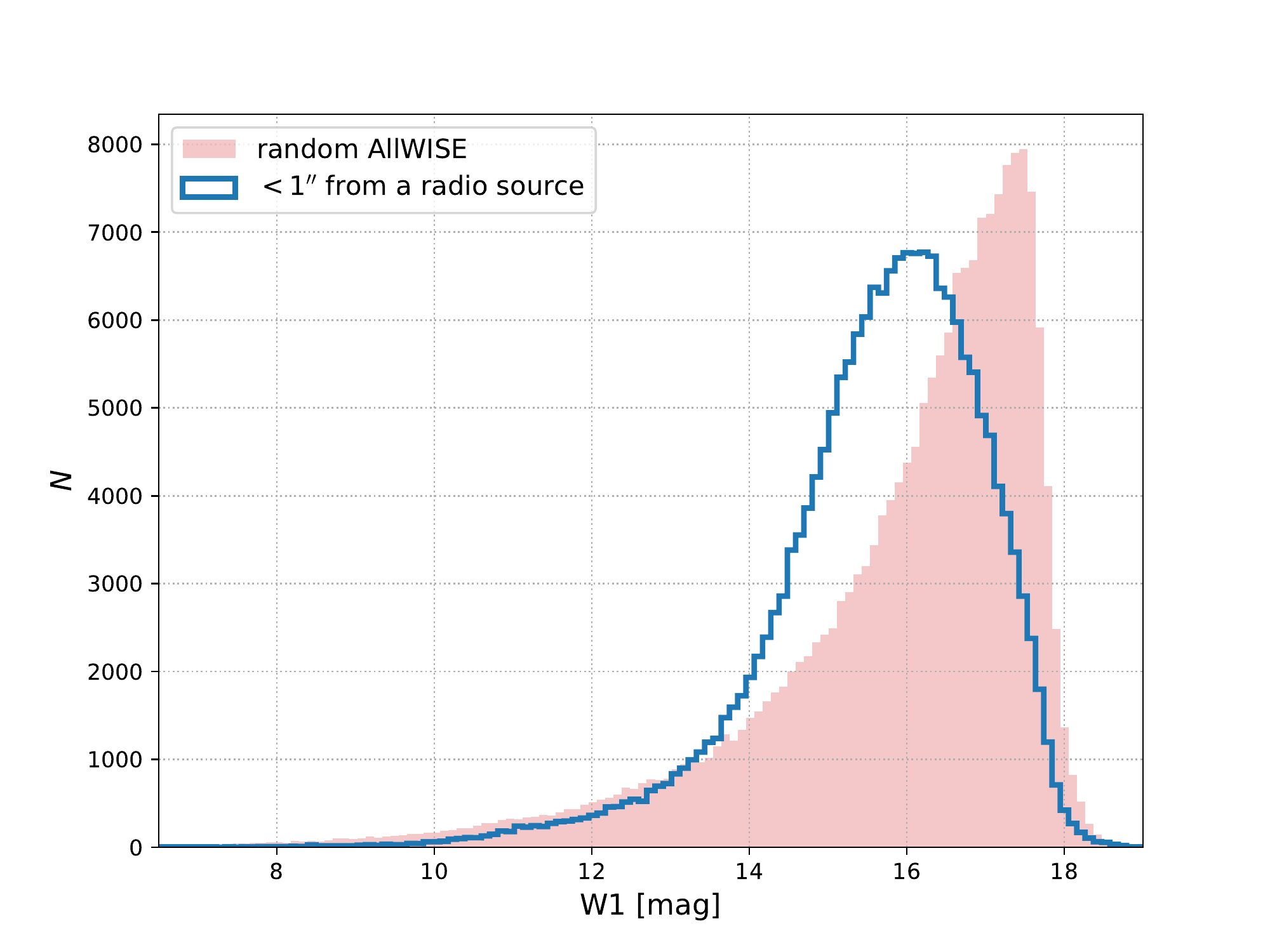}
    \caption{Distribution of W1-band magnitudes for AllWISE sources within $1''$ of a radio source (blue solid line), compared to a random sample of AllWISE sources (pink solid histogram).}
    \label{fig:magdists}
\end{figure}

\begin{figure}
    \centering
    \includegraphics[trim={0 0 0 0}, clip, width=0.95\columnwidth]{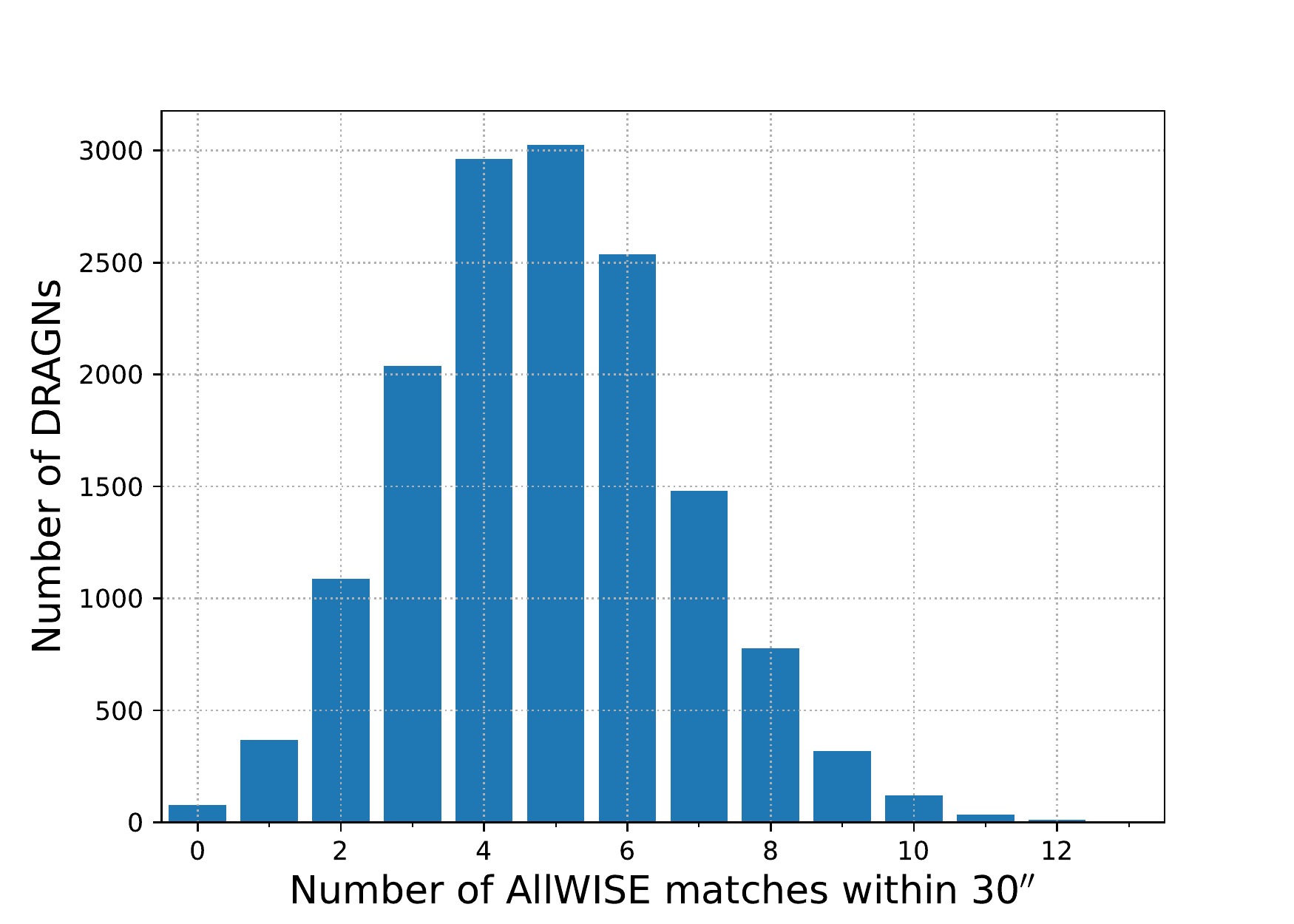}
    \caption{The distribution of the number of AllWISE matches found within $30''$ of the flux-weighted centroid of the DRAGNs.
    }
    \label{fig:hostcount}
\end{figure}

Taking the nearest AllWISE source to our DRAGNs may not be sufficient to identify the correct host.
Firstly, for any one radio/IR match, there is a (usually very small) possibility that the match is the result of a chance alignment of two unrelated sources.
It is therefore helpful to know for any one match how likely it is to be a genuine association, and this can be better constrained by using information about the match beyond just its angular offset.
For instance the hosts of radio sources are typically brighter than the background AllWISE source distribution.
We demonstrate this in Figure \ref{fig:magdists} which compares the W1 magnitudes of AllWISE sources within $1''$ of a VLASS source to those AllWISE sources within $1''$ of a random sky coordinate.
Secondly, the source density of AllWISE is such that there may be multiple host candidates for each of our DRAGNs. 
Indeed, we show the distribution of the number of candidate AllWISE matches for our DRAGNs in Figure \ref{fig:hostcount}.
For cases where multiple candidates are found, knowing the likelihood of each candidate to be the real match allows the best match to be selected.

One approach to finding the correct match between a radio source and an infrared (IR) source is to use the likelihood ratio \citep{Sutherland1992, McAlpine2012}.
In short, this is the ratio of the probability that a matched source is the correct association, to the probability of that match being made by chance.
These probabilities are determined using information (e.g. magnitudes, colors) about both the matched sources and background sources, as well as the angular separations between matches.
Furthermore, the likelihood ratio is especially useful where poor resolution radio observations have multiple IR counterparts \citep{McAlpine2012}, which, to first order, is how we can treat our DRAGNs.

To identify the most probable hosts for our radio sources (both DRAGNs and single-component), we determine the likelihood ratio for all possible matches using the W1-band magnitude information for the AllWISE sources.
Specifically, for this work, we define the likelihood ratio, $\text{LR}$ by:
\begin{equation}
    \label{eq:lr}
    \text{LR} = \frac{q(\text{W}1) f(r)}{n(\text{W}1)}.
\end{equation}
Here, $q(\text{W}1)$ is the probability that the radio source has an AllWISE counterpart with a given magnitude in the WISE W1-band, $f(r)$ is the radial separation probability distribution function for the cross match and $n(\text{W}1)$ is the sky distribution of AllWISE sources of a given W1-band magnitude.

To determine the LR, we adopt the approach detailed in Section 4 of \citet{Williams2019}, with the exception of how we deal with positional errors.
In section \ref{ssec:wisecandidates} we estimated the typical positional uncertainty of our DRAGNs to be $\sigma_{\text{pos}} = 6''$.
However, this large positional uncertainty is unlikely to be appropriate for single-component sources.
Here, the position of the host can be better constrained as, unlike for a pair radio lobes, the multiwavelength counterpart is generally coincident with the radio source.
In order to account for this, we estimate $P(r)$ for single-component sources in a similar fashion to how $P(r)$ was estimated for our DRAGNs (see Figure \ref{fig:hostsep} b).
For our DRAGNs we determined the positional accuracy from $P(r)=0.5$.
As genuine AllWISE matches are expected to be spatially coincident with unresolved radio sources, we take a more conservative approach and determine the positional accuracy at $P(r)=0.8$ to be $\sigma_{\text{pos}} = 1.''8$.

\begin{figure}
    \centering
    \includegraphics[width=\columnwidth]{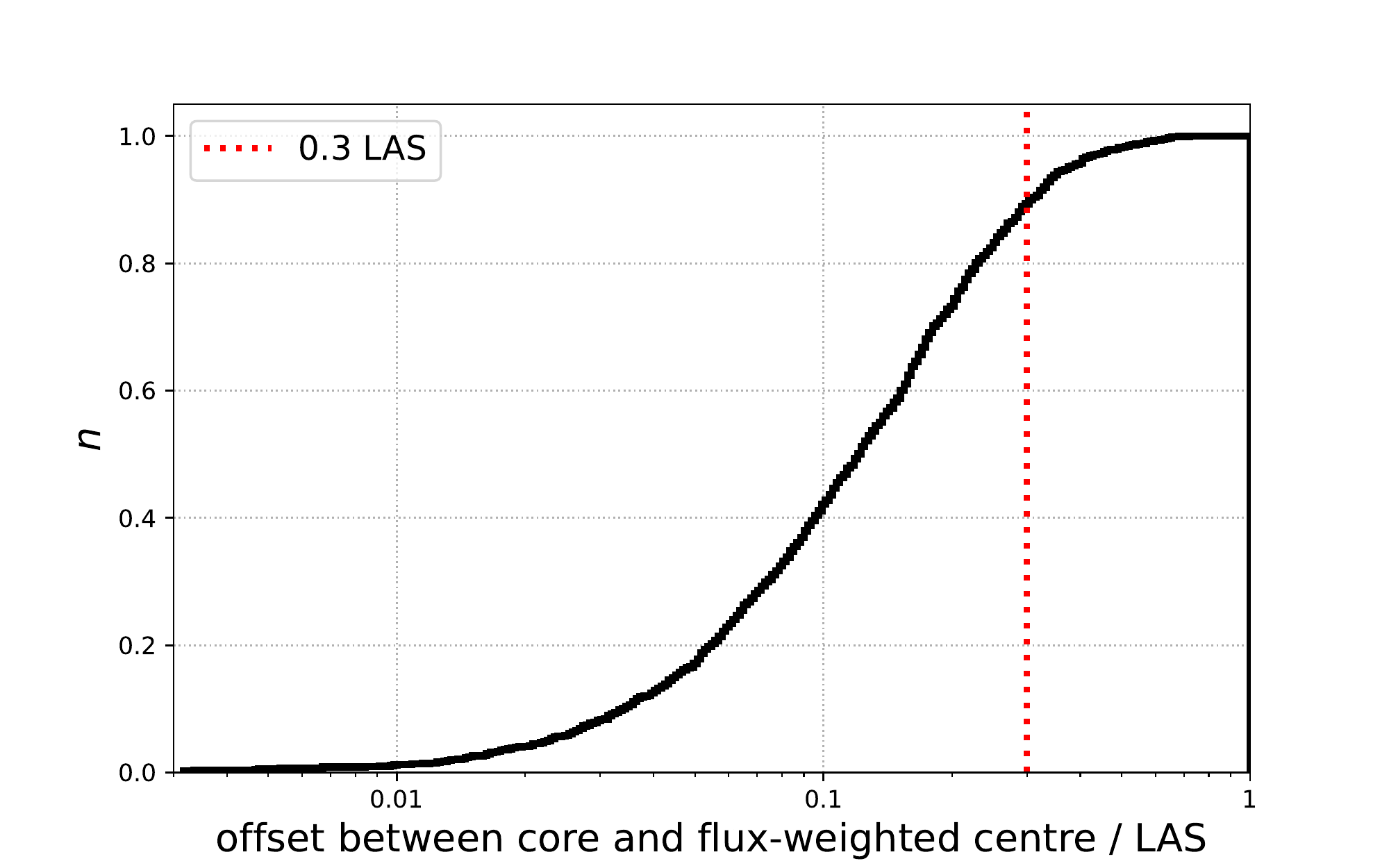}
    \caption{Cumulative distribution function (black solid line) of the angular separation of radio cores from the flux-weighted centroid of the two lobes normalised by the LAS for triple radio sources.
    The red dotted line shows the $0.3\,\text{LAS}$ upper limit we use when finding AllWISE Host IDs for our DRAGNs.}
    \label{fig:core_offsets}
\end{figure}

\begin{figure*}
    \centering
    \subfigure[]{\includegraphics[trim={0 0 15mm 15mm}, clip,width=0.99\columnwidth]{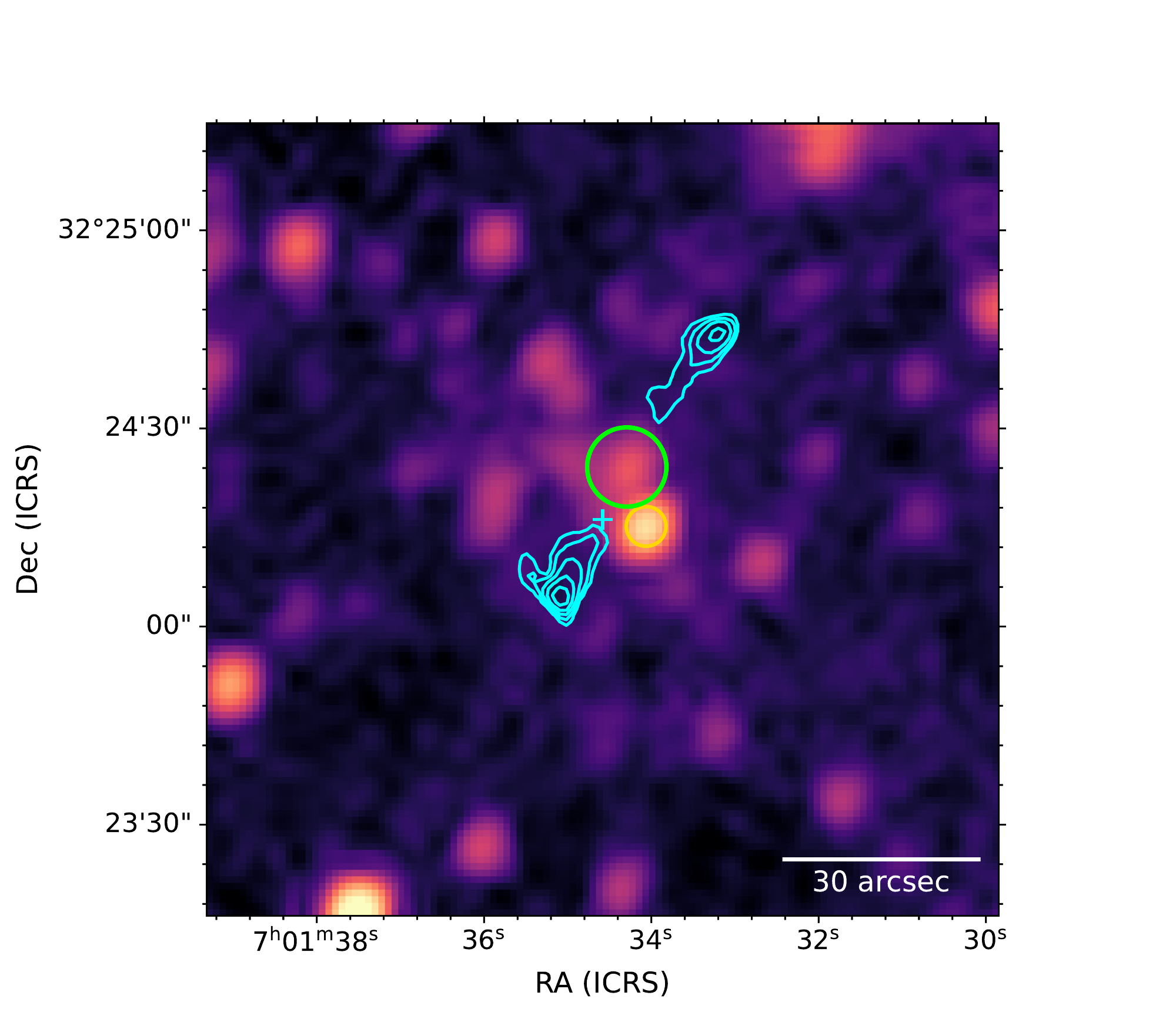}}
    \subfigure[]{\includegraphics[trim={0 0 15mm 15mm}, clip,width=0.99\columnwidth]{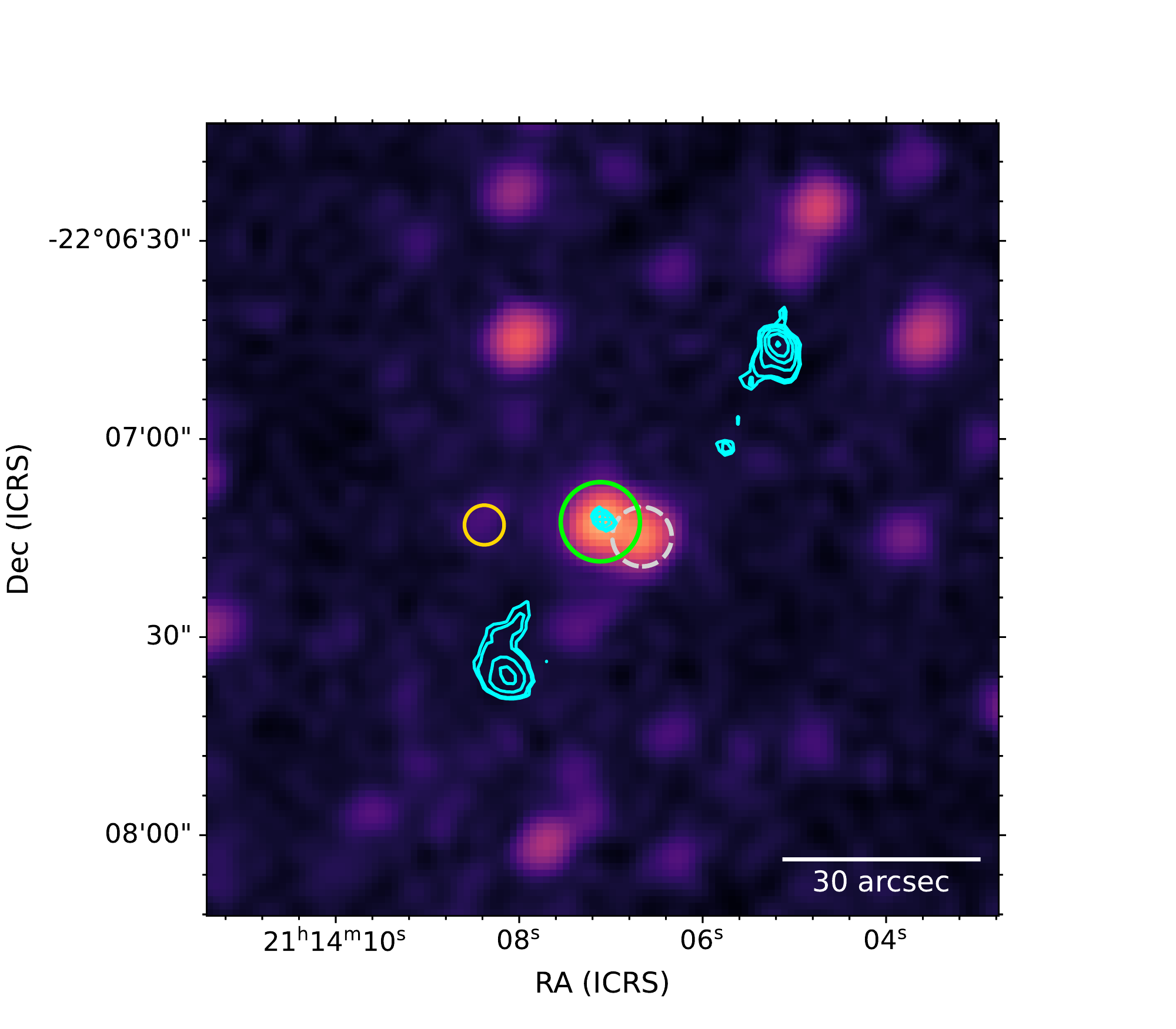}}
    \caption{Example unWISE W1-band images for DRAGNs where multiple host candidates are considered, showing the radio emission overlaid with cyan contours.
    Yellow circles are used to show AllWISE sources that are initially identified as being near the central position of the DRAGN (cyan cross) but are rejected as the host.
    The larger green circles show the adopted host.
    In panel a no radio core has been identified and the host ID is selected as most probable match from the likelihood ratio ($R=0.75$).
    Panel b shows an example DRAGN where a core has been identified and this information has been used to update the host ID.
    Here, the grey dashed circle shows the AllWISE source that the likelihood ratio identified as the most probable match ($R=0.59$) but that was replaced by the host candidate coincident with the radio core (green circle).}
    \label{fig:egmultihosts}
\end{figure*}

Using a search radius of $30''$ we query the AllWISE catalog to create a pool of likely matches to our radio sources.
For sources smaller than $\approx 1'$ this search radius can result in candidates being considered that are not located between the two lobes.
To counter this possibility we additionally only consider matches with an angular separation of $r < 0.3\,\text{LAS}$.
For triple sources, the core component is found within $0.3\,\text{LAS}$ of the flux-weighted centroid $90\,\%$ of the time (see Figure \ref{fig:core_offsets}), and we therefore consider that host candidates offset from the centroid of the DRAGN by more than $0.3\,\text{LAS}$ are unlikely to be realistic candidates.
For single-component sources, where $\sigma_{\text{pos}} > 0.3\,\text{LAS}$ we consider all matches with $r < \sigma_{\text{pos}}$ to be realistic.
Those matches we consider unrealistic are masked out from consideration before determining the LRs of the matches.

Knowing the LR for all possible AllWISE matches to a radio source, the reliability of any given match, $R_{i}$, is determined by:
\begin{equation}
    \label{eq:lrrel}
    R_{i} = \frac{\text{LR}_{i}}{\sum_{j=1}^{N}\text{LR}_{j} + (1-Q_{0})},
\end{equation}
for the match between the radio source and $i$th AllWISE candidate out of $N$ possible matches.
Here, $Q_{0}$ is an estimate of the fraction of radio sources with an AllWISE match \citep{Fleuren2012} and for any given radio source $\sum_{i=1}^{N} R_{i} = 1$.
For our sources we adopt as the host ID the AllWISE matches with reliability of $R>0.5$.
In panel a of Figure \ref{fig:egmultihosts} we show an unblurred WISE \citep[unWISE,][]{Lang2014} W1-band image for an example DRAGN with multiple AllWISE counterparts where the likelihood ratio has been useful in identifying the probable host.

\subsection{Additional Information from Radio Cores}
\label{ssec:hostcores}

For the purposes of host identification we have so far effectively treated our DRAGNs as though they were poorly resolved single-component sources.
However, of the DRAGNs for which host candidates are identified, $1,544$ also have a radio core.
Radio emission from a core will be spatially coincident with the AGN host galaxy.
Thus the core can be treated as an effective compact source and used to robustly identify the correct host.
With this additional information in hand, we can assess the host IDs obtained from the likelihood ratio, and update these where necessary.

There are three possible scenarios where both a core and host have been identified independently of one another for a DRAGN.
First, the core and host are co-located on the sky, which we define here as being separated by less than $1.''8$ (the same value as our adopted $\sigma_{\text{pos}}$ for single-component sources).
We find this to be the case for $1,144$ DRAGNs with both core and host identifications ($74\,$\%), and accept these host IDs as being correct.
Second, the core may be spatially coincident with an alternative host candidate, rather than the one with the highest likelihood ratio.
We find this to be the case for $29$ DRAGNs ($2\,$\%).
In such cases we update the host ID to the candidate determined by the core.
In panel b of Figure \ref{fig:egmultihosts} we show an example where the likelihood ratio would suggest an incorrect host for the DRAGN.
Here, the most likely candidate has $R=0.59$ (grey dashed circle in Figure \ref{fig:egmultihosts}b).
However, a radio core is coincident with one of the other host candidates ($R=0.40$, green circle in Figure \ref{fig:egmultihosts}b) allowing us to confidently adopt this AllWISE source as the host for the DRAGN.
Third, the core and host ID are not spatially coincident and the core is not co-located with an alternative host candidate, e.g., as a result of the real host being too faint to be detected in AllWISE.
For the $371$ ($24\,$\%) DRAGNs where this is the case we do not trust the host ID and consider the source to have no AllWISE counterpart.
It is also worth noting that the triple sources with misidentified hosts highlights the fact that the LR derived host IDs represent the \textit{most probable} host for each radio source and as such has a chance of being incorrect.
To aid others in using our data we include both the likelihood ratio and reliability for the host IDs in our catalog (see Appendix \ref{apx:data-model} for details). 

Taking into account that $24\,\%$ of the host IDs for triples are untrustworthy, reliable host IDs are found for $\approx 64\,\%$ of triple sources. 
Assuming a similar global host reliability for the entire catalog of DRAGNs suggests that $\approx55\,\%$ of our $\approx17\,000$ DRAGNs have robust host IDs.
The apparent improvement in the cross ID rate when a radio core is present is likely the result of having a more precise starting point when searching for host candidates. 
Recall from Section \ref{ssec:wisecandidates} that for triple sources the position of the radio core is used to search for host candidates, whereas for double sources the flux-weighted centroid of the two lobes is used.
The LR for any host candidate is a function of angular separation between the radio and IR sources--all other things being equal a larger angular offset will result in a lower LR.
One potential consequence of this is that a lower fraction of double sources than triple sources may have host candidates with $R>0.5$.
For example, in the case of a DRAGN with a radio core and three candidate hosts, if one of the host candidates is in fact the correct host it will have a very small angular offset from the radio position.
Consequently the correct host will likely have a substantially higher LR value than the two other candidates that are at larger angular offsets, leading to a situation where the correct host has $R$ close to unity and the other two candidates have $R\approx 0$.
However, if the radio core had not been detected, then the additional uncertainty in the radio position can lead to cases where the closest two host candidates are separated from the flux-weighted centroid adopted as the radio position by several arcseconds.
In cases where the host candidates also have similar magnitudes in the W1-band, this can result in multiple candidates having similar LR values such that even the most likely candidate has $R<0.5$.

\begin{figure}
    \centering
    \includegraphics[width=\columnwidth]{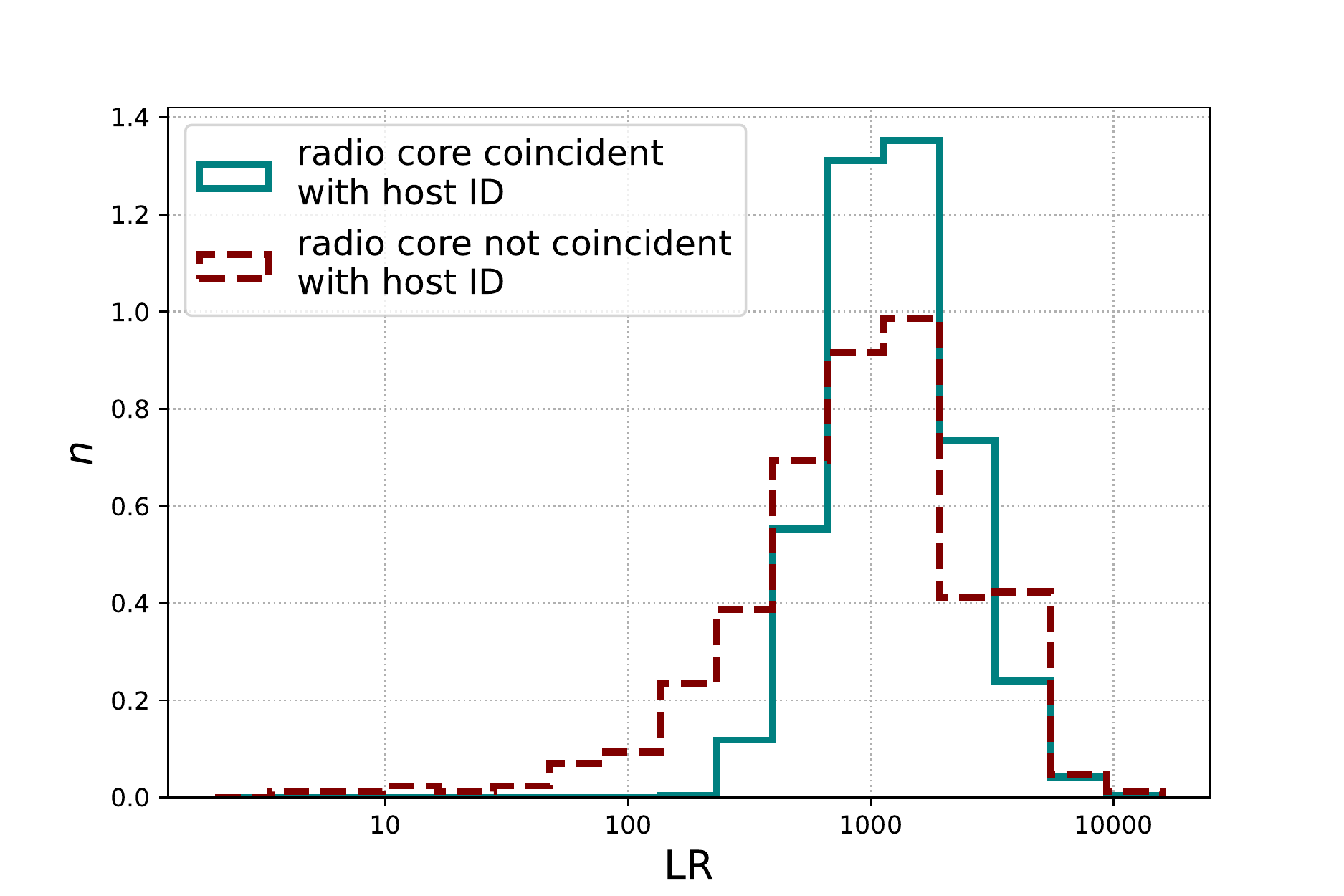}
    \caption{Comparison of the distributions of likelihood ratio values (LR) for triple sources where the host ID is spatially coincident with the radio core (teal solid line) and where the host ID is not co-located with the core (maroon dashed line).
    }
    \label{fig:corehostLR}
\end{figure}

Whilst the presence of a radio core can be used to confirm, update or reject the LR derived host IDs, this information was not used to determine the LR values.
One might, therefore, ask whether there is any difference in the LR values between the host IDs where the core confirms the host, and those where the core information leads us to reject the host ID.
In Figure \ref{fig:corehostLR} we compare the distributions of LR values for sources where the radio core and LR method agree (teal solid line) and disagree (maroon dashed line) on the host ID.
Although high values of LR are found for both populations, there is a tail to lower LR values seen for sources where the LR derived host ID and radio core are not spatially coincident that is not present for sources where these two approaches agree on the host.
The median LR for sources where the core confirms the host is $\approx 1,200$, compared to $\approx 1,000$ for those where the core refutes the nominal host. 
Even though the LR approach may sometimes misidentify the correct host ID from the available candidates, the LR itself may be lower in such cases.
Furthermore, it is notable that only for a small fraction of cases ($7\,\%$) where the LR host ID was shown to be incorrect was an alternative host candidate identified by the core, suggesting that IR imaging depth is driving the misidentifications.
Deeper IR (or optical) imaging relative to the radio depth is likely the key to improve the reliability of host identifications.  

After updating the host information where appropriate in DRAGNs with a radio core, we identify likely hosts for $12,950$ DRAGNs.
Furthermore, on rechecking the validation sample (see Section \ref{ssec:reliability}) after performing the host finding, we note that the probability of DRAGNs in our catalog being genuine is higher than the overall reliability of the catalog at $93.1_{-1.6}^{+1.1}\,\%$ when a host is identified.
This is probably the result of an IR counterpart between two real radio lobes being more likely than a random interloper between two associated radio sources at small angular separations.
An additional $234,033$ hosts are identified for the single-component sources.

\subsection{Redshifts}
\label{ssec:redshifts}

\subsubsection{Spectroscopic Redshifts}

In order to determine physical properties such as the linear size and luminosity of our DRAGNs, we must first determine the redshift, and, by extension, the distance of these sources.
Ideally, redshift is determined from spectroscopic observations of the host galaxy in order to get the most precise measurement.
To identify spectroscopic redshifts (spec-zs) for our radio sources (both single-component and DRAGNs), we cross match the host IDs with a number of legacy catalogs of spectroscopic data.
Namely, these are the Sloan Digital Sky Survey \citep[SDSS,][]{York2000, Blanton2017} Data Release 16 \citep[DR16,][]{Ahumada2020}, 
the third data release of the Galaxy and Mass Assembly survey \citep[GAMA,][]{Driver2011, Baldry2018},
the two-degree Field Galaxy Redshift Survey \citep[2dFGRS,][]{Colless2001}, 
the six-degree Field Galaxy Survey \citep[6dFGS,][]{Jones2004, Jones2005}, 
the WiggleZ survey \citep{Drinkwater2010, Drinkwater2018}, 
and the two-Micron All Sky Survey \citep[2MASS,][]{Skrutskie2006} redshift survey \citep[2MRS,][]{Huchra2012}.
Overall we find spec-zs for $32,761$ sources, $1,286$ of which are for DRAGNs \footnote{For sources where we don't find a spec-z in the legacy catalogs we check, smaller legacy catalogs or the wider literature may be able to provide spec-zs in some cases.}.
A full breakdown of how many spec-zs are obtained from each survey is given in Table \ref{tab:zfrom}.

\subsubsection{Photometric Redshifts}

While spectroscopic measurements are preferred for determining redshift, the time and expense of obtaining spectra means that the vast majority of radio galaxies--and more than $90\,\%$ of our DRAGNs--have not yet been observed in this manner.
In these cases, photometric redshifts (photo-zs) can provide an alternative to spec-z measurements.
As these are based solely on imaging data, photo-zs are often available for a much larger number of sources than spec-zs are, and are now frequently produced for wide-field imaging surveys \citep[e.g.,][]{Beck2016, Beck2021, Zhou2021}.

\begin{figure}
    \centering
    \includegraphics[width=\columnwidth]{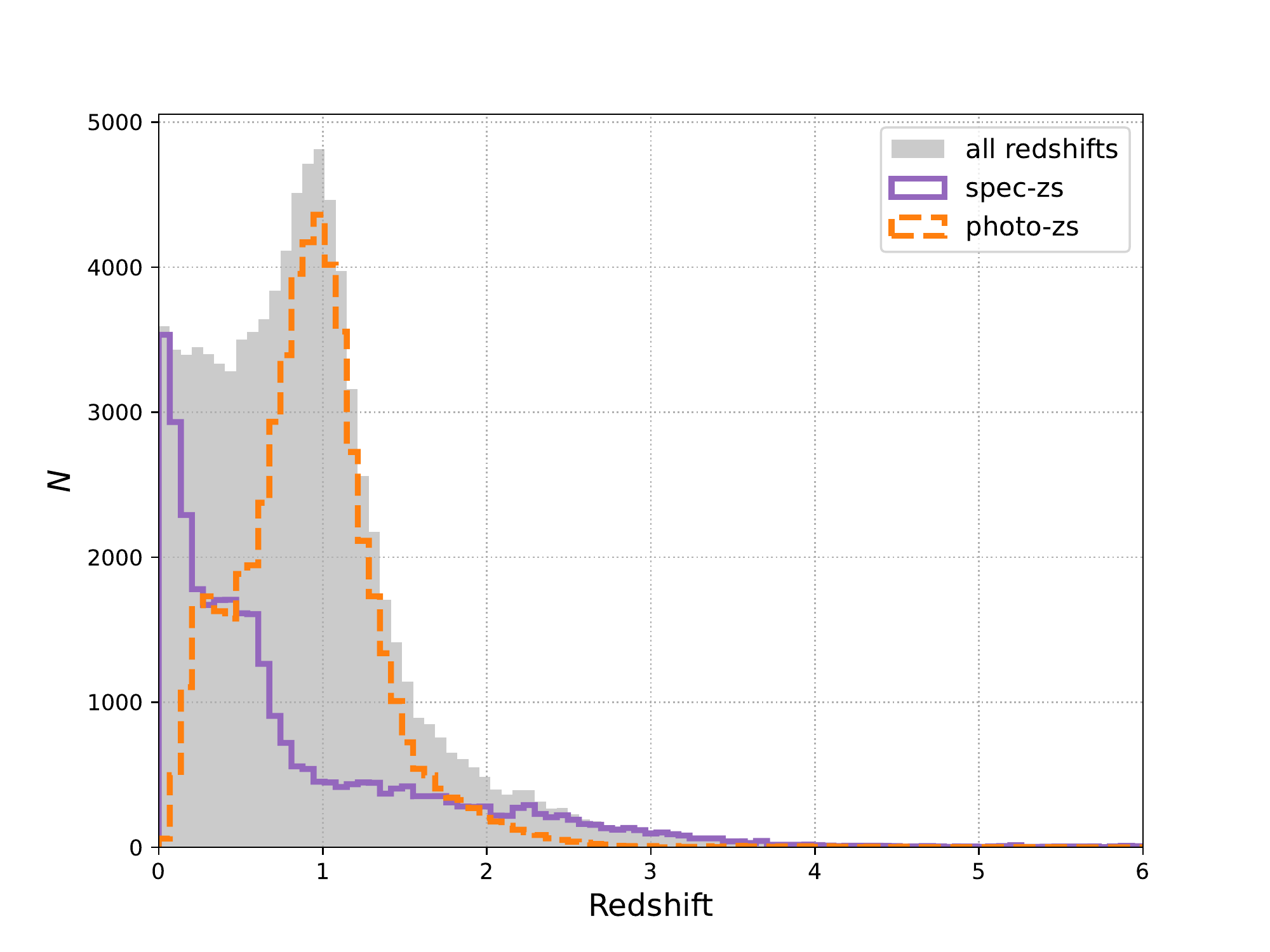}
    \caption{The redshift distribution of our radio sources (single-component and DRAGNs). The grey histogram shows all redshifts, spec-zs are shown by the purple solid line and the orange dashed line shows photo-zs.}
    \label{fig:zdists}
\end{figure}

In order to increase the number of DRAGNs in our sample with redshifts we cross match our hosts that do not have a spec-z with Data Release 8 of the Dark Energy Survey Spectroscopic Instrument (DESI) imaging Legacy Surveys \citep[LS DR8,][]{Dey2019} photo-z catalog \citep{Duncan2022}.
The \citet{Duncan2022} LS DR8 photo-zs are determined by Gaussian Mixture Models \citep[GMMs,][]{Bovy2011} making use of optical and infrared photometry from the $g$-, $r$-, $z$-, W1- and W2-bands.
Importantly for this work, the GMM derived photo-zs for LS-DR8 have been demonstrated to be more reliable for RLAGN than other alternative photo-zs for the DESI imaging Legacy Surveys \citep{Duncan2022}. 
Cross matching with the \citet{Duncan2022} LS DR8 photo-z catalog provides an additional $51,536$ and $2,552$ redshifts for our single-component sources and DRAGNs respectively.
In total, $83,011$ single-component sources and $3,838$ DRAGNs have either a spec-z or photo-z measurement available ($\approx 30\,\%$ of sources with a host).
The distributions of the collated redshifts are shown in Figure \ref{fig:zdists} split by redshift type (spec- or photo-z).

\begin{deluxetable}{lccc}
    \tabletypesize{\footnotesize}
    \tablecaption{The number of redshifts obtained from different redshift surveys.
    \label{tab:zfrom}}
    \tablewidth{0pt}
    \tablehead{
    \colhead{Redshift survey} & \colhead{$N_{\text{DRAGNs}}$} & \colhead{$N_{\text{single-component}}$} & \colhead{$N_{\text{total}}$}
    }
    \startdata
    SDSS DR16 & $1,150$ & $26,968$ & $28,118$\\
    6dFGS & $76$ & $2,513$ & $2,589$\\
    2MRS & $28$ & $1,205$ & $1,233$\\
    WiggleZ & $22$ & $356$ & $378$\\
    2dFGRS & $9$ & $341$ & $350$\\
    GAMA & $1$ & $92$ & $93$\\
    LS DR8 (photo-zs) & $2,552$ & $51,536$ & $54,088$\\
    \enddata
\end{deluxetable}

\section{Properties of the DRAGN population in VLASS}
\label{sec:dragnproperties}

\subsection{The Sizes and Luminosities of DRAGNs}
\label{ssec:sizelum}

\begin{figure}
    \centering
    \includegraphics[width=\columnwidth]{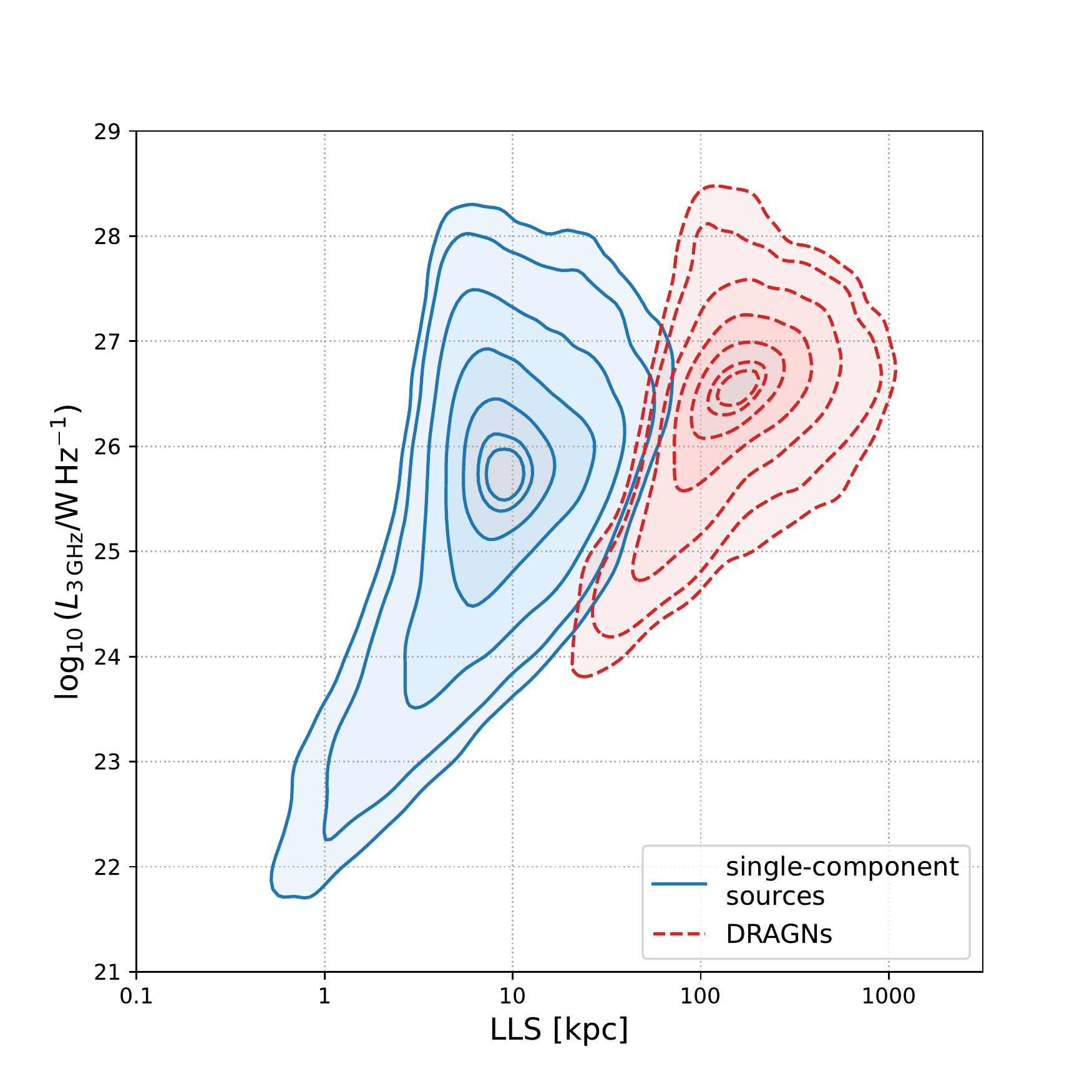}
    \caption{The linear-size versus luminosity plot (a \textit{P-D} diagram) for our DRAGNs (red dashed contours) and single-component sources (blue solid contours).
    The contour levels contain $5$, $10$, $25$, $50$, $75$, $90$ and $95\,$\% of each data set.
    }
    \label{fig:PD}
\end{figure}

We now turn to exploring the properties of the DRAGN population.
One of the most important properties of a radio galaxy is its luminosity. 
For the DRAGNs where we have obtained a redshift, we calculate their $3\,$GHz luminosity.
The flux density measurements used for this are scaled up by a factor of $1/0.87$ in order to account for the systematic flux under-measurement in the VLASS epoch 1 \textit{Quick Look} component catalog \citep[for a detailed description see Section 3.2 of][]{Gordon2021}.
Radio source luminosities are often compared to their projected largest linear size (LLS) on a power versus diameter (\textit{P-D}) diagram \citep[e.g.,][]{Baldwin1982, Blundell1999, An2012, Hardcastle2019, Mingo2022}.
We determine the LLS for our DRAGNs and show them on a \textit{P-D} diagram in Figure \ref{fig:PD}.
For comparison, we additionally show the linear size and luminosity distributions of single-component sources, excluding the unresolved `zero size' sources.

The DRAGNs are generally higher-power sources than their single-component counterparts by nearly an order of magnitude, with a median $3\,$GHz luminosity of $10^{26.5}\,\text{W}\,\text{Hz}^{-1}$ compared to $10^{25.7}\,\text{W}\,\text{Hz}^{-1}$ for single components.
On the \textit{P-D} plane, our DRAGNs occupy a region typically inhabited by FR II radio galaxies (see, e.g., Figure 7 of \citealp{Jarvis2019}, Figure 5 of \citealp{Mingo2019} or Figure 2 of \citealp{Hardcastle2020}).
It has been shown using LoTSS that some low luminosity FR IIs occupy the regions  of \textit{P-D} space classically dominated by FR Is \citep{Mingo2019}, suggesting that these populations are not cleanly segregated on the \textit{P-D} diagram.
LoTSS is a low frequency survey with high sensitivity to low surface brightness emission.
Conversely, VLASS is a high frequency survey, and the B- and BnA-configurations used by the VLA for VLASS observations lack the short baselines needed for sensitivity to diffuse emission.
The resultant selection effects inherent to VLASS, as well as \textsc{DRAGNhunter}'s strategy of requiring distinct components for each lobe, likely bias our DRAGNs towards those dominated by hotspots rather than the diffuse emission seen in FR Is.
While we have made no attempt at a more in-depth morphological classification of our DRAGNs in this work, the examples shown in Figure \ref{fig:egdoubles} would also suggest our DRAGNs mostly appear as FR IIs in VLASS.

The DRAGN population appears to be consistent with an extension of the single-component source population on the \textit{P-D} diagram such as would be expected from older and/or more intrinsically powerful radio jets \citep{Hardcastle2019, Gurkan2022}.
However, while this may be true \textit{on average} from a population perspective, inferring jet ages and powers for \textit{individual sources} on the \textit{P-D} diagram is complicated by factors such as host galaxy environment, and jet orientation effects \citep[e.g.,][]{An2012, Hardcastle2013, Hardcastle2014, Harwood2020}.
The absence of sources in the lower-right of the \textit{P-D} plane is not a real effect, and rather is driven by surface brightness limitations of the survey images  \citep{Hardcastle2016, Hardcastle2020}.

\subsection{Giant Radio Galaxies}

Giant Radio galaxies (GRGs) are some of the largest structures in the Universe, reaching projected linear sizes larger than $700\,$kpc, and GRGs up to $5\,$Mpc in length have been reported \citep{Willis1974, Bridle1976, IshwaraChandra1999, ODea2001, Dabhade2017, Oei2022}.
Identifying the largest radio galaxies is important in order to aid our understanding of the physics of jet propagation and ageing.
These galaxies typically have large angular extents, ranging from arcminute to degree scales \citep{Cotter1996, Schoenmakers2001, Kuzmicz2018}, and are often best identified in surveys at low frequency and that are sensitive to extended, low-surface brightness emission.
The survey design of VLASS, a relatively high frequency survey utilising the VLA's B- and BnA-configurations which lack short baselines, is thus not optimised for finding GRGs.
Nonetheless, given that we have linear size measurements for more than $3,000$ DRAGNs, we check our catalog for any GRGs that might have been found by VLASS.

\begin{deluxetable*}{lcccccc}
    \tabletypesize{\footnotesize}
    \tablecaption{Newly discovered giant radio galaxies in VLASS identified by \textsc{DRAGNhunter}.
    \label{tab:grgs}}
    \tablewidth{0pt}
    \tablehead{
    \colhead{Name} & \colhead{Flux density\tablenotemark{a}} & \colhead{LAS} & \colhead{Redshift} &
    \colhead{Redshift type} & \colhead{$\log_{10}L_{3\,\text{GHz}}$} & \colhead{LLS}\\
    \colhead{} & \colhead{[mJy]} & \colhead{[arcsec]} & \colhead{} &
    \colhead{} & \colhead{[$\text{W}\,\text{Hz}^{-1}$]} & \colhead{[kpc]}
    }
    \startdata
    J002506.84$-$342644.8 & 71.2 & 88 & 0.995 & photo & 26.78 & 700 \\
    J003022.33$-$090107.0 & 34.2 & 169 & 1.448 & photo & 26.92 & 1428 \\
    J003758.39$-$043651.5 & 23.3 & 118 & 0.524 & photo & 25.53 & 736 \\
    J010324.26$+$313216.6 & 45.4 & 92 & 1.193 & spec & 26.8 & 761 \\
    J011018.20$-$361711.5 & 35.8 & 92 & 1.005 & photo & 26.49 & 739 \\
    J013035.90$-$190120.1 & 73.0 & 90 & 1.036 & photo & 26.84 & 728 \\
    J013651.68$+$004055.7 & 55.2 & 94 & 0.83 & photo & 26.44 & 717 \\
    J013907.23$-$373323.1 & 99.2 & 117 & 0.772 & photo & 26.61 & 865 \\
    J015717.54$+$284734.8 & 108.7 & 137 & 0.841 & photo & 26.75 & 1043 \\
    J040701.36$-$315214.1 & 274.5 & 111 & 1.013 & photo & 27.38 & 894 \\
    J081740.34$+$294920.2 & 28.0 & 98 & 1.1 & photo & 26.49 & 798 \\
    J091452.88$+$225533.8 & 98.1 & 96 & 0.778 & photo & 26.62 & 714 \\
    J100749.11$-$045335.1 & 142.2 & 108 & 0.639 & photo & 26.54 & 744 \\
    J101718.07$+$393127.9 & 329.8 & 138 & 0.531 & spec & 26.69 & 869 \\
    J102214.84$+$174647.8 & 79.3 & 116 & 0.526 & spec & 26.06 & 728 \\
    J105304.35$+$312606.8 & 66.5 & 96 & 0.855 & photo & 26.56 & 736 \\
    J134817.65$+$055743.0 & 116.0 & 107 & 1.046 & photo & 27.05 & 867 \\
    J141622.01$+$590019.5 & 76.7 & 137 & 0.557 & spec & 26.12 & 882 \\
    J144925.52$+$221206.6 & 64.1 & 130 & 0.592 & photo & 26.11 & 862 \\
    J150558.82$-$061609.6 & 50.6 & 127 & 0.598 & photo & 26.02 & 847 \\
    J153230.42$+$241529.5 & 192.4 & 133 & 0.564 & spec & 26.53 & 865 \\
    J154057.76$+$171720.9 & 37.6 & 116 & 0.79 & photo & 26.22 & 867 \\
    J165037.20$+$324218.0 & 112.9 & 128 & 0.516 & photo & 26.19 & 796 \\
    J224402.55$-$095126.3 & 70.2 & 86 & 1.174 & photo & 26.97 & 711 \\
    J224430.84$+$265234.0 & 58.6 & 99 & 0.857 & photo & 26.51 & 759 \\
    J232458.76$+$280329.3 & 353.5 & 111 & 0.898 & photo & 27.35 & 863 \\
    J233451.89$+$080544.7 & 106.3 & 114 & 0.873 & photo & 26.79 & 882 \\
    J233753.38$-$143515.4 & 226.9 & 100 & 0.698 & photo & 26.85 & 711 \\
    J233855.71$-$105924.4 & 47.6 & 99 & 1.491 & photo & 27.1 & 840 \\
    J235725.34$-$113242.8 & 95.0 & 91 & 0.864 & photo & 26.73 & 705 \\
    J235811.70$-$083114.3 & 267.4 & 120 & 0.637 & photo & 26.81 & 828\\
    \enddata
    \tablenotetext{a}{The flux density measurements presented in this table are higher than the cataloged values by a factor of $1/0.87$ in order to account for the systematic underestimation of flux densities in the VLASS \textit{Quick Look} component catalog \citep[see Section 3 of][]{Gordon2021}. It is these values that have been used to estimate the radio luminosities.}
\end{deluxetable*}

For DRAGNs in our catalog listed as having \text{$\text{LLS} > 700\,$kpc,} we select only those with a likelihood ratio derived host reliability greater than $0.8$ or a radio core detection coincident with the host. 
These criteria select $43$ candidate GRGs, which we visually inspect to confirm their nature.
Of the $43$ DRAGNs selected as likely GRGs, we reject four ($9\,$\%) as being contaminants in our DRAGN sample.
Eight further DRAGNs ($19\,$\%) are rejected as the host ID is either incorrect or uncertain upon visual inspection.
Two of the DRAGNs rejected as GRGs have substantially lower likelihood ratios for their host IDs ($\text{LR} \sim 5$) than the rest of the GRG candidates ($\text{LR} \sim 1,000$).
For the other rejected candidates the likelihood ratio values were comparable to those confirmed by visual inspection.
The numbers of rejected candidate GRGs are unsurprising given the overall sample reliability (Section \ref{ssec:reliability}) and the expected failure rate of the host IDs (Section \ref{ssec:hostcores}).
This leaves us with $31$ GRGs, which we list in Table \ref{tab:grgs}.
Two of the GRGs, \text{J003022.33-090107.0} and J015717.54+284734.8, have projected linear sizes greater than $1\,$Mpc.

We cross match our $31$ GRGs with a number of existing GRG catalogs, namely:
\begin{itemize}
    \item A compilation of 349 GRGs from the literature by \citet{Kuzmicz2018}; 
    \item 272 GRGs identified by \citet{Kuzmicz2021} in the NRAO VLA Sky Survey \citep[NVSS][]{Condon1998} and SDSS;
    \item 162 GRGs identified by the Search and Analysis of Giant radio galaxies with Associated Nuclei (SAGAN) project \citep{Dabhade2020SAGAN}; 
    \item more than $2,200$ GRGs in LoTSS \citep{Dabhade2020LoTSS, Oei2022lotssgrgs};
    \item 55 GRGs in the ROGUE I (Radio sources associated with Optical Galaxies and having Unresolved or Extended morphologies I) catalog \citep{Koziel-Wierzbowska2020}.
\end{itemize}
A catalog of GRGs in RACS \citep{Andernach2021} contains an additional $178$ GRGs, but this catalog is limited to $\delta < -40^{\circ}$ and therefore does not overlap with VLASS.
In combination, these data sets provide a comprehensive list of all the previously reported GRGs.
None of our $31$ GRGs are identified in the above data sets, suggesting that these are indeed newly discovered giants. 
It is likely that these GRGs were not previously identified as such due to a lack of host IDs and/or redshift measurements.

These $31$ GRGs have been discovered despite neither VLASS nor \textsc{DRAGNhunter} being optimised to find sources with very large, multi-arcminute scale, angular extents.
In this work we have identified hosts and redshifts from existing survey data using an automated procedure, and more GRGs may be found by using the catalogued DRAGNs as a starting point for more thorough search.
For instance, using our adopted cosmology, $700\,$kpc will always correspond to $\text{LAS}>80''$.
In our catalog $576$ DRAGNs have such large angular sizes, but we have only identified redshift measurements for $124$ of these.
It is possible that at least some of the $452$ remaining GRGs with $\text{LAS}>80''$ may have redshift measurements available from legacy data we have not searched, either from the literature or additional surveys such as the Panoramic Survey Telescope and Rapid Response System (Pan-STARRS) survey \citep{Chambers2016}.
Alternatively, further GRGs may be identified by relaxing the size criteria used by \textsc{DRAGNhunter} to reduce contamination from spurious `double source' detections.
Such a dedicated search for GRGs is beyond the scope of this work, but presents tantalising opportunities for future studies of the largest radio galaxies.

\subsection{The Host Galaxies of DRAGNs}

Knowing the AllWISE counterparts to our radio sources provides information on the galaxies themselves that host the AGN.
A common diagnostic plot for IR sources is the WISE color-color diagram that compares the W1$-$W2 color to W2$-$W3 color, where the WISE W1, W2 and W3 filters are centered on wavelengths of $3.4\,\mu$m, $4.3\,\mu$m and $12\,\mu$m respectively.
To this end, we select radio sources with reliable WISE magnitudes, i.e., those with $S/N > 3$ in the W1, W2 and W3 bands. 
Additionally, so as not to include objects where the photometry may suffer from blending in the high-density galactic plane, we exclude sources with low galactic latitudes, $|b|<10^{\circ}$ (approximately $8\,\%$ of our sources with host IDs lie this close to the galactic equator).

The single-component radio source sample may also contain blazars and star-forming galaxies, as well as the smaller angular scale radio galaxies with which we wish to compare our DRAGNs.
Likely blazars are removed by only selecting sources with $\text{LAS}>3''$, i.e. those that are clearly resolved by VLASS. 
Contamination from likely star-forming galaxies is addressed by only including sources with $L_{3\,\text{GHz}} > 10^{23}\,\text{W}\,\text{Hz}^{-1}$.
Assuming a typical spectral index of $\alpha = -0.7$, in order to produce $3\,$GHz luminosities higher than $10^{23}\,\text{W}\,\text{Hz}^{-1}$ in the absence of an AGN the host galaxies would require star-formation rates in excess of $100\,\text{M}_{\odot}\,\text{yr}^{-1}$ \citep{Bell2003}.
Consequently, we can be confident this population is dominated by RLAGN.

\begin{figure}
    \centering
    \subfigure[DRAGNs]{\includegraphics[width=0.99\columnwidth]{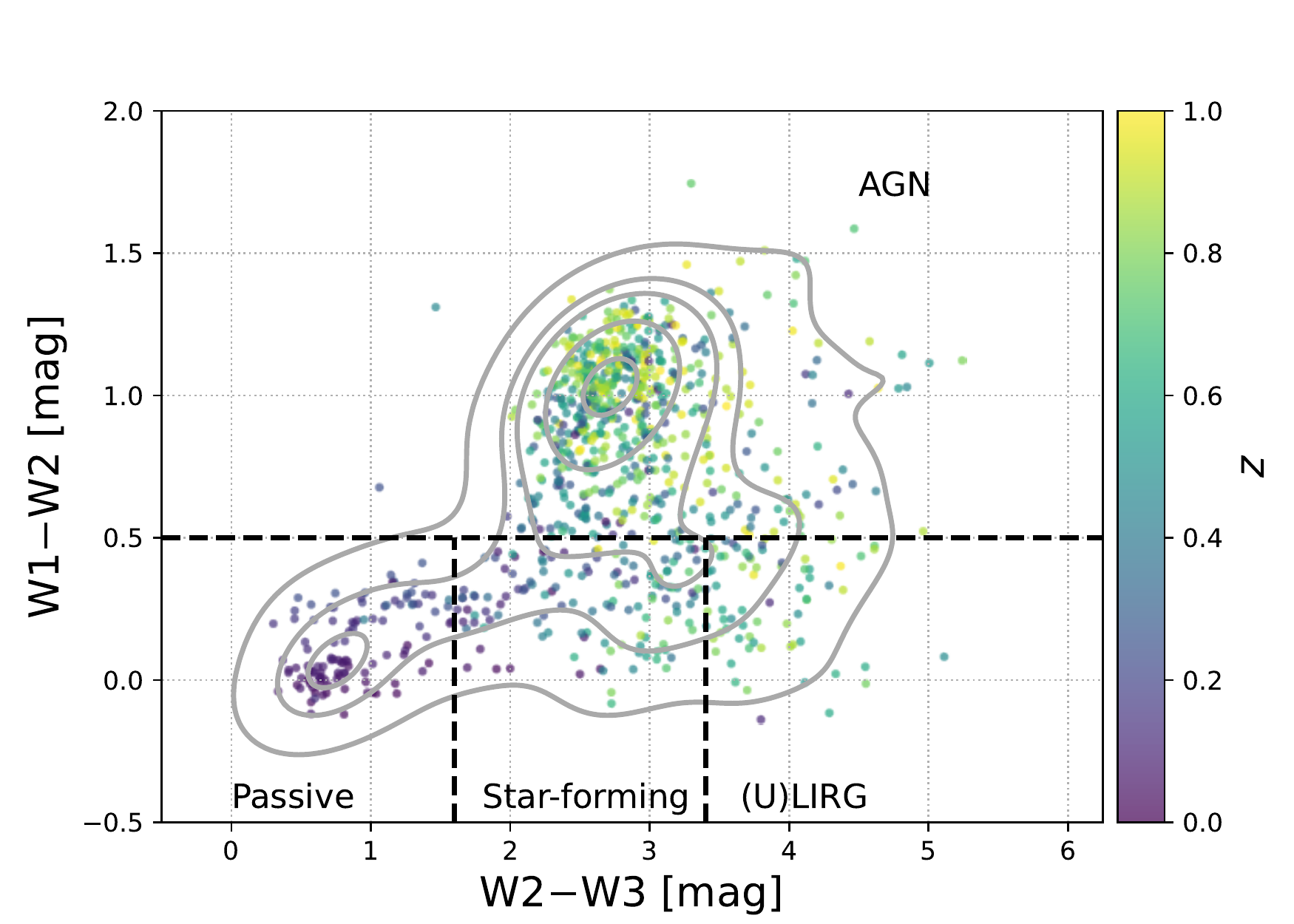}}
    \subfigure[Single-component sources ($\text{LAS}>3''$)]{\includegraphics[trim={0 0 0 1cm}, clip, width=0.99\columnwidth]{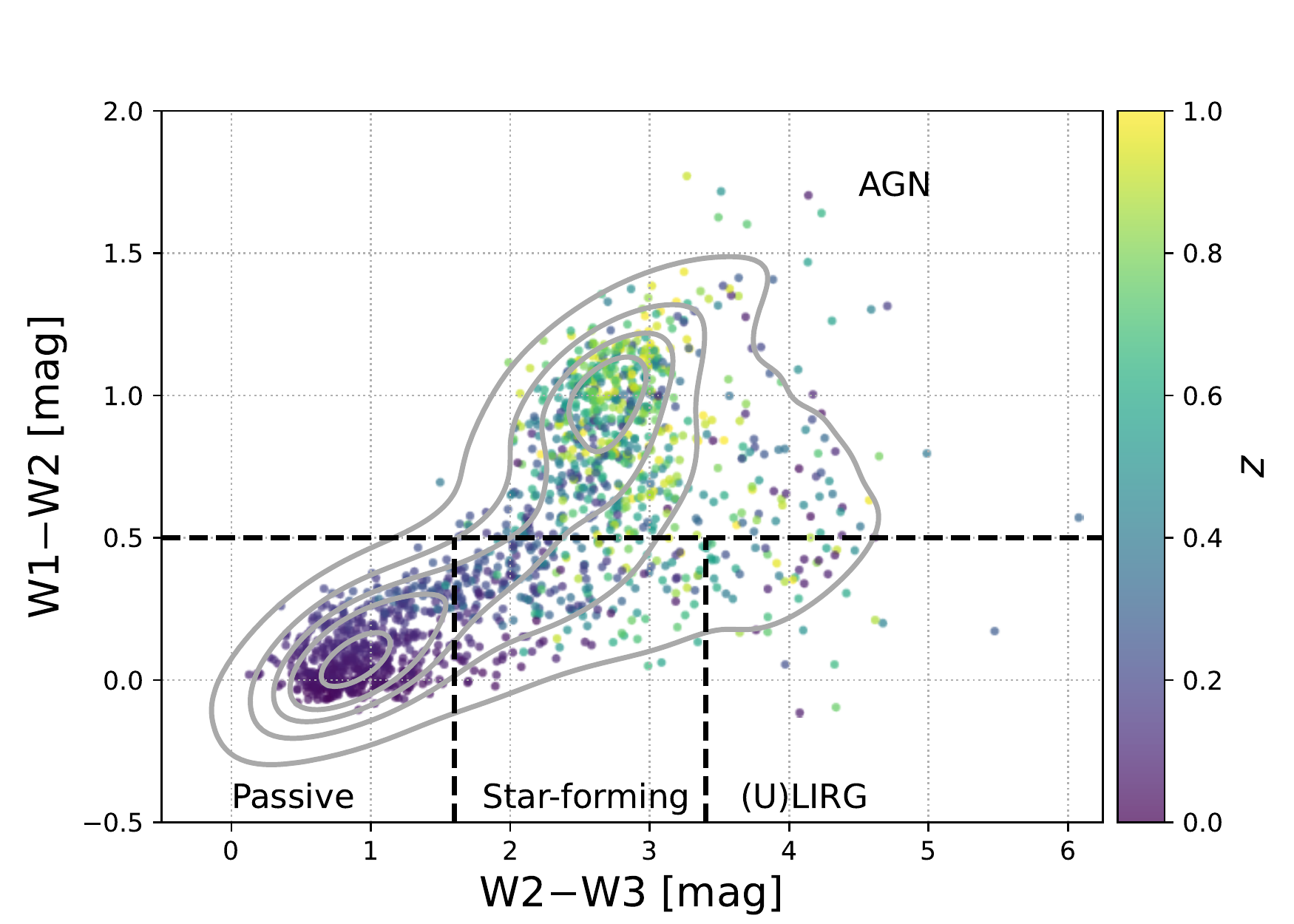}}
    \caption{The WISE color/color distributions for the hosts of DRAGNs (panel a) and extended (LAS$>3''$) single-component radio sources (panel b).
    The contour levels contain $95$, $75$, $50$, $25$ and $5\,$\% of the data points, and all points are colored by the redshift of the host galaxy.
    Only galaxies with $S/N > 3$ in W1, W2 and W3 are included.}
    \label{fig:wisecolors}
\end{figure}

The WISE color-color diagram is most useful at $z<1$ as the WISE bands start to trace different parts of the host galaxy SED at higher redshifts \citep{Donley2012, Assef2013}.
Limiting our sample to those sources at $z<1$, we identify $888$ DRAGNs and $1,422$ single-component radio sources that satisfy our selection criteria.
A further $316$ DRAGNs and $697$ single-component sources have either poor $S/N$ or a lower limit in their W3 magnitude measurement and are not included in this analysis.
We plot the WISE colors of the hosts of these RLAGN in figure \ref{fig:wisecolors}, showing DRAGNs in the upper panel and single-component sources in the lower panel.
For our comparison of the IR colors of extended single-component radio galaxies and DRAGNs, we adopt the \citet{Mingo2016} classification of WISE host galaxies. Broadly, these criteria provide two different diagnostics.
First, whether the IR colors are dominated by the AGN ($\text{W1}-\text{W2} > 0.5$) or the host galaxy ($\text{W1}-\text{W2} < 0.5$).
Second, for those sources where the host galaxy dominates the IR colors, the $\text{W2}-\text{W3}$ color provides information on the host galaxy type:
\begin{itemize}
    \item passive and elliptical galaxies generally have $\text{W2}-\text{W3} < 1.6$,
    \item galaxies with $1.6 < \text{W2}-\text{W3} < 3.4$ are typically disk dominated and have more active star-formation,
    \item and sources where $\text{W2}-\text{W3} > 3.4$ are usually starburst galaxies, often (Ultra) Luminous Infrared Galaxies([U]LIRGs).
\end{itemize}

\begin{figure}
    \centering
    \includegraphics[width=0.99\columnwidth]{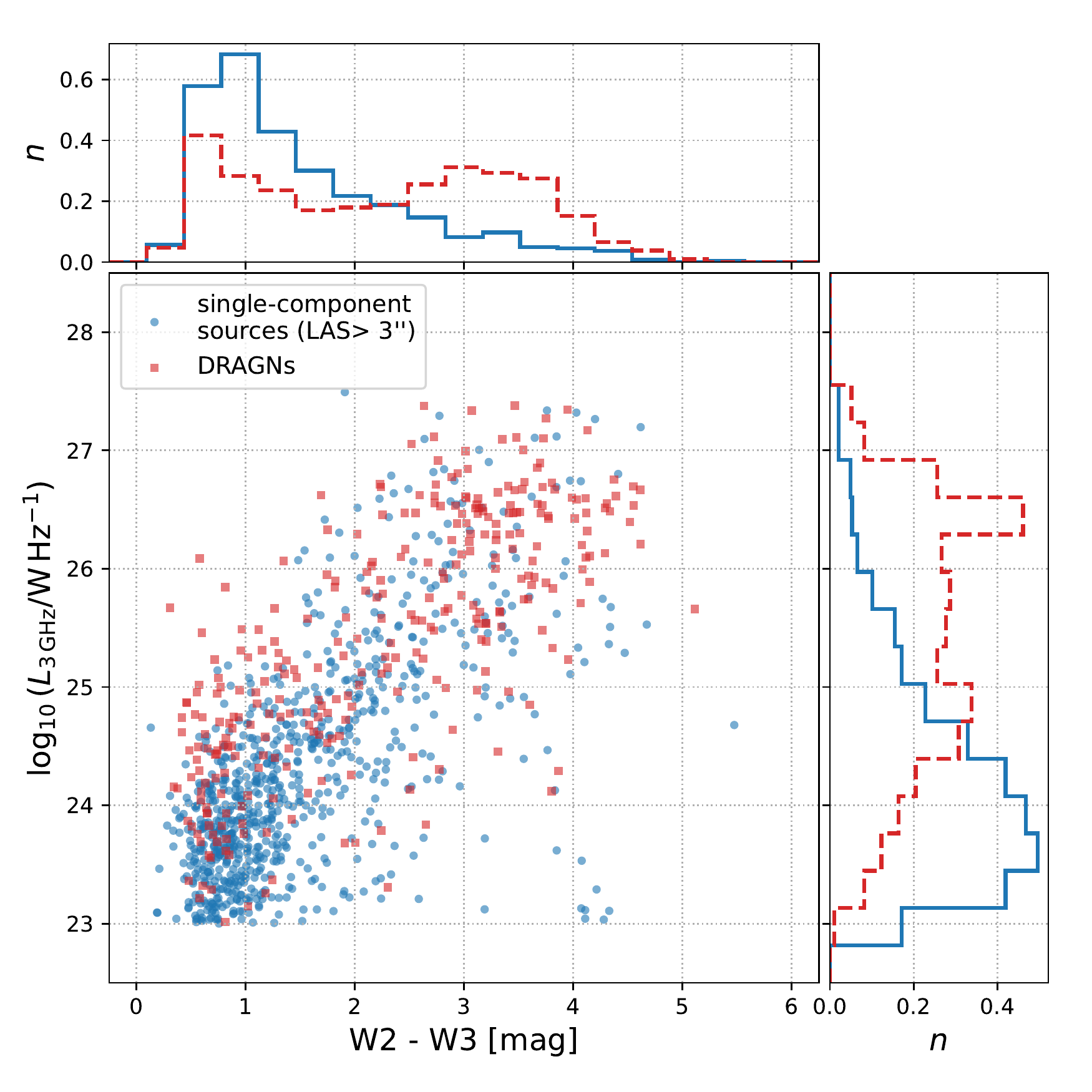}
    \caption{Comparison of the W2$-$W3 colors and radio luminosities of DRAGNs (red) and extended (LAS$>3''$) single-component radio sources (blue) where \text{$\text{W1}-\text{W2}<0.5$}.
    Area normalised histograms of the two distributions are shown along the $x$- and $y$-axes using the same color scheme as the scatter plot.}
    \label{fig:w2-w3vlum}
\end{figure}

Figure \ref{fig:wisecolors} shows some differences between the IR color distributions of the DRAGNs and extended single-component radio sources.
For the DRAGNs, $65.2\pm1.6\,\%$ have hosts with AGN-like colors, $12.3_{-1.0}^{+1.2}\,\%$ are passive, $14.6_{-1.1}^{+1.3}\,\%$ have star-forming colors, and $7.9_{-0.8}^{+1.0}\,\%$ are (U)LIRGs.
This dominance of AGN IR colors with a near even mix of passive and star-forming hosts when the AGN does not dominate the IR closely resembles the WISE colors seen in previous works studying powerful extended radio galaxies \citep[e.g.][]{Gurkan2014, Banfield2015, Mingo2019}.
However, in the case of the single-component sources, when the IR colors are not AGN-like there is a bias towards passive hosts.
Here, $45.3\pm1.3\,\%$ have hosts where the AGN dominates the IR colors,  $34.7_{-1.2}^{+1.3}\,\%$ have passive colors, $16.7_{-0.9}^{+1.0}\,\%$ are star-forming galaxies, and for $3.2_{-0.4}^{+0.5}\,\%$ the hosts have IR colors associated with (U)LIRGs.
The scatter points in both panels of Figure \ref{fig:wisecolors} are colored by redshift, and this shows that the passive/elliptical host galaxies are typically at lower redshift than the star-forming and (U)LIRG hosts.
Recall that in Section \ref{ssec:sizelum} we showed that our single-component sources are typically less luminous than our DRAGNs.
It is therefore prudent to check if the excess of passive/elliptical hosts for the single-component sources is simply the effect of better sampling the low-luminosity population of this sample.
We compare the $\text{W2}-\text{W3}$ colors and $3\,$GHz luminosities for DRAGNs and single component radio sources that have host-dominated IR colors ($\text{W1}-\text{W2}<0.5$) in Figure \ref{fig:w2-w3vlum}.

It is clear from Figure \ref{fig:w2-w3vlum} that the different luminosity distributions of the two radio source samples is the dominant cause of the WISE color differences we observe.
Qualitatively, at $L_{3\,\text{GHz}} \gtrsim 10^{25}$ the single-component sources appear to have a similar $\text{W2}-\text{W3}$ distribution to the DRAGNs.
To test this in a more quantified manner, for each of our DRAGNs we randomly select a single-component source controlled for on both redshift and luminosity.
This is achieved by requiring $\Delta L_{3\,\text{GHZ}} <0.2\,\text{dex}$ and $\Delta z < 0.01$, where $\Delta L_{3\,\text{GHZ}}$ and $\Delta z$ are the difference in radio luminosity and redshift respectively between a DRAGN and a randomly selected single-component source.
We then perform a two-sample Kolmogorov-Smirnov (KS) test on the resultant $\text{W2}-\text{W3}$, distributions which returns a $p$-value of $\approx 0.3$.
Thus, when accounting for luminosity and redshift differences in the DRAGNs and single-component radio sources, we find no statistically significant differences in the $\text{W2}-\text{W3}$ color distributions of DRAGNs and single-component radio sources.

The shift to bluer WISE colors with increasing radio luminosity is likely linked to accretion modes of the AGN.
Radiatively efficient AGN are more often found in galaxies with relatively young stellar populations than radiatively inefficient AGN \citep[e.g.][]{Best2012, Janssen2012, Butler2018, Williams2018, Kondapally2022}.
At low radio luminosities, radiatively efficient AGN only constitute a few percent of RLAGN.
However, at higher luminosities the fraction of AGN that are radiatively efficient increases, and accounts for approximately half of the RLAGN population at $L_{1.4\,\text{GHz}}\sim 10^{25.5}\,\text{W}\,\text{Hz}^{-1}$ \citep{Best2012}.
Moreover, \citet{Mingo2022} show in their Figure 5 that radiatively efficient AGN in their data have blue $\text{W2}-\text{W3}$ colors.
Although we have made no effort to identify our AGN as either radiatively efficient or inefficient in this work, it seems likely that the blue WISE colors of our DRAGNs may be an indicator of their accretion mode.

\section{Triple Source Statistics}
\label{sec:triples}


A third component is present in just over $10\,\%$ of the DRAGNs ($1,836$ objects).
The additional component in these associations provides an opportunity for further analysis, for a couple of key reasons.
First, the presence of a radio core is a robust indicator of the location of the AGN central engine, particularly for those sources without a host ID.
Second, the relatively small number of triple sources means that a post-hoc visual inspection of the triples to remove spurious detections is practical.

Each of these three-component sources is inspected by eye to identify contaminants that are not in fact DRAGNs, $245$ spurious detections are found. 
These are available in a machine readable table, (see Table \ref{tab:triples} for the first five rows), to aid scientists who wish to create reliable samples of triples from the VLASS Epoch 1 \textit{Quick Look} catalog.
However, we do not perform the visual inspection prior to the automated pipeline used to identify host IDs in order to maintain compatibility with future versions of the catalog (e.g., from subsequent VLASS epochs).
In the remainder of this section we present statistics on the basic radio geometry and symmetry of these triple sources.

\begin{deluxetable}{cc}
    \tablecaption{Results of the visual inspection of triple sources identified by \textsc{DRAGNhunter}.
    \label{tab:triples}}
    \tablewidth{0pt}
    \tablehead{\colhead{ Name} & \colhead{Artifact\_flag}}
    \decimalcolnumbers
    \startdata
    J000105.36$-$165940.3 & 1 \\
    J000108.78$-$123309.6 & 0 \\
    J000324.49$+$534446.2 & 0 \\
    J000402.24$+$332009.7 & 0 \\
    J000511.26$-$075558.4 & 0 \\
    ... & ... \\
    ... & ... \\
    \enddata
    \tablecomments{The first five rows are shown here with the full table available in the online version of the journal article.
    Columns: (1) Name of the triple source, (2) flag set 1 if visual inspection shows the triple to be a spurious detection.}
\end{deluxetable}

\subsection{Flux Ratios and Arm Lengths of Radio Lobes}
\label{ssec:fluxsize_symmetry}

From their location on the \textit{P-D} diagram (Figure \ref{fig:PD}), our DRAGNs are likely dominated by FR II morphologies.
This is supported by the number of FR IIs seen when assessing the reliability of our catalog (e.g. see Figure \ref{fig:egdoubles}).
All things being equal, the lobes of FR IIs should have similar brightnesses. 
One key factor that might conflate this from the observer's point of view, is that of relativistic beaming.
In this event, one would expect the brighter lobe to appear closer to the central AGN than the fainter lobe as a result of increased hotspot prominence \citep{Magliocchetti1998, Harwood2020}.

\begin{deluxetable}{lc}
    \tabletypesize{\footnotesize}
    \tablecaption{Key statistics for lobe flux and arm length ratios in DRAGNs with a host ID coincident with a radio core.
    \label{tab:symstats}}
    \tablewidth{0pt}
    \tablehead{
    \colhead{\textbf{Statistic}} & \colhead{\textbf{Value}}
    }
    \startdata
    Spearman rank coefficient ($\rho$) & $-0.12$\\
    $p$-value & $4\times10^{-6}$\\
    flux ratio $68\,\%$ spread & $1/2.2 < S_{1}/S_{2} < 2.2$\\
    flux ratio $95\,\%$ spread & $1/4.9 < S_{1}/S_{2} < 4.9$\\
    arm length ratio $68\,\%$ spread & $1/1.9 < r_{1}/r_{2} < 1.9$\\
    arm length ratio $95\,\%$ spread & $1/3.6 < r_{1}/r_{2} < 3.6$\\
    \enddata
\end{deluxetable}

\begin{figure}
    \centering
    \includegraphics[width=0.99\columnwidth]{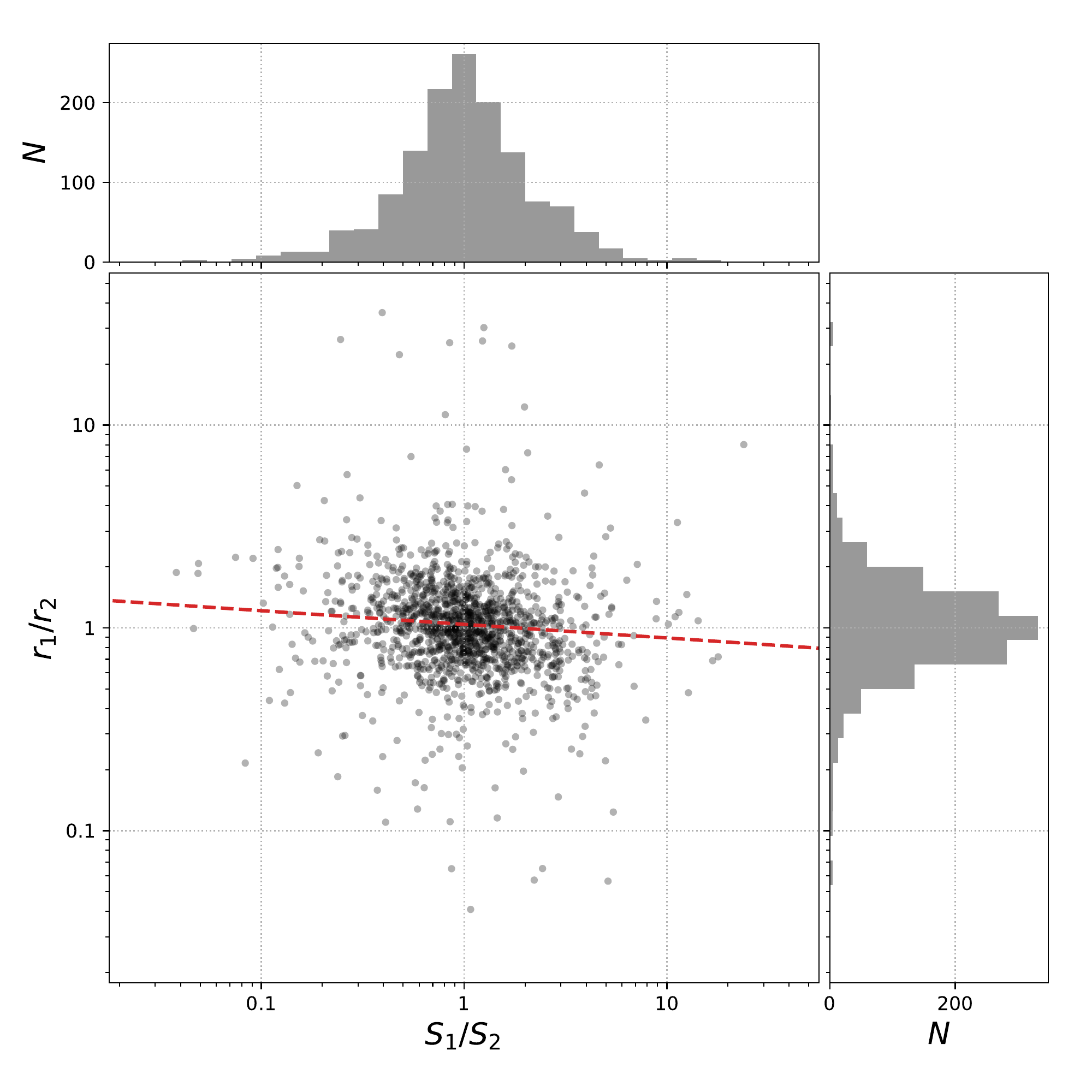}
    \caption{The lobe flux density ratio versus arm-length ratio for triple radio sources.
    Only the $1,383$ with relative errors of less than $10\,\%$ in both $S_{1}/S_{2}$ and $r_{1}/r_{2}$ are included.
    The red dashed line highlights the weak correlation between flux and geometric symmetry of the lobes of DRAGNs.}
    \label{fig:symmetry}
\end{figure}

Each DRAGN has an `arm length' $r_{i}$ from its core to each of its lobes, $i$, that have a flux density, $S_{i}$.
Of our DRAGNs with cores, $1,522$ have a relative error in both arm length ratio, $r_{1}/r_{2}$, and lobe flux ratio, $S_{1}/S_{2}$, of less than $10\,\%$.
For these sources we list key statistics for the distributions $S_{1}/S_{2}$ and $r_{1}/r_{2}$ in Table \ref{tab:symstats}, and plot these variables against each other in Figure \ref{fig:symmetry}.
In our sample, we find a weak but significant correlation between $S_{1}/S_{2}$ and $r_{1}/r_{2}$, with a Spearman rank correlation coefficient of $\rho = -0.12$ and $p$-value of $4\times 10^{-6}$.
We highlight this correlation in Figure \ref{fig:symmetry} with a red dashed line showing a least-squares fit to the data with a slope of $-0.07\pm0.02$.
This correlation shows a weak trend for the brighter component to be closer to the radio core than the fainter component, consistent with the findings of \citet{delaRosa2019}.
The median LAS of this sample is $33''$, and we split our sample into two subsets of small ($\text{LAS}<33''$) and large ($\text{LAS}\geq 33''$) DRAGNs.
Here we find that the correlation holds for large DRAGNs ($\rho = -0.18$, $p=4 \times 10^{-7}$), but is not found for small DRAGNs ($\rho = -0.04$, $p=0.24$).
This likely results from the larger relative uncertainties in arm length measurements for smaller sources.

\subsection{Jet Bending Angles in Triple Sources}

\begin{figure}
    \centering
    \includegraphics[width=\columnwidth]{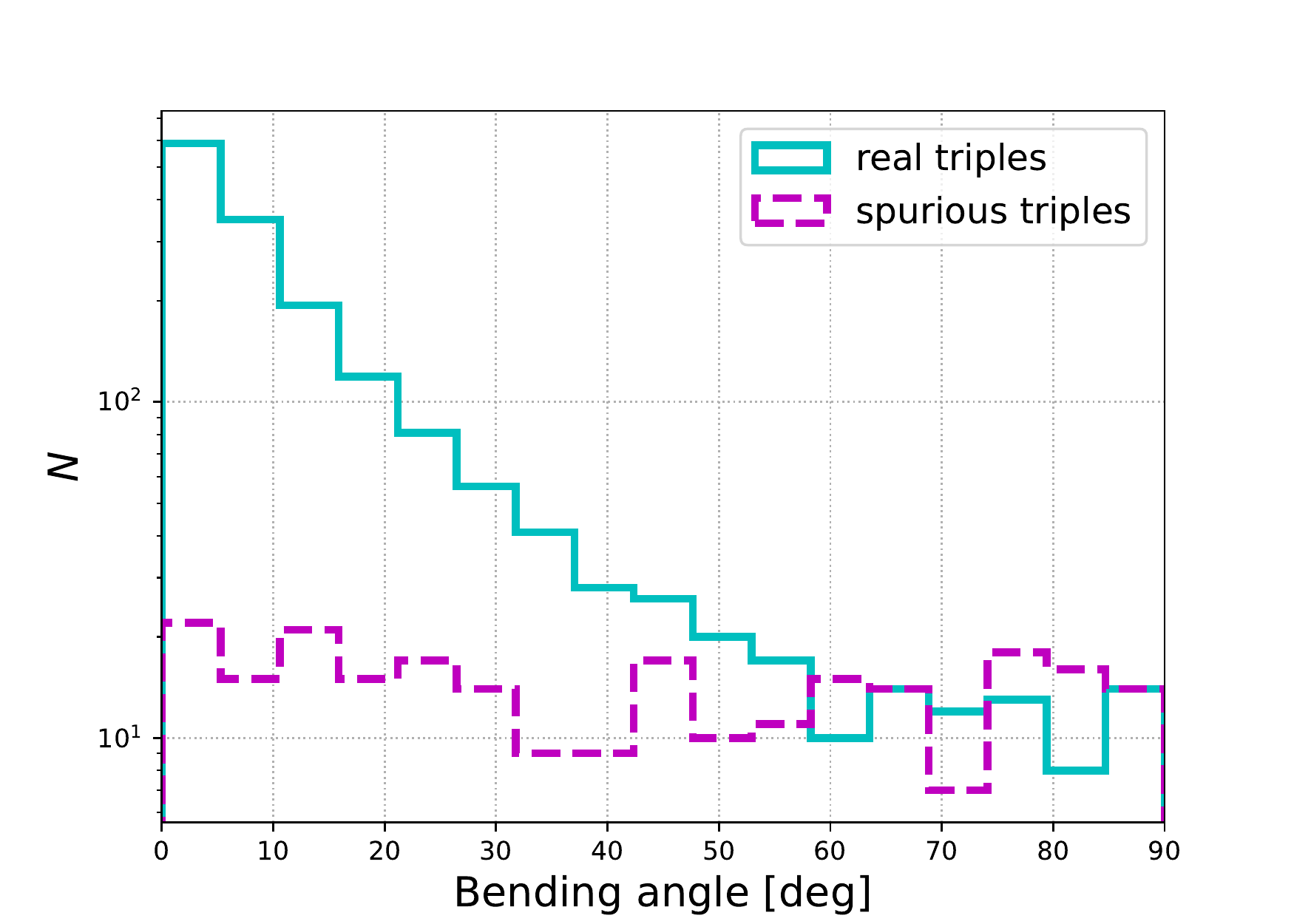}
    \caption{Distribution of bending angle measurements for triple sources.
    Real triple sources are shown by the cyan solid line, with the measurements for spurious triples shown by the magenta dashed line.}
    \label{fig:opening_angles}
\end{figure}

The bending angle of a DRAGN is a measure of its deviation from a perfectly straight geometry.
Using the component positions we measure the bending angles of our triples, which range from 0 to 90 degrees.
A great majority have small bending angles, indicating that most 3-component DRAGNs are straight or modestly bent.
The fraction of contaminants depends heavily upon bending angle, dominating the catalog entries at high bending angles but reaching only a few percent at small bending angles (see Figure \ref{fig:opening_angles}).
This trend can be understood by noting that heavily bent real radio sources are relatively rare, while associations of artifacts in the VLASS quick look images often are distributed over a wide range in azimuth around bright sources. 

In this section we have presented statistics on the flux and armlength symmetry, and jet bending angles of the triple sources identified by \textsc{DRAGNhunter}.
In doing so we have taken no account of the host galaxy environment--an important factor that can impact all of these variables \citep{Hardcastle2013, Hardcastle2014, Garon2019}.
For instance, \citet{Rodman2019} find that shorter lobe extents are found in denser environments based on a small sample of $16$ FR IIs.
The observations of \citet{Rodman2019} are supported by \citet{Yates-Jones2022} who, using numerical simulations, additionally find that dense environments are expected to produce brighter lobes.
A number of studies have shown that bent-jet radio sources are more likely in dense environments with cluster winds acting to distort the morphology of the radio source \citep[e.g.][]{Blanton2000, Garon2019, Moravec2020, Morris2022}.
A follow-up to this work will analyse the relationship between galaxy environment and the bending angle of DRAGNs in VLASS (K. Achong et al, 2023, in prep.)
While taking account of the host galaxy environment is beyond the scope of this work, the statistics that we report here are likely an interesting representation of the \textit{global} population of DRAGNs.

\section{Summary and Future Work}
\label{sec:summary}

We have defined an algorithm, \textsc{DRAGNhunter}, to search for double-lobed radio sources using only survey component catalogs, and used this to construct a catalog of $>17,000$ DRAGNs in VLASS.
This catalog has a reliability of $\approx 89\,\%$, rising to $93.5\,\%$ if a host galaxy is identified, and $>97\,\%$ if selecting those DRAGNs with signal-to-noise in their angular size measurement $> 20$ and a flux density ratio between the two lobes of $< 10$.
Although it is difficult to estimate the completeness of our sample without a `ground truth' catalog of DRAGNs in VLASS, comparisons with FIRST suggest that we identify $\gtrsim 45\,\%$ of DRAGNs with $S_{3\,\text{GHZ}}>20\,$mJy and $\gtrsim 85\,\%$ at $S_{3\,\text{GHZ}}>100\,$mJy.
In addition to identifying the DRAGNs, we have used the likelihood ratio method to identify the probable hosts for more than $70\,\%$ of our DRAGNs.
Complementary to this we identify hosts for more than $230,000$ single-component radio sources.
The catalog of these data will be made publicly available following publication of this paper via the Canadian Initiative for Radio Astronomy Data Analysis (CIRADA)\footnote{\url{www.cirada.ca}}, the CDS VizieR service\footnote{\url{https://vizier.cds.unistra.fr/index.gml}}\citep{Ochsenbein2000} and as machine readable tables in the electronic version of this journal article (see Appendix \ref{apx:data-model}).

The DRAGNs cataloged in this work have properties consistent with being a larger, more powerful extension of the single-component radio galaxy population.
On the radio \textit{P-D} diagram our DRAGNs reside in the region traditionally occupied by FR IIs.
Although no additional morphological classification is attempted in this work, visual inspection of a random sample of these DRAGNs is also suggestive that our catalog consists predominantly of FR IIs.
Exploring the extreme linear size regime of our catalog, we identify $31$ previously undiscovered giant radio galaxies.
The IR colors of the host galaxies of our DRAGNs are found to be consistent with the hosts of single-component sources when accounting for luminosity and redshift.

The VLASS \textit{Quick Look} images used in this work are a rapidly produced data product and are known to have quality limitations.
To enable rapid production, these images are only subject to shallow cleaning, and are not self-calibrated \citep{Lacy2019}.
Consequently, while components are detected down to $S_{\text{peak}}\approx 1\,$mJy/beam in the \textit{Quick Look} images, components fainter than $S_{\text{peak}} \approx 3\,$mJy/beam can suffer from unreliable measurements and a higher than expected contamination from spurious detections \citep{Gordon2021}.
In the future, \textit{Single Epoch} images that are less affected by these limitations will be available for the entire VLASS footprint \citep{Lacy2022}.
For each epoch this will allow catalogs of DRAGNs in VLASS to be produced using components with $S_{\text{peak}}\gtrsim 600\,\mu$Jy/beam, and a three-epoch stack should allow for components down to $S_{\text{peak}}\gtrsim 350\,\mu$Jy/beam to be used \citep{Lacy2020}.
In addition to the added usable depth, the \textit{Single Epoch} images will also provide spectral index information derived from the coefficients of the image Taylor-terms.
These spectral indices can be used in determining whether a component is likely to be a lobe or a core, and may be useful in improving the reliability of the cataloged DRAGNs.

It is our hope that this catalog as-is will prove a useful resource for the astronomical community.
For instance, the large size and high reliability of this catalog make it a potential training set for machine learning algorithms designed to identify DRAGNs.
However, as with all work there is scope for improvement.
Currently DRAGNs are selected by frequentist cuts to the input data.
Thus, a clear direction for improving \textsc{DRAGNhunter} is to take a Bayesian approach to identifying DRAGNs. 
For example, where a component is associated with multiple component-pairs, rather than just taking the closest pairing, a probability of being the correct association can be assigned to each pair based various parameters such as the ratio of component flux densities or the presence of a host galaxy candidate between the components.
This Bayesian philosophy will be the long-term focus for making \textsc{DRAGNhunter} as useful as possible for the coming generation of radio continuum surveys.


 \section*{}\noindent 
The authors thank the anonymous referee for their helpful report.
Y.A.G. is supported by US National Science Foundation (NSF) grant AST 20-09441.
Partial support for L.R. comes from US NSF grant AST 17-14205 to the University of Minnesota.
H.A. benefited from grants CIIC 90/2020, 174/2021, and 138/2022 of Universidad de Guanajuato.
L.K.M. was supported by the Medical Research Council [MR/T042842/1].
C.P.O., S.A.B. and A.N.V. are supported by NSERC, the Natural Sciences and Engineering Research Council of Canada.
K.A. and C.B. acknowledge support from the NSF under Cooperative Agreements No. 1647375 and 1647378, including the Radio Astronomy Data Imaging and Analysis Lab (RADIAL) Research \& Training Experience program.
K.A. and C.B. further acknowledge support from the Alfred P. Sloan Foundation’s Creating Equitable Pathways to STEM Graduate Education program and NSF grant AST 21-50222.
The Canadian Initiative for Radio Astronomy Data Analysis (CIRADA) is funded by a grant from the Canada Foundation for Innovation 2017 Innovation Fund (Project 35999) and by the Provinces of Ontario, British Columbia, Alberta, Manitoba and Quebec, in collaboration with the National Research Council of Canada, the US National Radio Astronomy Observatory and Australia’s Commonwealth Scientific and Industrial Research Organisation.

%

\vspace{5mm}
\facilities{
This work made use of the cross-match service \citep{Boch2012, Pineau2020} and the VizieR catalogue access tool \citep{Ochsenbein2000} provided by CDS, Strasbourg, France.
Additionally used the facilities of the Canadian Astronomy Data Centre (CADC), an organisation operated by the National Research Council of Canada with the support of the Canadian Space Agency.
Observations from the VLA, WISE, SDSS, 2dFGRS, 6dFGS, 2MASS and LS DR8 were used in this work.\\
\indent The VLA is operated by NRAO, a facility of the National Science Foundation operated under cooperative agreement by Associated Universities, Inc.\\
\indent WISE is a joint project of the University of California, Los Angeles, and the Jet Propulsion Laboratory/California Institute of Technology, and NEOWISE, which is a project of the Jet Propulsion Laboratory/California Institute of Technology. WISE and NEOWISE are funded by the National Aeronautics and Space Administration.\\
\indent Funding for the SDSS IV has been provided by the Alfred P. Sloan Foundation, the U.S. Department of Energy Office of Science, and the Participating Institutions.
SDSS acknowledges support and resources from the Center for High-Performance Computing at the University of Utah. The SDSS web site is \url{www.sdss.org}.
SDSS is managed by the Astrophysical Research Consortium for the Participating Institutions of the SDSS Collaboration including the Brazilian Participation Group, the Carnegie Institution for Science, Carnegie Mellon University, Center for Astrophysics | Harvard \& Smithsonian (CfA), the Chilean Participation Group, the French Participation Group, Instituto de Astrofísica de Canarias, The Johns Hopkins University, Kavli Institute for the Physics and Mathematics of the Universe (IPMU) / University of Tokyo, the Korean Participation Group, Lawrence Berkeley National Laboratory, Leibniz Institut für Astrophysik Potsdam (AIP), Max-Planck-Institut für Astronomie (MPIA Heidelberg), Max-Planck-Institut für Astrophysik (MPA Garching), Max-Planck-Institut für Extraterrestrische Physik (MPE), National Astronomical Observatories of China, New Mexico State University, New York University, University of Notre Dame, Observatório Nacional / MCTI, The Ohio State University, Pennsylvania State University, Shanghai Astronomical Observatory, United Kingdom Participation Group, Universidad Nacional Autónoma de México, University of Arizona, University of Colorado Boulder, University of Oxford, University of Portsmouth, University of Utah, University of Virginia, University of Washington, University of Wisconsin, Vanderbilt University, and Yale University.\\
\indent 2MASS is a joint project of the University of Massachusetts and the Infrared Processing and Analysis Center/California Institute of Technology, funded by the National Aeronautics and Space Administration and the National Science Foundation.\\
\indent 
The Legacy Surveys consist of three individual and complementary projects: the Dark Energy Camera Legacy Survey (DECaLS; Proposal ID \#2014B-0404; PIs: David Schlegel and Arjun Dey), the Beijing-Arizona Sky Survey (BASS; NOAO Prop. ID \#2015A-0801; PIs: Zhou Xu and Xiaohui Fan), and the Mayall z-band Legacy Survey (MzLS; Prop. ID \#2016A-0453; PI: Arjun Dey). 
DECaLS, BASS and MzLS together include data obtained, respectively, at the Blanco telescope, Cerro Tololo Inter-American Observatory, NSF’s NOIRLab; the Bok telescope, Steward Observatory, University of Arizona; and the Mayall telescope, Kitt Peak National Observatory, NOIRLab. 
Pipeline processing and analyses of the data were supported by NOIRLab and the Lawrence Berkeley National Laboratory (LBNL). 
The Legacy Surveys project is honored to be permitted to conduct astronomical research on Iolkam Du’ag (Kitt Peak), a mountain with particular significance to the Tohono O’odham Nation.\\
\indent 
NOIRLab is operated by the Association of Universities for Research in Astronomy (AURA) under a cooperative agreement with the National Science Foundation. 
LBNL is managed by the Regents of the University of California under contract to the U.S. Department of Energy.\\
\indent 
This project used data obtained with the Dark Energy Camera (DECam), which was constructed by the Dark Energy Survey (DES) collaboration. 
Funding for the DES Projects has been provided by the U.S. Department of Energy, the U.S. National Science Foundation, the Ministry of Science and Education of Spain, the Science and Technology Facilities Council of the United Kingdom, the Higher Education Funding Council for England, the National Center for Supercomputing Applications at the University of Illinois at Urbana-Champaign, the Kavli Institute of Cosmological Physics at the University of Chicago, Center for Cosmology and Astro-Particle Physics at the Ohio State University, the Mitchell Institute for Fundamental Physics and Astronomy at Texas A\&M University, Financiadora de Estudos e Projetos, Fundacao Carlos Chagas Filho de Amparo, Financiadora de Estudos e Projetos, Fundacao Carlos Chagas Filho de Amparo a Pesquisa do Estado do Rio de Janeiro, Conselho Nacional de Desenvolvimento Cientifico e Tecnologico and the Ministerio da Ciencia, Tecnologia e Inovacao, the Deutsche Forschungsgemeinschaft and the Collaborating Institutions in the Dark Energy Survey. 
The Collaborating Institutions are Argonne National Laboratory, the University of California at Santa Cruz, the University of Cambridge, Centro de Investigaciones Energeticas, Medioambientales y Tecnologicas-Madrid, the University of Chicago, University College London, the DES-Brazil Consortium, the University of Edinburgh, the Eidgenossische Technische Hochschule (ETH) Zurich, Fermi National Accelerator Laboratory, the University of Illinois at Urbana-Champaign, the Institut de Ciencies de l’Espai (IEEC/CSIC), the Institut de Fisica d’Altes Energies, Lawrence Berkeley National Laboratory, the Ludwig Maximilians Universitat Munchen and the associated Excellence Cluster Universe, the University of Michigan, NSF’s NOIRLab, the University of Nottingham, the Ohio State University, the University of Pennsylvania, the University of Portsmouth, SLAC National Accelerator Laboratory, Stanford University, the University of Sussex, and Texas A\&M University.\\
\indent 
BASS is a key project of the Telescope Access Program (TAP), which has been funded by the National Astronomical Observatories of China, the Chinese Academy of Sciences (the Strategic Priority Research Program “The Emergence of Cosmological Structures” Grant \# XDB09000000), and the Special Fund for Astronomy from the Ministry of Finance. 
The BASS is also supported by the External Cooperation Program of Chinese Academy of Sciences (Grant \# 114A11KYSB20160057), and Chinese National Natural Science Foundation (Grant \# 12120101003, \# 11433005).\\
\indent 
The Legacy Survey team makes use of data products from the Near-Earth Object Wide-field Infrared Survey Explorer (NEOWISE), which is a project of the Jet Propulsion Laboratory/California Institute of Technology. 
NEOWISE is funded by the National Aeronautics and Space Administration.\\
\indent 
The Legacy Surveys imaging of the DESI footprint is supported by the Director, Office of Science, Office of High Energy Physics of the U.S. Department of Energy under Contract No. DE-AC02-05CH1123, by the National Energy Research Scientific Computing Center, a DOE Office of Science User Facility under the same contract; and by the U.S. 
National Science Foundation, Division of Astronomical Sciences under Contract No. AST-0950945 to NOAO.}


\software{The work carried out in this paper made use of of the following software packages and tools:
APLpy \citep{Robitaille2012},
AstroPy \citep{Astropy2013, Astropy2018, Astropy2022},
Astroquery \citep{Ginsburg2019},
Matplotlib \citep{Hunter2007},
NumPy \citep{Harris2020},
Pandas \citep{McKinney2010, Pandas2020},
SAOImage DS9 \citep{Joye2003},
SciPy \citep{Virtanen2020},
Seaborn \citep{Waskom2021}
and TOPCAT \citep{Taylor2005}.}



\pagebreak

\appendix
\section{Catalog Data Model}
\label{apx:data-model}
The catalog described within this paper will be released as two separated tables: i) \texttt{Source and Host information}, and ii) \texttt{DRAGN properties}.
The \texttt{Source and host information} table lists main properties of all sources identified (single component and DRAGNs) as well as information on the AllWISE host and its redshift where available.
The two catalog tables share a number of columns beyond what is necessary to enable table joining so as to maximise the standalone utility of each table.
Tables \ref{tab:sources-meta} and \ref{tab:dragns-meta} give the column descriptions for the \texttt{Source and Host information} and \texttt{DRAGN properties} tables respectively.
The full data tables are available in the online version of the journal article, as well as via CIRADA and the CDS VizieR service.
Future versions of this catalog (e.g. using data from later VLASS epochs and \textit{Single Epoch} images) will released by CIRADA.

\begin{deluxetable}{lllr}
    \tabletypesize{\footnotesize}
    \tablecaption{Source \& host information table column descriptions
    \label{tab:sources-meta}}
    \tablewidth{0pt}
    \tablehead{
    \colhead{Column number} & \colhead{Label} & \colhead{Description} & \colhead{units}
    }
    \startdata
    1 & Name\tablenotemark{a} & Name of the source &  \\
    2 & RA &  R.A. of the source & deg \\
    3 & DEC &  Decl. of the source & deg \\
    4 & Flux\tablenotemark{b} &  Total flux density of the source & mJy \\
    5 & E\_Flux &  Uncertainty in \textit{Flux} & mJy \\
    6 & LAS &  Estimate of the Largest Angular Size of the source & arcsec \\
    7 & E\_LAS &  Uncertainty in \textit{LAS} & arcsec \\
    8 & Type\tablenotemark{c} &  Type of source &  \\
    9 & Source\_flag\tablenotemark{d} &  Source quality flag &  \\
    10 & AllWISE & Name of the AllWISE host ID &  \\
    11 & RA\_AllWISE & R.A. of the AllWISE host & deg \\
    12 & DE\_AllWISE & Decl. of the AllWISE host & deg \\
    13 & Sep\_AllWISE & Angular separation between radio source and AllWISE host ID & arcsec \\
    14 & LR & Likelihood ratio of host ID &  \\
    15 & Rel & Probabilty that the host is correct &  \\
    16 & Host\_flag\tablenotemark{e} &  Host ID flag &  \\
    17 & W1mag &  Vega magnitude of AllWISE host in the W1 band & mag \\
    18 & E\_W1mag &  Uncertainty in \textit{W1mag} & mag \\
    19 & W2mag &  Vega magnitude of AllWISE host in the W1 band & mag \\
    20 & E\_W2mag &  Uncertainty in \textit{W2mag} & mag \\
    21 & W3mag &  Vega magnitude of AllWISE host in the W1 band & mag \\
    22 & E\_W3mag &  Uncertainty in \textit{W3mag} & mag \\
    23 & W4mag &  Vega magnitude of AllWISE host in the W1 band & mag \\
    24 & E\_W4mag &  Uncertainty in \textit{W4mag} & mag \\
    25 & z & Host redshift &  \\
    26 & z\_err &  Uncertainty in \textit{z} &  \\
    27 & z\_type &  Redshift type &  \\
    28 & z\_survey &  Survey that the redshift was obtained from &  \\
    \enddata
    \tablecomments{This table contains $595,375$ rows and is provided in machine readable format in the electronic version of this journal article.}
    \tablenotetext{a}{For single-component sources this is the Julian \textit{Component\_name} from the VLASS Quick Look component catalog \citep{Gordon2021} to allow easy joining with that catalog. For DRAGNs the \textit{Name} is a Julian name of the format Jhhmmss.ss$\pm$ddmmss.s.}
    \tablenotetext{b}{This is the sum of the cataloged fluxes of all associated components. The flux-scaling correction of $1/0.87$ recommended in \citet{Gordon2021} has \textbf{not} been applied to these values and is left to the discretion of the end-user.}
    \tablenotetext{c}{`S' is a single-component source; `D' is a DRAGN.}
    \tablenotetext{d}{Set to $1$ if $\text{Type}=\text{`D'}$ and \textbf{either} $\text{Lobe\_flux\_ratio} < 0.1$ \textbf{or} $\text{Lobe\_flux\_ratio} > 10$ \textbf{or} $\text{LAS}/\text{E\_LAS}<20$.
    For all other sources this flag is set to $0$.}
    \tablenotetext{e}{Set to $-2$ if the LR identified host of a DRAGN is co-located with a radio core, $-1$ if the LR identified host of a DRAGN has been replaced by a host candidate coincident with a core, $0$ for single component-sources and DRAGNs without a radio core, and $1$ for DRAGNs with a radio core that is not co-located with a host candidate.}
\end{deluxetable}

\begin{deluxetable}{lllr}
    \tabletypesize{\footnotesize}
    \tablecaption{DRAGN properties table column descriptions
    \label{tab:dragns-meta}}
    \tablewidth{0pt}
    \tablehead{
    \colhead{Column number} & \colhead{Label} & \colhead{Description} & \colhead{units}
    }
    \startdata
    1 & Name & Julian name of the source &  \\
    2 & RA &  R.A. of the source & deg \\
    3 & DEC &  Decl. of the source & deg \\
    4 & Flux\tablenotemark{b} &  Total flux density of the source & mJy \\
    5 & E\_Flux &  Uncertainty in \textit{Flux} & mJy \\
    6 & Core\_prom &  Fraction of \textit{Flux} associated with \textit{Core} &  \\
    7 & E\_Core\_prom &  Uncertainty in \textit{Core\_prom} &  \\
    8 & Lobe\_flux\_ratio &  Ratio of the flux from \textit{Lobe\_1} to the flux from \textit{Lobe\_2} &  \\
    9 & E\_Lobe\_flux\_ratio &  Uncertainty in \textit{Lobe\_flux\_ratio} &  \\
    10 & LAS &  Estimate of the Largest Angular Size of the source & arcsec \\
    11 & E\_LAS &  Uncertainty in \textit{LAS} & arcsec \\
    12 & Misalign\_1\tablenotemark{f} &  Relative misalignment of \textit{Lobe\_1} & deg \\
    13 &  E\_Misalign\_1 &  Uncertainty in \textit{Misalign\_1} & deg \\
    14 & Misalign\_2\tablenotemark{f} &  Relative misalignment of \textit{Lobe\_2} & deg \\
    15 & E\_Misalign\_2 &  Uncertainty in \textit{Misalign\_2} & deg \\
    16 & Mean\_misalign &  Mean value of \textit{Misalign\_1} and \textit{Misalign\_2} & deg \\
    17 & E\_Mean\_misalign &  Uncertainty in \textit{Mean\_misalign} & deg \\
    18 & Lobe\_1 &  Component name of \textit{Lobe\_1} &  \\
    19 & Lobe\_2 &  Component name of \textit{Lobe\_2} &  \\
    20 & Core &  Component name of \textit{Core} if identified &  \\
    21 & RA\_core &  R.A. of \textit{Core} & deg \\
    22 & DEC\_core &  Decl. of \textit{Core} & deg \\
    23 & RA\_median &  Median R.A. of two lobes & deg \\
    24 & DEC\_median &  Median Decl. of two lobes & deg \\
    25 & RA\_fw &  Flux-weighted central R.A. of two lobes & deg \\
    26 & DEC\_fw &  Flux-weighted central Decl. of two lobes & deg \\
    27 & Source\_flag\tablenotemark{d} &  Source quality flag &  \\
    28 & AllWISE & Name of the AllWISE host ID &  \\
    29 & Sep\_AllWISE & Angular separation between radio source and AllWISE host ID & arcsec \\
    30 & LR & Likelihood ratio of host ID &  \\
    31 & Rel & Probabilty that the host is correct &  \\
    32 & Host\_flag\tablenotemark{e} &  Host ID flag &  \\
    \enddata
    \tablecomments{This table contains $17,724$ rows and is provided in machine readable format in the electronic version of this journal article.}
    \tablenotetext{b}{This is the sum of the cataloged fluxes of all associated components. The flux-scaling correction of $1/0.87$ recommended in \citet{Gordon2021} has \textbf{not} been applied to these values and is left at the discretion of the end-user.}
    \tablenotetext{d}{Set to $1$ if  \textbf{either} $\text{Lobe\_flux\_ratio} < 0.1$ \textbf{or} $\text{Lobe\_flux\_ratio} > 10$ \textbf{or} $\text{LAS}/\text{E\_LAS}<20$.
    For all other sources this flag is set to $0$.}
    \tablenotetext{e}{Set to $-2$ if the LR identified host of a DRAGN is co-located with a radio core, $-1$ if the LR identified host of a DRAGN has been replaced by a host candidate coincident with a core, $0$ for single component-sources and DRAGNs without a radio core, and $1$ for DRAGNs with a radio core that is not co-located with a host candidate.}
    \tablenotetext{f}{Components with low aspect ratios (nearly circular geometry) can have large uncertainties in their measured misalignments.
    Users are advised to make use of the appropriate uncertainty measurements provided (E\_Misalign\_n) if using these values.}
    
\end{deluxetable}




\newpage
\clearpage
\newpage

\bibliography{VLASS_doubles.bib}{}
\bibliographystyle{aasjournal}



\end{document}